\documentclass{article}
\usepackage{graphicx}
\usepackage{authblk}
\usepackage{verbatim}
\usepackage{array, tabularx}
\usepackage{subfigure}

\usepackage[a4paper, total={6in, 8in}]{geometry}
\usepackage{xr}
\usepackage[utf8]{inputenc}
\usepackage[T1]{fontenc}
\usepackage{amsthm}
\usepackage{amsmath}
\usepackage{amssymb}
\usepackage{siunitx}

\usepackage{url}
\usepackage{booktabs} 
\usepackage{xspace}
\usepackage{float}
\usepackage{multirow}
\usepackage{alltt}
\usepackage{color}
\usepackage{pifont}
\usepackage{makecell}
\usepackage{multicol}
\usepackage[super]{nth}

\newcommand\cparagraph[1]{\vspace{0.6mm}\noindent\textbf{#1.}}

\let\origref\ref
\def\ref#1{\textnormal{\origref{#1}}}

\newcommand{\cmark}{\ding{51}}

\title{Upgrading Democracies with Fairer Voting Methods}

\author[1]{Evangelos Pournaras}
\author[1]{Srijoni Majumdar}
\author[1]{Thomas Wellings}
\author[2]{Joshua C. Yang}
\author[1,3]{Fatemeh B. Heravan}
\author[4]{Regula H{\"a}nggli Fricker}
\author[2,5]{Dirk Helbing}

\affil[1]{School of Computer Science, University of Leeds, Leeds, UK, E-mails: \{E.Pournaras,S.Majumdar,T.Wellings\}@leeds.ac.uk}

\affil[2]{Computational Social Science, ETH Zurich, Zurich, Switzerland, E-mails: \{joyang,dhelbing\}@ethz.ch}

\affil[3]{Department of Computer Science, University of Huddersfield, Huddersfield, UK, E-mail: f.banaieheravan@hud.ac.uk}

\affil[4]{Department of Communication and Media Research, University of Fribourg, Fribourg, Switzerland, E-mail: regula.haenggli@unifr.ch}

\affil[5]{Complexity Science Hub Vienna, Metternichgasse 8, 1030 Vienna, Austria}
\date{}
\begin{document}
	\maketitle
	
	\begin{abstract}
		\footnotetext[1]{Corresponding author: Evangelos Pournaras, School of Computing, University of Leeds, Leeds, UK, E-mail: e.pournaras@leeds.ac.uk}
		Voting methods are instrumental design elements of democracies. Citizens use them to express and aggregate their preferences to reach a collective decision. However, voting outcomes can be as sensitive to voting rules as they are to people's voting choices. Despite significance and interdisciplinary scientific progress, several democracies keep relying on outdated voting methods that do not fit modern, pluralistic societies well, while lacking social innovation. Here, we demonstrate how one can upgrade real-world democracies, namely by using alternative preferential voting methods such as cumulative voting and the method of equal shares designed for a proportional representation of voters' preferences. We rigorously evaluate the striking voting outcomes of these fair voting methods in a new participatory budgeting approach applied in the city of Aarau, Switzerland, including past and follow-up evidence. Results show more winning projects with the same budget. They also show broader geographic and preference representation of citizens by the elected projects, in particular for voters who used to be under-represented. We provide causal evidence showing that citizens prefer proportional voting methods, which possess strong legitimacy without the need of very specialized technical explanations. We also reveal strong underlying democratic values exhibited by citizens who support fair voting methods such as altruism and compromise. These findings come with the momentum to unleash a new and long-awaited participation blueprint of how to upgrade democracies globally. 
	\end{abstract}
	
	\section{Introduction}\label{sec:introduction}
	
	\cparagraph{Reaching collective decisions that benefit the broader society often remains an unfulfilled promise} While voting methods have been instrumental to govern societies with citizens' participation, they are often based on outdated approaches to foster collective intelligence~\cite{Kearns2009,Arnold2017,Mann2017}. This may put democracies at risk instead of empowering them~\cite{Emerson2020}. Standard majoritarian voting methods, for example, over-prioritize popular preferences over those of under-represented groups. Disregarding alternative voices and their value often results in lost opportunities for all. Over-relying on limited-choice majoritarian voting frequently results in promoting competition instead of cooperation, in marginalization of minorities instead of inclusion, in few winners and many losers~\cite{Santos2024,Emerson2020,Pournaras2020,Helbing2023,Arnold2017,Mann2017,Hanggli2021}. Polarization, exacerbation of inequalities, and radicalization of the public sphere~\cite{Lorenz2023,Bruter2017} are prominent implications of voting methods that fail to foster fairness by design. Therefore, the lack of fair, proportional representation of the different preferences in a society is a persistent setback to address.
	
	\cparagraph{The challenge of adopting democratic innovations} Proportional voting methods and collective decision making have attracted a broad interest by interdisciplinary research communities~\cite{Pukelsheim2017,Lewis2018,Fisher2018}. This includes the fields of social choice, computational social science, economics, complexity theory, and political science. Nevertheless, adopting these methods in practice and changing the status quo is itself a long-standing and multi-faceted challenge~\cite{Madhavan2017}. It often comes as a `chicken and egg' problem, (i) with difficulties in communicating alternative, fairer voting methods effectively and (ii) with a lack of evidence of how they function in the real world, which mutually enforce each other. This causes risk aversion at the political level and limited will for change, which further maintains the lack of evidence, also with regard to the legitimacy perception of voters~\cite{Cho2018,Park2023,Noveck2017,Arnold2017,Bruter2017}. Such unprecedented complexity for democratic innovations requires rigorous scientific investigation~\cite{Noveck2017,Madhavan2017,Bruter2017,Eliassi2020}. A blueprint for shifting to fairer voting methods is urgently needed. 
	
	\cparagraph{Participatory budgeting as a blueprint for democratic innovations} In recent years, there has been quite a number of noteworthy ``\emph{democratic innovations}". These include citizen assemblies, government labs (``GovLabs"), open innovation (open access/data/source, creative commons), crowd sourcing/sensing/funding, hackathons, fablabs, makerspaces, and citizen science. While related to some of these recent trends, our article focuses on novel interventions in participatory budgeting and their broader impact. This is a process designed to distribute some public budget of a city council in a bottom-up way by engaging citizens in collaborative brainstorming sessions to eventually propose their own projects, solutions, and ideas, on which they later vote~\cite{Fairstein2023}. Citizens may also be involved in the implementation of the selected projects. 
	
	\cparagraph{How participatory budgeting generalizes} Participatory budgeting offers a great potential for demonstrating grassroots democratic innovations in the real world, which can be a blueprint that can be scaled up and wide later in other (collective) decision-making processes. For instance, the selection of projects to counter environmental problems, the distribution of research funds, the use of pensions funds, and entrepreneurial or stock market investments are all timely and pertinent budgeting processes that can benefit remarkably from budget-aware voting methods and, hence, from democratic innovations. Initiated in Porto Alegre back in 1989, participatory budgeting has since been adopted in many places across the globe, from remote communities in Kenya, Indonesia, and the Philippines to major cities such as Barcelona, Warsaw, Helsinki, Paris, and New York~\cite{Wampler2021,Dias2018}. Its uptake has been particularly significant in Europe, with a surge of participatory initiatives in recent years~\cite{Fairstein2023}. By 2019, Europe accounted for more than half of all participatory budgeting schemes worldwide, with more than 5,000 initiatives~\cite{Dias2018}. Despite the profound importance of voting methods for the quality of the outcomes~\cite{Arrow1964}, the impact of fairer voting methods has, so far, been under-explored.
	
	\cparagraph{Collective choice made fairer} We demonstrate a breakthrough for democracies by applying alternative voting methods designed for a more expressive and proportional representation of citizens in real-world settings. This has been achieved with a novel approach for field-testing and systematic assessing a new participatory budgeting process in the city of Aarau, Switzerland, called ``Stadtidee'' (``City Idea''). The participatory budgeting process there was so successful that it is planned to repeat. It was also followed by similar implementations elsewhere (e.g. in Poland), which support our innovative approach and conclusions. The outcomes in Aarau are particularly stunning as compared to a similar ``Stadtidee'' format in Zurich, where the conventional majoritarian voting method was applied (see Section~\ref{subsec:generalizing}). Therefore, our field test is documented as a methodological breakthrough that is of real-world political relevance. 
	
	We used the ballot format of \emph{``Cumulative Voting''}~\cite{Cooper2007,Skowron2025} allowing voters to distribute 10 points over at least 3 out of 33 project ideas. Compared to other prominent input voting methods such as approval, ranking, or knapsack voting~\cite{Goel2019}, cumulative voting enables voters to express their preferences more effectively by weighting their support for different options. This expression of preferences is more inclusive and can provide a more proportional representation of minority groups. Complementary, we used the \emph{``Method of Equal Shares''}~\cite{Peters2021} to aggregate the points allocated by citizens and elect the winning project ideas. This voting method promotes a more proportional representation in comparison with majoritarian voting. Proportionality is the favorable aspect of fairness we study in this article. It avoids an undesirable feature of majoritarian voting approaches, which can sometimes lead to a "tyranny of the majority", where "the winner takes it all", i.e. monopolizes all the budget~\cite{Nyirkos2018}. In contrast, proportional representation is achieved by assigning to each voter the same decision power, which is given by an equal share of the budget to be distributed. This manifests both, in a better reflection of voter preferences and in a fairer geographical distribution of winning projects~\cite{Arnold2017}. In practice, the method of equal shares is resourceful, as it is likely to select a larger number of projects, which are cost-efficient, while satisfying the preferences of more people and increasing overall voter satisfaction~\cite{Maharjan2024,Yang2024} as shown in Section~\ref{subsec:voting-outcomes}. Given the principle of proportionality that we applied, we refer to these alternative decision-making approaches here as ``\emph{fair voting methods}''. As it turns out, they also promote legitimacy and democratic values (see Section~\ref{subsec:legitimacy-gain} and~\ref{subsec:explaining-factors} respectively). Our study provides universal findings, beyond Aarau, as we also study a large number of past participatory budgeting processes and recent follow-ups adopting fair voting methods. 
	
	\cparagraph{Overcoming barriers of shifting to fair voting methods} Combining cumulative voting (for casting the vote) and equal shares (for aggregating the points and electing the winning projects) in a real-world participatory budgeting process allows us to unravel: 
	\begin{itemize}
		\item[(i)] how to systematically adopt fair voting methods; apparently, for policy-makers and a general public familiar with elections, the method of equal shares is pretty different from the widespread and longstanding standard method of utilitarian ``greedy'' (i.e., majoritarian) voting, which simply sums up the votes and selects the most popular project ideas fitting the budget in a majoritarian way; 
		\item[(ii)] how the method of equal shares influences participation in the project ideation phase, allowing for the consideration of additional criteria during the voting process; importantly, project costs become instrumental, as more expensive projects require broader support, which may discourage expensive projects; this allows for a larger number of more resource-efficient projects which are fairer distributed in space~\cite{Maharjan2024}; 
		\item[(iii)] the legitimacy of the voting outcomes; although the method of equal shares is designed to promote fairness, sacrificing popular expensive projects may be perceived as unfair by their supporters, who may be a considerable share of the population; whether the selection of more alternative projects can compensate for the potential dissatisfaction is key for the perceived legitimacy of fair voting methods; 
		\item[(iv)] causal factors that promote a shift to fair voting methods; the crucial question is, whether there are any personal traits or societal values that support and accelerate democratic upgrades based on fair voting methods. 
	\end{itemize}
	
	\cparagraph{A systematic and rigorous approach to upgrade democracies with fair voting methods} We co-designed the participatory budgeting process called ``Stadtidee'' (``City Idea'') in collaboration with the city of Aarau, Switzerland. The process was implemented in the year 2023.  It was characterized by the following innovations (see Section~\ref{sec:campaign-design} for more information): 
	\begin{itemize}
		\item[(i)] a real-world application of the method of equal shares in combination with cumulative voting as the ballot format; combined with each other, these methods enable a better, proportional representation of citizen preferences; 
		\item[(ii)] an inclusive participation fostered throughout the entire process; the citizens proposed 161 ideas, out of which 33 were voted on by 1703 citizens, women in majority and children included, with 17 projects selected for implementation~\cite{CityIdeaReport2025}, see Table~\ref{table:field-test} for an outline; 
		\item[(iii)] a survey study was conducted before and after voting with a control condition; this allowed us to rigorously assess the new process in depth over its entire lifetime; we were also able to assess the legitimacy of fair voting methods before and after voting and to profoundly understand the dynamics of voting outcomes that these methods generate, including causal and explanatory factors; 
		\item[(iv)] a secure and privacy-preserving voting using and extending the Stanford open-source platform for participatory budgeting~\cite{SPBStadtidee2023,Wellings2023}; this ensures transparency and trust when using fair voting methods. 
	\end{itemize}
	
	\cparagraph{A comprehensive spatio-temporal evidence} To confirm and generalize the qualities of voting outcomes reached by fair voting methods, we highlight the continuum of evidence we provide: 
	
	\begin{itemize}
		\item[(i)] \textbf{Local comparison}: The actual use of equal shares in Aarau vs. the hypothetical use of the standard majoritarian method in Aarau (shown in Section~\ref{subsec:voting-outcomes});
		\item[(ii)] \textbf{Global comparison}: The actual use of equal shares in Aarau vs. the actual use of the standard majoritarian method in past elections (shown in Section~\ref{subsec:generalizing});
		\item[(iii)] \textbf{Confirmation on earlier evidence}: The hypothetical use of equal shares in past elections vs. the actual use of the standard majoritarian method in past elections (shown in Section~\ref{subsec:generalizing});
		\item[(iv)] \textbf{Confirmation on follow-up evidence}: The actual use of equal shares in Aarau vs. the actual use of equal shares in follow-up elections of Aarau (shown in Section~\ref{subsec:generalizing});
	\end{itemize}
	
	These four aforementioned innovations, in combination with comprehensive spatio-temporal comparisons with earlier and follow up real-world elections, enabled a rigorous test of fair voting methods and offers a strong foundation for the validity of our conclusions. This provides a promising blueprint for innovative upgrades of democracies in the future.

	\section{Results}\label{sec:results}
	
	This article presents three key results: 
	
	\begin{itemize}
		\item[(1)] Fair ballot aggregation results in voting outcomes with more winners, while achieving better representation of voter preferences and diversity of project ideas, also in space (Section~\ref{subsec:voting-outcomes} and~\ref{subsec:generalizing}). 
		
		\item[(2)] Voters find the adopted ballot aggregation method of equal shares more legitimate, preferring it over the majoritarian standard method and finding it fairer (Section~\ref{subsec:legitimacy-gain}). 
		
		\item[(3)] The shift to a fairer voting method benefits from democratic values and promotes them. In fact, democratic values such as altruism, support for more expressive and inclusive ballot formats, and opposition to costly popular projects causally \textit{explain} the preference for more proportional voting methods (Section~\ref{subsec:explaining-factors}). 
	\end{itemize}

	\subsection{Fair voting methods show multi-faceted qualities}\label{subsec:voting-outcomes}
	
\begin{figure}[!htb]
		\centering
		\includegraphics[width=1.0\textwidth]{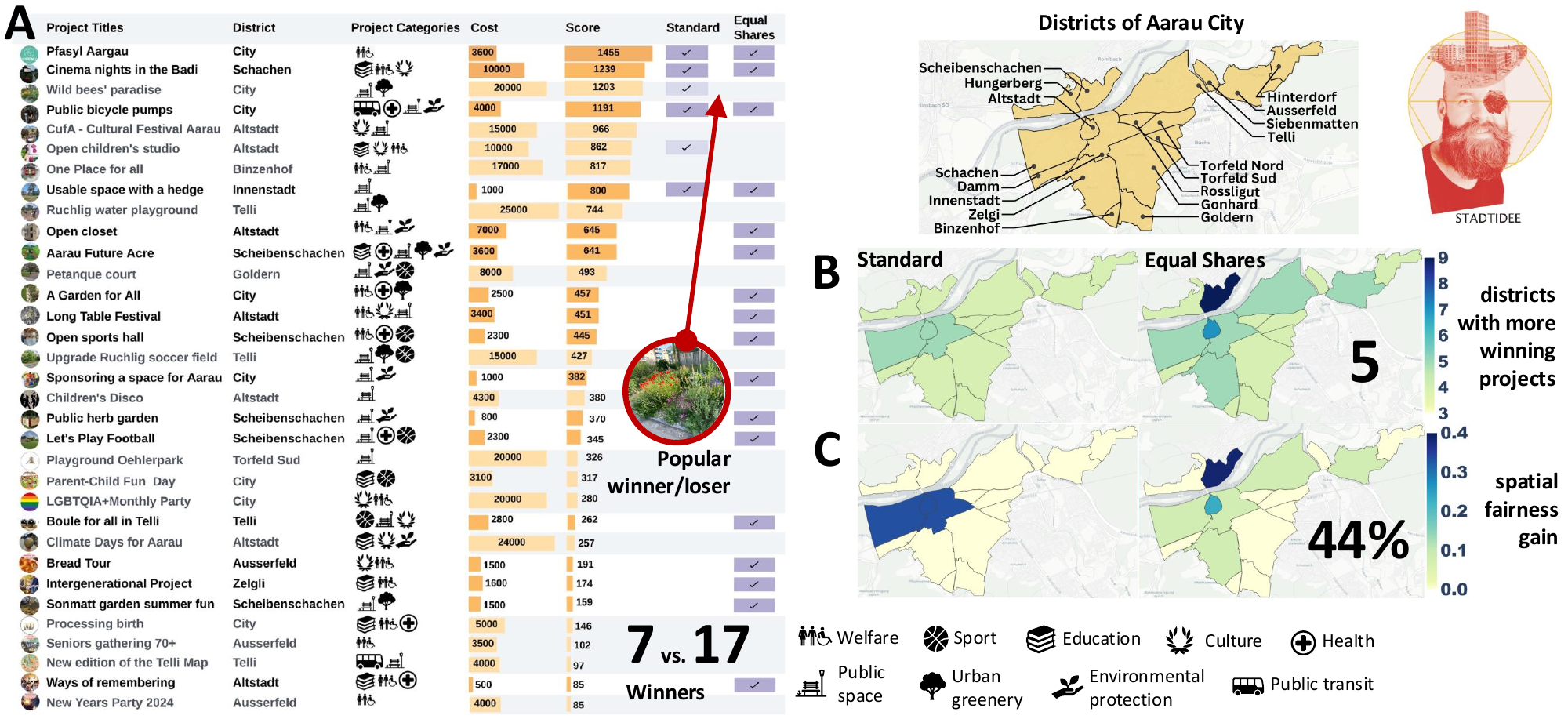}
		\caption{\textbf{Compared to the utilitarian greedy ("Standard") method, which is majoritarian in nature, the number of winning projects resulting from the method of equal shares (see column "Equal Shares") increases dramatically, with more than half of the proposed projects winning, while the citizens' proposed project ideas are better represented geographically.} (A) Equal shares (s)elects 17 winning projects as compared to 7 with the majoritarian standard method. The project `Wild bees' paradise', requiring 40\% of the total budget, is popular winner under the standard method, but a loser under equal shares. This allows more low-cost projects to win, which, in combination, represent the voter preferences better. (B) Under equal shares, five districts receive more winning projects as compared to the standard majoritarian method, thereby strengthening the periphery of the city. (C) Spatial fairness increases by 44\% under equal shares. This is measured by the relative reduction in the index of dispersion for the number of winning projects relative to the proposed ones for each district.}
		\label{fig:results}
	\end{figure}

	\cparagraph{Why the method of equal shares results in more winners}
	Figure~\ref{fig:results}a shows the 33 projects proposed for voting (very left column) and the related voting results (right columns under "Standard" and "Equal Shares"). The following key observations can be made:
	\begin{itemize}
		\item[(i)] Equal shares selects 17 winning projects, with more than half of the proposed projects winning; in contrast, the standard majoritarian utilitarian greedy method would select 7 projects only, which would imply 26 losing projects, thereby creating a significant deficit in representing the broad range of citizen preferences; 
		\item[(ii)] The project entitled ``Wild bees' paradise'' is an example of an expensive popular loser using the method of equal shares, which would win under the standard majoritarian method; with 33.64\% of voters selecting this project, it is the 3rd most popular project, but it would consume 40\% of the overall available budget. Similarly, the project``Open Children Studio" is the other expensive popular loser consuming 18.2\% of the budget. The method of equal shares selects several other, lower-cost projects instead, each of which may be less popular, but together they better address the preferences of a broader population: 75\% representation compared to 50\% that ``Wild bees' paradise'' and ``Open Children Studio" achieve together. 
		\item[(iii)] The mean budget share of the proposed projects is 14.33\%, which is 48.37\% lower than a typical proposed project within 345 past participatory budgeting processes conducted with the standard majoritarian method~\cite{pabulib}; this illustrates the impact of ballot aggregation design already in the project ideation phase: it is more resourceful by encouraging lower-cost projects of different and often novel nature. This has the potential for synergetic impact creation due to the greater variety of cost-effective projects~\cite{Maharjan2024}.
	\end{itemize}
	
	\cparagraph{Equal shares improves the voters' geographic representation, especially for city peripheries} Figure~\ref{fig:results}b illustrates that the method of equal shares results in 5 districts (27.77\%) with more winning projects compared to the standard majoritarian method, while no district receives a lower number of projects. This particularly benefits the periphery of the city that would otherwise be under-represented. Moreover, Figure~\ref{fig:results}c measures the relative number of winning as compared to the number of proposed projects in each district. Using the method of equal shares, the spatial fairness increases by 44\% when measured by the reduction in the index of dispersion, which demonstrates that more proposed project win in each district. These results confirm that the periphery of Aarau is better represented by the method of equal shares, both in terms of the voter preferences and the number of proposed project ideas selected. 
	
	\subsection{Generalizing fair voting methods as a distinguishing paradigm}\label{subsec:generalizing}
	
\begin{figure}[!htb]
		\centering
		\includegraphics[width=1.0\textwidth]{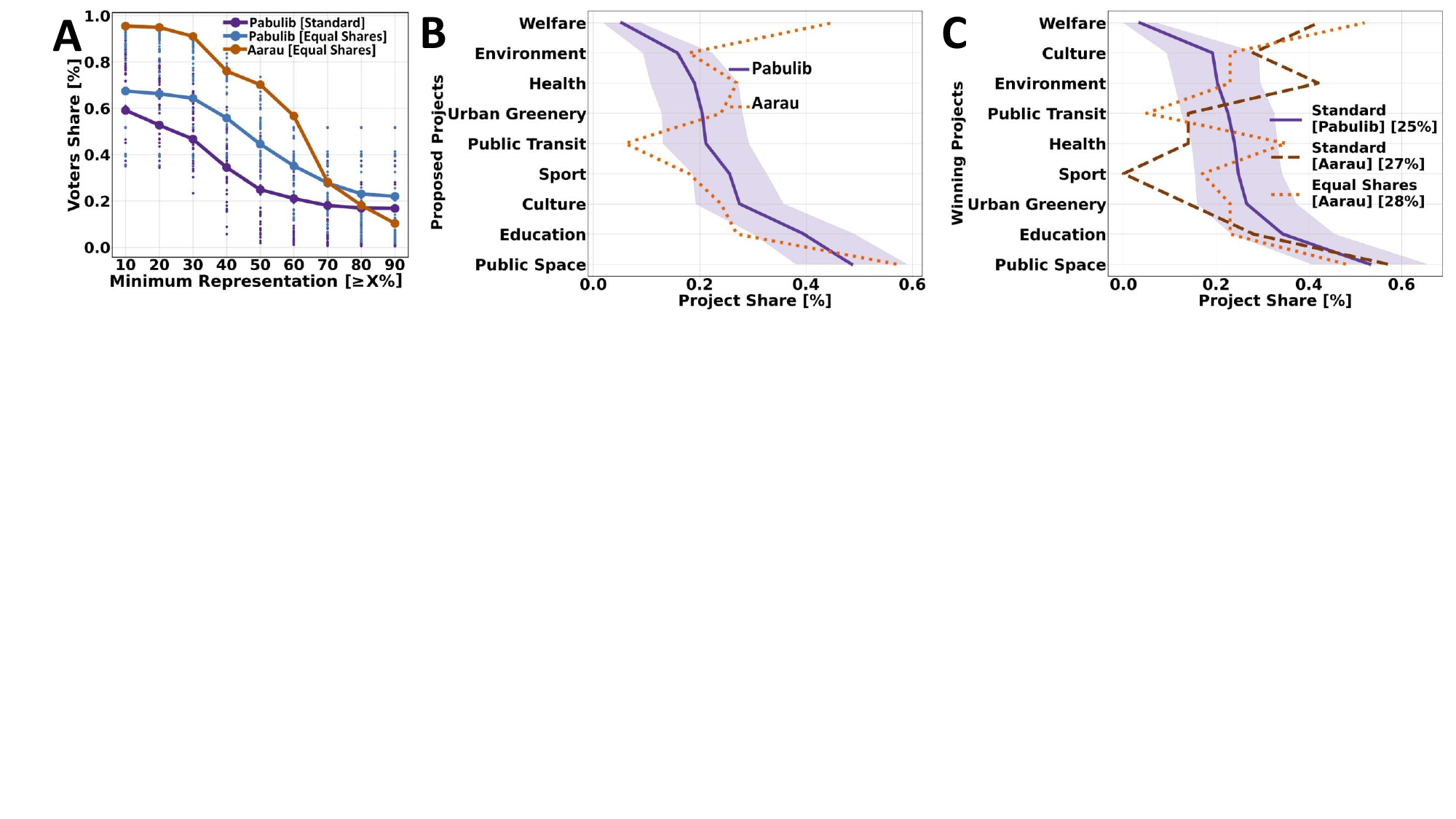}
		\caption{\textbf{Compared to earlier participatory budgeting experiences based on the standard utilitarian greedy method, the method of equal shares improves the representation of voter preferences and project categories, both in the project ideation and the project selection phase.} (A) 27.8 \% more voters on average get any minimum representation level using the method of equal shares as compared to earlier participatory budgeting processes based on the standard utilitarian greedy method. (B) The projects proposed for voting under equal shares promote project categories that have traditionally been under-represented, such as welfare and health, while projects that usually come with high costs, such as public transit, lose appeal. (C) When using the method of equal shares, some otherwise under-represented projects are winning. On average, project categories show a 3\% higher representation.}
		\label{fig:representation}
	\end{figure}

	\cparagraph{The method of equal shares represents voters better} Figure~\ref{fig:representation}a shows the share of voters for different minimum levels of the mean representation of their preferences (i.e. what share of voters have at least 0.1, 0.2,...,0.9 representation). Representation is measured by the relative number of projects preferred (i.e. selected) by a voter among those that win. The mean is calculated first for the voters of each election and then among elections (Figure~\ref{fig:pabulib-representation-voters} is generated by averaging all voters among all elections). This figure reveals whether we have few or many voters that are very satisfied or unsatisfied (high/low representation). To generalize our findings we compare the following:
	\begin{itemize}
		\item[(i)] The standard method of 38/345 past participatory budgeting processes~\cite{Faliszewski2023,pabulib} (baseline) that share similar characteristics with the process in Aarau (matched, see Section~\ref{sec:choice-ballot-aggregation});
		\item[(i)] The method of Equal shares applied hypothetically to these 38 past elections;
		\item[(i)] The method of Equal shares as applied in Aarau.
	\end{itemize}
	
	Strikingly, much more voters (difference of 57.7\% to 80.7\%) are well-represented to a minimum extent with equal shares in Aarau, even when considering up to the minimum representation of 70\% (Figure~\ref{fig:representation}a). In contrast, a larger voters share for more than 90\% representation is observed for the past elections using the standard majoritarian method. Such so high representation levels imply less capacity for compromise, leaving a larger part of the population with lower representation levels. This representation gain is not so high (difference of 49.6\% to 56.7\%) if the method of equal shares is applied to the past 39 past elections. This can be explained by (i) the different nature of the proposed projects when using equal shares, coming with lower cost (48.4\% lower budget share) and (ii) the larger number of winning projects (23.8\% relative increase in the number of winning projects). Table~\ref{tab:representation-gain} also confirms that participatory budgeting processes with the project and budget characteristics (matching criteria) of Aarau result in the highest representation gain. This strengthens further the evidence for the approach of Aarau as a blueprint for democratic updates based on fair voting methods.
	
	\cparagraph{The method of equal shares encourages proposed projects that are novel and win} Figure~\ref{fig:representation}b illustrates the share of the proposed project categories in the participatory budgeting in Aarau compared to projects proposed in earlier participatory budgeting processes~\cite{Faliszewski2023,pabulib}. We find a significant change in the nature of the proposed projects: welfare and health projects, which were usually under-represented before, are more prominently featured, while public transit projects are less prominent. This shift is reflected by the costs of projects in two categories: the average costs of public transit projects in the participatory budgeting process of Aarau amount to 8\% of the overall budget, while welfare-related projects amount to 12.25\%. In earlier participatory budgeting processes, these numbers were 18.19\% and 7.52\%, respectively. Figure~\ref{fig:representation}c presents the share of the winning project categories in the participatory budgeting process of Aarau for the method of equal shares and for the standard majoritarian method. It also shows the share in earlier participatory budgeting experiences that relied on the standard method. Health and sports would be disadvantaged by the standard method, as it would reduce their share in the winning set by 19.13\% and 18.05\%, respectively.
	
	\cparagraph{Regional comparison: Aarau [equal shares] vs. Zurich [standard]} To further demonstrate the impact and validity of the innovative voting approach used in Aarau, we compare it with the previous participatory budgeting process in Zurich, which used the standard majoritarian voting method. Zurich (which is 38 km away from Aarau, i.e. located in rather close proximity) ran the ``City Idea'' format in 2021~\cite{ZurichPBWeb,ZurichPBReport}. Although data of individual voters are not available for analysis to us, a comparison based on aggregate measures is a particularly revealing `crash-test' because of the following interesting facts:
	
	\begin{itemize}
		\item[(i)] With a population size of 423,193>22,013, a larger pool of citizens can be attracted to participate in the ``City Idea'' process in Zurich as compared to Aarau; 
		\item[(ii)] The winning/proposed projects originated from all quarters of Zurich, which aimed to encourage a broad engagement by the citizens of Zurich in the participatory budgeting process: West: 16/33, North: 14/24, East: 14/29, South: 17/49; 
		\item[(iii)] The available budget for all regions was around 540,000 CHF, which was much more than the 55,000 CHF provided by the smaller city of Aarau. 
	\end{itemize}
	
	Despite those distinguishing factors, which are in favor of  Zurich, the approach in Aarau turned out to be more effective than the one in Zurich. Even though Zurich had a larger size and a more diverse geographical origin of its proposed projects, the process attracted only the votes of 1,804 citizens, which was approximately the same number as in Aarau, corresponding to a 17 times higher turnout in Aarau. Moreover, many of the proposed projects in Zurich claimed the maximum amount of 10,000 CHF, while the average cost was 6.6\% of the total budget, which is even lower than the 15\% in Aarau.
	
	Accordingly, the projects in Aarau came with different cost modalities and higher diversity. Although the standard majoritarian voting method used in Zurich prioritized popularity, there were 7 projects in Zurich that were not selected because they each fell short by just 1 to 4 additional votes. This creates several origins of dissatisfaction for the voters in Zurich, in contrast to the process in Aarau, where the fair voting methods create several new ways for voters to have their preferred projects winning as showed in Section~\ref{subsec:voting-outcomes}.
	
	\cparagraph{Broader evidence beyond the case of Aarau} To strengthen our findings further, we also highlight the case of two other participatory budgeting processes that used the method of equal shares during 2023 in two cities of Poland: (i) Świecie and (ii) Wieliczka (``Green Million''). In the first case of Świecie that was a follow-up of Aarau, 17 of 22 projects are selected for funding, while the majoritarian standard method would have resulted in only 9 winning projects. In addition, the representation of voters' preferences increased from 61\% to 80\% by the method of equal shares. The top-2 project in terms of voter approvals was not funded, which would have required 36.6\% of the budget, while it was approved by 24.4\% of the population. In the second case of Wieliczka, 30 of 64 projects were selected for funding, while the majoritarian standard method would have resulted in 23 winning projects. Here, the representation of voters' preferences increased from 61\% to 71\%. The top-2 project in voters' approvals was also not funded, which would have required 10\% of the budget and was approved by 8.9\% of the population. Both cases confirm the distinguishing voting outcomes we found for the method of equal shares: proportional representation is achieved by funding more projects, while a few expensive popular projects are sacrificed. However, this increases the overall satisfaction. In contrast to these two elections, further novel data were collected during the participatory budgeting process in Aarau, which allow us to rigorously explore the legitimacy of fair voting methods and the citizen values that explain the shift to those. 
	
	\subsection{Adopting fair voting improves legitimacy}\label{subsec:legitimacy-gain}
	
\begin{figure}[!htb]
		\centering
		\includegraphics[width=1.0\textwidth]{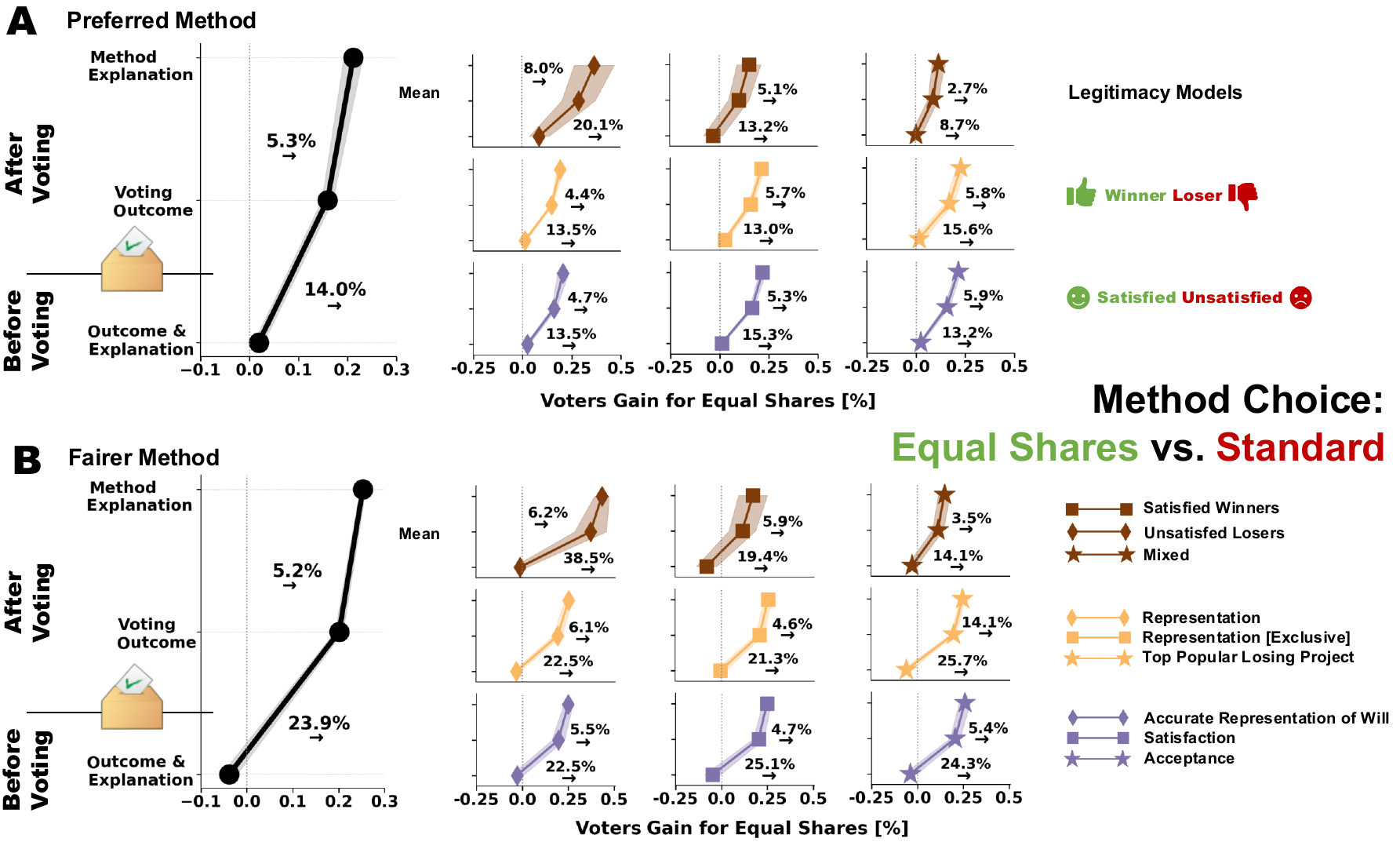}
		\caption{\textbf{In the exit survey after voting, the method of equal shares was found to be more preferred and fairer than the standard majoritarian ballot aggregation method. Strikingly, before voting and adopting equal shares, there was no clear preference for any of the two methods, while the standard method was found fairer. The voting outcomes themselves were sufficient for individuals to determine equal shares as their preferred and fairer ballot aggregation method. (The explanation of the methods does not significantly contribute to shifting the choice to equal shares.)} Furthermore, the legitimacy of equal shares was assessed based on (A) the voters' preference and (B) their perceived fairness of equal shares over the standard method. The x-axis determines the difference (gain) of the voters proportion between those choosing equal shares and those choosing the standard method. Legitimacy is assessed based on 24 combinations of winners/losers and satisfied/unsatisfied voters that are grouped into (i) satisfied winners, (ii) unsatisfied losers and (iii) mixed. Winners/losers are determined in three different ways: 1st and 2nd quartile of representation score for the winning projects that are (i) non-exclusive and (ii) exclusive between the two ballot aggregation methods. (iii) Non-supporters and supporters of the top popular losing project. Satisfied/unsatisfied voters are determined by three post-voting survey questions that assess: (i) satisfaction, (ii) acceptance of the outcome and (iii) accurate representation of will.}
		\label{fig:legitimacy}
	\end{figure}

	\cparagraph{Rigorous evaluation of a new ballot aggregation method} Figure~\ref{fig:legitimacy} illustrates which ballot aggregation method voters prefer and which they find fairer in the entry survey before voting and in the exit survey after voting. Section~\ref{sec:choice-ballot-aggregation} also shows additional results of which method voters find more convenient. Section~\ref{sec:equal-shares-explanation} further illustrates how the two methods, based on which our measurements were made, were presented to the voters, in particular the voting outcomes and the explanations. The shift from the majoritarian standard method to the method of equal shares becomes obvious by the gain in the share of voters, who \emph{prefer} equal shares after voting and find it \emph{fairer} (rigorously measured on the same matched population). We also assessed which method voters found more convenient (see Figure~\ref{fig:convenience}). The legitimacy of the two ballot aggregation methods was determined by measuring the satisfaction of voters with the voting outcomes of each method. Remarkably, it increased even if the voter is classified as a `loser' (based on different measurements); see Section~\ref{sec:legitimacy} for more information. To assess legitimacy thoroughly, the voters' gain (from before to after voting) was disaggregated further down to all 24 combinations of winners/losers and satisfied/unsatisfied voters (see Figure~\ref{fig:legitimacy_fairness_preference_heat}). For simplicity, in Figure~\ref{fig:legitimacy} we grouped those into three groups: satisfied winners, unsatisfied losers and mixed, each consisting of 8 combinations. Three ways to measure each of the winners/losers and satisfied/unsatisfied are illustrated in Section~\ref{sec:legitimacy}.  
	
	\cparagraph{The adoption of the method of equal shares exhibits a shift to be the most preferred and fairer method} The results of Figure~\ref{fig:legitimacy} show that, there is a significant gain in voters who prefer the method of equal share and find it fairer in the exit survey after voting when compared to the entry survey before. This is particularly prominent for unsatisfied losers who chose the standard method as fairer before voting but switched to equal shares after voting (see Figure~\ref{fig:legitimacy_fairness_preference_heat}). More specifically, 54\% of the population preferred equal shares before voting, while after voting, supporters of equal shares raised to 80\% (p=0.0034). Similarly, 45\% of voters found equal shares fairer before voting and 89\% fairer after voting (p=0.028). 
	
	\cparagraph{The method of equal shares shows stronger legitimacy} The significant legitimacy of equal shares is reflected in several ways (see also Figure~\ref{fig:legitimacy_fairness_preference_heat}): (i) Losers, determined as supporters of the top popular losing project (`Wild bees' paradise'), prefer the method of equal shares and find it fairer. (ii) Remarkably, the method of equal shares is even found fairer and preferred to a higher extent among losing voters than among winning ones, who often prefer the standard majoritarian voting method. (iii) The shift to equal shares after voting is found fairer among those who do not accept the voting outcome, which is a strong proxy of legitimacy~\cite{Weatherford1992,Johnson2006}. However, we also found some cases where more voters shifted to the standard majoritarian voting method after voting (see Section~\ref{sec:choice-ballot-aggregation}).
	
	\cparagraph{Voting outcome itself vs. method explanation} 
The explanation of the voting method has an additional contribution of 5.3\% to the shift towards equal shares as the \textit{preferred} method, while it has an additional contribution of 5.2\% to the shift towards equal share as the \textit{fairer} method. Accordingly, the method explanation has a small but statistically significant contribution to the shift to equal shares (with the exception of the mixed setting for the preferred method, p=0.07). These results demonstrate that the voting outcome itself is sufficient to establish a change in favor of the method of equal shares. 

\subsection{Democratic values explain voters' shift to the fairer voting method}\label{subsec:explaining-factors}

\begin{figure}[!htb]
	\centering
	\includegraphics[width=0.9\textwidth]{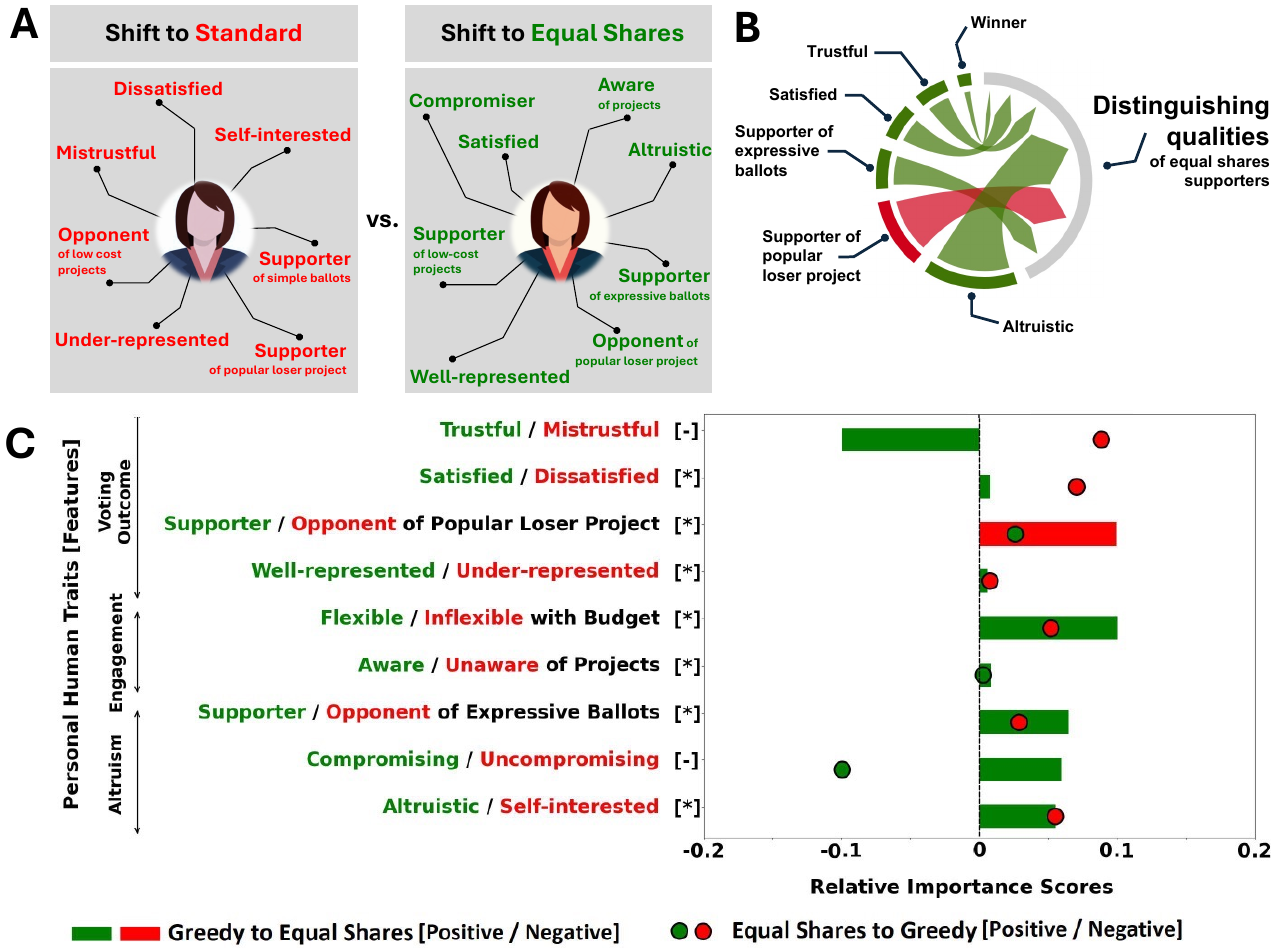}
	\caption{\textbf{The shift to equal shares as the preferred method after voting is explained by distinguishing human traits that reflect the promotion of strong democratic values such as being altruistic, compromisers, and supporting more expressive, inclusive ballots. In contrast, shifting to the standard majoritarian method is found among voters showing mistrust, under-representation, dissatisfaction, and self-interest. These voters also support the popular costly loser project.} 
	A classification model (with 85.92\% accuracy) is designed with four classes representing the preferred ballot aggregation method before and after voting (with explanation of methods). The independent variables are human traits of voters that explain the observed transitions. (A) Distinguishing human traits of voters who shift to the standard method and equal shares after voting. These are prominently associated with democratic backsliding on the one hand, and promotion of democratic values on the other hand. (B) Summary of distinguishing qualities for those who prefer the method of equal shares over the standard majoritarian method after voting. Altruism, support of expressive, inclusive ballots and opposition to the popular, costly loser project of equal shares are qualities with the highest prominence for those voters preferring equal shares. (C) Human traits (features) that explain the preference of each voter before and after voting, in particular the transitions from the standard majoritarian method to equal shares and from equal shares to the standard method (see Figure~\ref{fig:legitimacy}). The `*' indicates that a feature is statistically significant (p<0.05) in explaining the voters' choice of the method before and after voting.}
\label{fig:traits}
\end{figure}

\cparagraph{Clash between democratic values and democratic backsliding}
Figure~\ref{fig:traits} identifies human factors that causally explain the shift from one ballot aggregation method before voting to another one after voting (with the explanation of method), see also Figure~\ref{fig:legitimacy} and Section~\ref{sec:explanations}. The classification model is illustrated in Section~\ref{sec:shifts-methods} and comes with an accuracy of 85.92\%. Strikingly, the group of voters who shift from the standard majoritarian voting method to equal shares exhibits democratic values that explain this transition, see Figure~\ref{fig:traits}c. These voters come with distinguishing qualities (Figure~\ref{fig:traits}b) such as being altruistic, being compromisers, and supporting more expressive, inclusive ballots. They are aware of the proposed projects and flexible to select low-cost projects, which also makes them more likely to oppose a popular costly project. This group is well-represented in the voting outcome. In contrast, the voters that shift from equal shares to the standard majoritarian method do not show these democratic values. Instead they are driven by self-interest, mistrust and dissatisfaction. They support simpler ballot formats, are not well-represented in the voting outcomes, and support the popular loser project under equal shares. 

\cparagraph{Distinguishing qualities of voters who shift preference to equal share}
Figure~\ref{fig:traits}b summarizes the distinguishing qualities of voters, who shift to equal shares by taking the difference of the human traits for the two transitions (standard to equal shares -- equal shares to standard). Altruism, opposition to the popular (costly) loser project and support of the expressive, inclusive ballot formats are the most prominent distinguishing qualities observed for voters who shift to equal shares. Figure~\ref{fig:traits}c shows the importance of features that explain the classification of voters based on which of the two transitions they make for their preferred ballot aggregation method before and after voting (see Tables~\ref{tab:accuracy_results},~\ref{tab:feature-loss} and~\ref{tab:trait-survey-map} in Section~\ref{sec:shifts-methods}). Further results for the transitions to methods that are found fairer and more complex are shown in Figures~\ref{fig:ES_G_G_ES},~\ref{fig:ES_G},~\ref{fig:Convenience_Explain} in Section~\ref{sec:shifts-methods}.

\section{Discussion}\label{sec:discussion}

\cparagraph{A blueprint to upgrade democracies} In our study, we have shown how fair voting methods can better foster collective intelligence that is based on higher inclusion and diversity, with strong evidence provided by (i) a rigorous participatory budgeting process in the city of Aarau in Switzerland as well as (ii) extensive comparisons with other participatory processes. 
Subsequent processes using fair voting methods confirm our findings~\cite{equalsharesmethod}. Accordingly, the standard majoritarian approach of voting has a number of deficits, such as the selection of few projects only, which often lack cost-effectiveness, leaving a large portion of good project ideas unrealized. As projects selected by majoritarian voting are typically implemented in the city center, they largely fail to achieve spatial fairness or inclusion. In the past, this issue has been addressed by dividing a city into multiple districts and running multiple budgeting processes in parallel, as in the case of Zurich in 2021. In contrast, combining cumulative voting with the method of equal shares proves to be resourceful, as it supports more winning projects. It also promotes fairness, satisfaction, inclusion, trust, and democratic values. Accordingly, this approach offers a powerful way of upgrading democracies. Implementing this approach, using digital means, is also a value-sensitive approach to promoting digital democracy by design~\cite{Helbing2023}.

\cparagraph{Success of project implementation} By the end of 2024, 9 of 17 projects selected by the ``City Idea'' process in Aarau were already implemented and 5 projects got an extension, while 3 projects could not be implemented mainly due to feasibility and budgetary constraints~\cite{CityIdeaReport2025}. These were among the top-11 projects. Two thirds of these projects would also have been winners if the standard majoritarian method had been used. These results demonstrate that having more winning projects using the method of equal shares does not negatively affect the overall capacity of citizens to implement them. In contrast, the standard majoritarian voting method applied in Zurich for the ``City Idea'' process did not produce so satisfactory results, and the press critically discussed the problem of cost-inefficient projects~\cite{ZurichPBReport}.

\cparagraph{Participation levels} The participatory budgeting process in Aarau achieved significant participation of citizens at different levels. During the project ideation phase, 161 ideas were proposed, which exceeded by far the initial goal of the campaign to attract 100 ideas~\cite{CityIdeaReport2025}. In comparison, 176 ideas were proposed and 135 of them presented for voting in Zurich in 2021 with a 19 times larger population and a 10 times larger total budget. The total of 1,703 participating voters in Aarau (as compared to 1,804 in Zurich) is a significant number for such a campaign in a town of 14,310 residents with voting rights in 2023. Nevertheless, the turnout did not reach the level of official elections such as referenda and national elections, yet. A 22\% of voters showed interest in participating in the ``City Idea'' process even though they were not official residents of Aarau, while 9\% of voters were not Swiss citizens. The average age of voters was 43, and there were more female than male voters.  Children over 15 years were allowed to vote. Among the low number of citizens who abstained but participated in the surveys (296), the top-5 reasons for not voting were "no time" (17.1\%), "forgetting" (12.8\%), "lack of information" (9.9\%), "unappealing projects" (7.7\%) or some "other reason" (7.0\%), see Question EB.7 in Table~\ref{table:exit-digital}. 

\cparagraph{Upgrading democracies requires upgrading digital tools} We created and used a novel combination of open-source and value-sensitive digital tools to meet a broad range of expectations regarding transparency, security, privacy, reproducibility, explainability and fairness. This required a significant implementation effort, a mobilization of various communities, a strong commitment and the building of shared values with the city authorities, which is a process that has not been systematically pursued to this extent before. A highlight of this effort was the privacy-preserving verification of citizens' eligibility when voting with the social security number, which ensured a trustworthy voting procedure. The integration of our open-source implementation of cumulative voting~\cite{SPBStadtidee2023} into the Stanford Participatory Budgeting platform is a stepping stone towards participatory tools that are more standardized and interoperable. This is also relevant for other large open-source communities such as Decidim and Consul as well as the ones created by the computational social choice community such as Pabutools~\cite{pabutools} and Preflib~\cite{Mattei2013}. 

\cparagraph{Explaining and adopting fair voting methods} 53.17\% of voters understood how equal shares calculates the winning projects (MB.3.3), while 55.95\% understood why popular projects are not easily funded by equal shares (MB.3.6), and 41.26\% found this fair (MB.3.7). Strikingly, the results of Figure~\ref{fig:legitimacy} demonstrate that explaining the expected difference in the voting outcomes reached by equal shares is sufficient for voters to prefer this method. A more detailed explanation of how the method works increases the tendency towards preferring equal shares over the standard method only slightly. This result has significant implications for overcoming the long-standing hesitation with regard to democratic innovations, since citizens show a readiness to accept more complex decision-making methods as long as they understand the anticipated impact on the decision outcome. In addition, the experience of the different results helps citizens to shape a practical understanding of the method. Moreover, given that equal shares is designed for a proportional representation of voter preferences and is known for its stability~\cite{Fairstein2023}, it is more unlikely to cause controversies over voting outcomes. Nevertheless, the additional effort required to explain the method of equal shares should not be under-estimated. As shown in Section~\ref{sec:equal-shares-explanation}, the introduction of equal shares requires a systematic communication effort over the lifetime of the participatory budgeting process, using creative media and visualizations suited for the broader public. At the project ideation phase, awareness of the method is required to mobilize the population for ideas that may not have been considered under the majoritarian rule. As the costs of the proposed projects play a key role for selecting them, citizens need to be resourceful and prioritize different kinds of ideas, which requires a focus on cost-effective projects that are supported by a sufficiently large group of the population. Instead, large and costly projects may be more risky to propose as they require the support of a broader population in order to get selected. 

\cparagraph{Fair voting methods to foster a novel impact creation process} The influence of the chosen voting method on the proposed and selected project ideas can transform the process of creating real-world impact. For instance, novel welfare projects gain momentum, as they appeal to several interest groups, while they may also be easier to implement than costly infrastructural projects~\cite{Maharjan2024}. We foresee that, on the long term, proportional voting methods could enable an incremental grass-roots implementation of sustainable development projects over several participatory budgeting campaigns, which could also support a fairer transition to more sustainable future. 

\cparagraph{Fair voting methods are a better-than-optimal paradigm} Over-simplifying, averaging or overriding people's preferences while optimizing societies for goals that people may not even explicitly agree on is a significant risk that democracies face in the era of artificial intelligence (AI) run by Big Tech. In this context, the notion of `optimal' solutions is actually misleading and displaced from the reality of people, as demonstrated by the inferiority of popular expensive projects or projects focused on city centers that are typically proposed in participatory budgeting processes under standard majoritarian voting.

In contrast, proportional voting methods work well for pluralistic settings, where voters may not agree on one and the same goal, but decisions are based on collective intelligence. This approach excels when goals are complementary, but is also suited to handle situations of competitive or even incompatible goals, or when significant minorities are involved. Small compromises, such as excluding a popular expensive project, open up opportunities for large collective returns with benefits for a great majority of people. For instance, the inclusion of otherwise under-represented health projects or projects in the city periphery can reduce social tensions and benefit a broader population. The demonstrated approach of combining cumulative voting with the method of equal shares delivers a powerful blueprint for upgrading democracies. It allows societies to achieve:
\begin{itemize}
\item[(1)] a larger number of realized projects within the available budget; 
\item[(2)] a greater resourcefulness of novel projects and a multi-faceted impact; 
\item[(3)] more participation and inclusion; 
\item[(4)] greater transparency and trust in the voting process; 
\item[(5)] better representation of citizen preferences and higher satisfaction; 
\item[(6)] more (spatial) fairness; 
\item[(7)] higher legitimacy thanks to a greater acceptance of the collective outcomes; 
\item[(8)] strengthened democratic values of voters. 
\end{itemize}
We call these impacts ``better than optimal'' as they tend to be a win-win for all, while they are distinguished from what existing optimization and data-driven decision-making approaches typically offer. 

\cparagraph{Fair voting methods in the era of artificial intelligence} We would like to complete this discussion on a remark addressing the use of data-driven AI systems for decision making. While we consider AI a very powerful technology with numerous promising application areas, how to use it in the context of governing democracies must be studied very carefully. AI safeguards for democracy are critical and yet to be created. Existing decision-making methods supported by AI solutions show alarming vulnerabilities and are still lacking fairness and legitimacy~\cite{Saetra2022,Helbing2015,Koster2022}. For instance, voting methods with a deficit on fairness can exacerbate biases and inconsistencies by unconsolidated use of large language models~\cite{Majumdar2024}. So far, it is not clear how to address some existing AI risks. Therefore, alternative approaches such as the one presented here are urgently needed. 

\section{Methods}\label{sec:methods}

We illustrate the designed participatory budgeting campaign of ``City Idea'', the evaluation approach of legitimacy and the models for causal explanations for the shifts between ballot aggregation methods before and after voting.

\subsection{The participatory budgeting process of ``City Idea'': Field test design}\label{sec:campaign-design}

Table~\ref{table:field-test} outlines the participatory budgeting campaign of ``City Idea''. It consists of six phases over the passage of nine months. An entry and exit survey were conducted before and after voting respectively. The entry survey was launched via physical invitation letters to all citizens sent by city council (10,000 A5 flyers~\cite{CityIdeaReport2025}). This resulted in 3592 respondents. Some of these filled in their email to voluntarily participate in the exit survey and this resulted in 808 respondents. The surveys included questions about demographics and political opinion (Table~\ref{table:entry-demographics}-\ref{table:exit-demographics}), engagement and digital skills (Table~\ref{table:entry-digital}-\ref{table:exit-digital}), the ``City Idea'' voting (Table~\ref{table:entry-city idea}-\ref{table:exit-city idea}), the voting outcome preference (Table~\ref{table:entry-voting outcome}-\ref{table:exit-voting outcome}) and the voting input and aggregation method (Table~\ref{table:entry-voting input}-\ref{table:exit-voting input}). Survey questions encode variables, which are observed for changes before and after voting. The overall study has received ethical approval from University of Fribourg (\#2021-680). 

\begin{table}[!htb]
\caption{\textbf{The rigorously designed participatory budgeting campaign of ``City Idea'' with 6 phases over a period of 9 months in 2023}.  Participants of the entry survey, voting and exit survey are linked using an anonymous identifier (hash) of the citizens' social security number. This allows the study of several choice variables before and after voting. }\label{table:field-test}
\centering
\resizebox{\textwidth}{!}{%
\begin{tabular}{lllllllllll}\toprule
	\multicolumn{1}{l}{Stage} & Start  & End & Scale & \multicolumn{7}{l}{Shared Participants in Stages}  \\\midrule
	1. Entry survey & January  & March & 3592 participants  & \cmark& \cmark & \cmark & & &  \cmark&  \cmark\\
	2. Project ideation& 13th of February  & 18th of March & 161 projects &  & & & & & &\\
	3. Deliberation & 20th of March  & 11th of June & 33 projects  & & & & & & &\\
	4. Projects campaigning &  5th of June & 11th of June & 73 participants, 1682 comparisons   & & & & & & &\\
	5. Voting (All) &  \multirow{2}{*}{12th of June}  &  \multirow{2}{*}{25th of June} & 1703 participants  & & \cmark & & & \cmark &  & \cmark \\
	5. Voting (Aarau) &   &  & 1330 participants  & \cmark  & & & \cmark & &  \cmark & \\
	6. Exit survey & July  & September & 808 participants & & & \cmark & \cmark & \cmark &  \cmark & \cmark
	\\\midrule
	&  & & Matched population: & 505 & 505 & 804 & 252  & 252 & 252 &252\\\midrule
	Implementation & July 2023  & December 2024 & 9 implemented, 5 extended & & & & & & & \\\bottomrule
\end{tabular}
}
\end{table}

\cparagraph{Study of ballot formats} The preference over different ballot formats is studied via Question MB.2 in exit survey (see Table~\ref{table:exit-voting input}). The transitivity over different ballot formats is studied with a hypothetical voting scenario in the entry survey with 5 costed projects (Question MA.6 to MA.10 in entry survey) and a budget of 50,000 CHF, in which participants voted using 5 different ballot formats in sequence, see Figure~\ref{fig:ballot-formats}. The involved ballot formats include: single choice, approval, score, Knapsack, and ranking.

\cparagraph{Study of ballot aggregation} The preference, fairness and simplicity of the standard method versus the method of equal shares are assessed (Questions~MA.1, MA.3, MA.4 in entry survey and Question~MB.4, MB.6, MB.7, MB.9, MB.10 in exit survey). Before voting, participants assess the voting outcome of a simple hypothetical voting scenario depicted in Figure~\ref{fig:ballot-aggregation}. Post voting, participants assess the legitimacy of aggregation methods on the real-world ballots after exposure to the winning projects elected via ballot aggregation methods followed by the explanation of the methods (see Figure~\ref{fig:q2224graphic}). With this two-stage assessment we can distinguish the choice of an aggregation method because of its voting outcomes vs. the method explanation. 

\cparagraph{Project ideation, deliberation and campaigning} Citizens proposed more than 161 project ideas out of which 33 projects are selected to put for voting~\cite{CityIdeaReport2025}. Projects are eliminated or similar project ideas are combined to strengthen their chances to win. Project ideation involved support and guidance with planning, implementation, economic and legal feasibility by a municipal advisory board of 7 people, a core team of 3 people and ambassadors. The city council distributed posters and flyers to various municipal institutions and locations. It also organized two press releases that resulted in 17 media reports. Moreover, the interactive information campaign `\emph{know your projects}' was launched in the city that engaged 73 participants in 1682 random pairwise comparisons of projects (see Table~\ref{tab:pairwise_winners} for further information) via the Stanford Participatory Budgeting platform ~\cite{Wellings2023}. Awareness about the method of equal shares and what projects are expected to win was raised, emphasizing the challenge of electing projects that cover a large share of the budget. Section~\ref{sec:equal-shares-explanation} provides more information about how the method of equal shares was communicated and explained to citizens. 

\cparagraph{Voting} The city of Aarau has a total of 22,593 citizens, out of which 17,428 are Swiss citizens. A total of 1703 voters participated in the "City Idea" campaign, out of which 1330 are citizens of Aarau. This also allows individuals to have a voice, who are not living in Aarau, but have a connection to the city, such as workers and broader family members. Children over 15 years old could also vote. Voting was digital, using the open-source Stanford Participatory Budgeting platform~\cite{Wellings2023}. However, voters with low digital literacy were supported in virtual voting centers around the city. The voting platform was extended to support cumulative voting by distributing 10 points over at least 3 projects~\cite{SPBStadtidee2023,Wellings2023}. Voters are securely verified in a privacy-preserving way by providing a unique (hash) code that was generated by the city council using the social security number of each citizen (AHV card). The same platform was used for project campaigning by using the pairwise project comparisons feature. The winning projects were calculated using the method of equal shares and the Add1U completion method implemented within the Pabutools (the ones of Świecie and Wieliczka used the Add1 completion method, while Add1U would yield 1 and 3 more winners)~\cite{pabutools}. Voting outcomes with the utilitarian completion method were also calculated and assessed in the exit survey (Table~\ref{table:exit-voting outcome}). 

\subsection{Evaluation of legitimacy}\label{sec:legitimacy}

To assess the legitimacy of ballot aggregation methods, we rely on well-established theory of political legitimacy~\cite{Weatherford1992} and social choice~\cite{Arrow2010,Brandt2016}. Among the three notions of \emph{input} (voting participation), \emph{throughput} (the voting process) and \emph{output} (voting outcome) legitimacy~\cite{Scharpf1999,Schmidt2013}, we mainly focus in Section~\ref{subsec:legitimacy-gain} on output legitimacy. However, we also assess the choice of ballot aggregation method before and after voting with equal shares as well as the effect of explaining equal shares, both capturing throughput legitimacy. 

As such, by legitimacy we refer to voters who remain satisfied and accept the voting outcome of a ballot aggregation method (explained or not) even if they are among the losers, i.e. the did not vote for a winning project~\cite{Hanggli2024}. By considering all 24 combinations of winner vs. loser voters who are satisfied vs. unsatisfied (see Figure~\ref{fig:legitimacy}), we have a rigorous full factorial design to assess the legitimacy of ballot aggregation methods: which method voters prefer (Question MB.4 and MB.7 in Table~\ref{table:entry-voting input}), find fairer (Question MB.6 and MB.9) and more convenient (MB.10).

To distinguish winner and loser voters who are satisfied by the voting outcome from those who are not, we combine (i) satisfaction proxies based on survey questions with (ii) actual measurements of voters' representation by the voting outcomes. 

\cparagraph{Proxies of satisfied vs. unsatisfied voters} We use different satisfaction proxies based on questions of the exit survey in Table~\ref{table:entry-digital} (OB.1). In particular, we assess whether voters are satisfied with the voting outcome of ``City Idea'' (OB.1.1), whether they accept this outcome (OB.1.2) and whether they feel the outcome accurately represents the will of Aarau citizens (OB.1.7). Based on 11-point likert-scale responses, we split the populations two quartiles, the first one for the satisfied voters and the second one for the unsatisfied ones. 

\cparagraph{Determining winner vs. loser voters} To distinguish winner and loser voters by the voting outcome of equal shares, we measure the mean representation of voters' preferences in the voting outcomes. The representation of a voter under cumulative voting is determined by the total points assigned by a voter to projects that win over the total points the voter assigns to projects. Winners represent the first quartile, while losers the second one. We also measure the representation of voters' preferences by exclusive winning projects, which are the ones that win using the equal shares, but loose using the standard method. Exclusive representation reflects more prominently for voters the representation gain of equal shares over the standard method. We also distinguish voters to those who assigned points to the popular loser project of `Wild bees' paradise', which was the 3rd most popular project that is not elected under equal shares but it would be elected under the standard method. This allows us to assess the legitimacy of a particular effect of equal shares: `sacrificing' a more expensive popular project, to elect instead several other projects that increase in overall the proportional representation of voters. Whether supporters of such exclusive losing project, which is popular and expensive, can accept this compromise for public good is a key question we explore in Section~\ref{subsec:legitimacy-gain}.

\subsection{Explaining shifts between ballot aggregation methods}\label{sec:explanations}

In this section, we present a causal analysis of how personal traits influence shifts in individual voter preferences for ballot aggregation methods, based on the pre-voting survey, the actual voting process and the post-voting survey.

\cparagraph{Prediction modeling for causal explanations} We explore a broad spectrum of factors (independent variables) that explain shifts in the choices of the ballot aggregation methods (dependent variable) as illustrated in Figure~\ref{fig:legitimacy} and Section~\ref{sec:legitimacy}. The independent variables include personal human traits that can be found in the surveys related to political engagement and interest, altruistic behavior, voting outcome preferences, project preferences, and demographics. Different prediction models are constructed for shifts with and without method explanations. The shifts between methods model the dependent variable A and they are represented by a four-class classification problem, capturing transitions between aggregation methods: (i) standard to equal shares, (ii) standard to standard, (iii) equal shared to standard and (iv) equal shares to equal shares. 

\cparagraph{Prediction algorithms} Prediction models are constructed using tree-based gradient boosting algorithms~\cite{prokhorenkova2018catboost} and neural networks with three hidden layers~\cite{singh2019study}. Different groups of personal human traits are assessed for the capacity to causally explain shifts between different ballot aggregation methods. The selection of these algorithms aligns well with how they handle categorical data (e.g. CatBoost~\cite{hancock2020catboost} and neural networks~\cite{wu2018development} known to perform well even on smaller datasets). Further details about how we account for imbalances of features, their collinearity and hyperparameter optimization of the models are illustrated in Section~\ref{sec:shifts-methods}.

\cparagraph{Approach for causal explanations} Once predictions are made, we derive the relative importance of the personal traits via a two-step explainable framework. Initially, an ablation study~\cite{mohseni2021multidisciplinary} is performed to systematically evaluate the {\em relative} predictive contribution of each personal human trait. In the second step, we employ model-agnostic Shapley Additive Explanations (SHAP)~\cite{nohara2019explanation} to assess the {\em individual} contribution of each trait to the prediction performance. More details regarding the performance of the explanatory framework can be found in Tables~\ref{tab:accuracy_results},~\ref{tab:feature-loss},~\ref{tab:trait-survey-map} and Figures~\ref{fig:ES_G_G_ES},~\ref{fig:ES_G},~\ref{fig:Convenience_Explain} in Section~\ref{sec:shifts-methods}.

\section*{Data Availability}

All collected and generated data along with the software code are available at: \\\url{http://doi.org/10.5281/zenodo.15162222} (with restricted access until publication).


\section*{Code Availability}

The improved Stanford participatory budgeting platform is available at:\\ \url{https://github.com/TDI-Lab/SPB-Stadtidee}. 


\section*{Author Contributions}

\textbf{EP}: Conceptualization, Methodology, Validation, Formal analysis, Investigation, Resources, Writing - Original Draft, Visualization, Supervision, Project administration, Funding acquisition. \textbf{SM}: Software, Validation, Formal analysis, Investigation, Data Curation,  Writing - Review \& Editing, Visualization. \textbf{TW}: Software, Validation, Investigation, Data Curation, Writing - Review \& Editing. \textbf{JCY}: Conceptualization, Methodology. \textbf{FBH}: Software, Validation, Data Curation. \textbf{RHF}: Conceptualization, Methodology, Data Collection, Investigation, Resources,  Supervision,  Writing - Review \& Editing, Project lead and administration, Funding acquisition \textbf{DH}: Conceptualization, Investigation, Resources, Writing - Input to Original Draft, Writing - Review \& Editing, Supervision, Project administration, Funding acquisition.

\section*{Competing Interests}

The authors declare no competing interests.

\section*{Additional Information}

Supplementary information: The online version contains supplementary material available at \url{https://doi.org/}. Correspondence and requests for materials should be addressed to Evangelos Pournaras.

\section*{Acknowledgments}

This project received support by the SNF NRP77 `Digital Transformation' project ``Digital Democracy: Innovations in Decision-making Processes'', \#407740\_187249 and the UKRI Future Leaders Fellowship (MR\-/W009560\-/1): ``\emph{Digitally Assisted Collective Governance of Smart City Commons--ARTIO}''. The authors would like to thank the representatives from the city of Aarau for supporting and running the ``Stadtidee'' [translated: "City Idea"] campaign. The authors would also like to thank Abhinav Sharma for supporting the team at University of Leeds to collect the voting data, Sajan Maharjan for his support with the Pabulib data, as well as Michael Buess from Demoscope for his support with the surveys. The authors are grateful to Lodewijk Gelauff and Ashish Goel for their invaluable support they provided us to extend the Stanford Participatory Budgeting platform.

	\bibliography{bibliography}

    \bibliographystyle{naturemag}
	
\makeatletter\@input{yy.tex}\makeatother
\end{document}


\maketitle
	
	\footnotetext[1]{Corresponding author: Evangelos Pournaras, School of Computer Science, University of Leeds, Leeds, UK, E-mail: e.pournaras@leeds.ac.uk}
	
	\tableofcontents

	\newpage	
	\section{Voters' Profile: Demographics and Political Opinion}\label{sec:demographics}

	\begin{table}[!htb]
		\caption{Survey questions for the entry phase – demographics and political opinion}\label{table:entry-demographics}
	\centering
	\resizebox{0.99\textwidth}{!}{%
		\begin{tabular}{p{0.15\textwidth}p{0.62\textwidth}p{0.19\textwidth}p{0.78\textwidth}}
			
			\toprule
			\textbf{ID} & \textbf{Question} & \textbf{Type}& \textbf{Options} \\
			\midrule
			PA.1 & Language of survey & Single Choice & 2 [English, German] \\\midrule

			PA.2 & Gender & Single Choice & 3 [Male, Female, Other] \\\midrule

			PA.3 & Age & String  & Age in Years  \\\midrule
			
			PA.4 & Birth month and year & String & Month and Year \\\midrule

			PA.5 & Do you have children? &  Single Choice & 3 [No, Yes, No answer] \\\midrule
			
			PA.6 & What is the highest education you have completed so far? & Single Choice & 15 [Did not complete school, Primary school, High school, Apprenticeship, Vocational School, Diploma Middle school, 
			Business school, Seminar for teachers, Specialist Training, Matura school, Federal Training, University of Applied Sciences or University of Education, University or Swiss Federal Institute of Technology, Don't know, No answer] \\\midrule
			
			PA.7 & Civil status  & Single Choice & 3 [Married, Single, Divorced]  \\\midrule
			
			PA.8 & Were you born in Switzerland? &  Single Choice  & 4 [No, Yes, Don't know, No answer] \\\midrule
			
			PA.9 & Did your parents migrate to Switzerland? & Single Choice & 5 [Yes both parents were born here, Only one parent was born here , No both parents immigrated, Don't know, No answer] \\\midrule

			PA.10 & Canton & Single Choice & 27 [ZH, BE, LU, UR, SZ, OW, NW, GL, ZG, FR, SO, BS, BL, SH, AR, AI, SG, GR, AG, TG, TI, VD, VS, NE, GE, JU, Keine Antwort] \\\midrule
			PA.11 & Nationality  & String & Country \\\midrule
			
			PA.12 & Residential status & String &   String\\\midrule
			
			PA.13 & How many people live in your household? & String  & Number of people \\\midrule
			
			PA.14 & What is your household type? & Single Choice & 4 [Single-person household, Multi-person household with children, Multi-person household without children, Couple/two-person household]\\\midrule

			PA.15 & District & Single Choice  & 18 [Innenstadt, Zelgli, Telli, Goldern, Hungerberg, Scheibenschachen, Gonhard, Rossligut, Ausserfeld, Siebenmatten, Binzenhof, Altstadt, Torfeld Sud, Schachen, Brunnbach, Tannengut, Hinterdorf] \\\midrule
			
			PA.16 & Roughly, how many years have you lived in Aarau? & Single entry & Number  \\\midrule
			
			PA.17 & Are you entitled to vote in Switzerland? & Single Choice & 2 [Yes, No] \\\midrule

			PA.18 & How interested are you in politics? & Group of questions & 5 questions\\
			\qquad PA.18.1 & \qquad Politics in general  & Ratio Scale & 6 [Not interested at all, Rather not interested, Somewhat interested, Very interested, Don't know, No answer]   \\ 
			\qquad PA.18.2 & \qquad Local politics in Aarau  & Ratio Scale & 6 [Not interested at all, Rather not interested, Somewhat interested, Very interested, Don't know, No answer]\\ 
			\qquad PA.18.3 & \qquad Cantonal politics in Argovia  & Ratio Scale & 6 [Not interested at all, Rather not interested, Somewhat interested, Very interested, Don't know, No answer]\\
			\qquad PA.18.4 & \qquad National politics  & Ratio Scale & 6 [Not interested at all, Rather not interested, Somewhat interested, Very interested, Don't know, No answer] \\
			\qquad PA.18.5 & \qquad International politics & Ratio Scale & 6 [Not interested at all, Rather not interested, Somewhat interested, Very interested, Don't know, No answer] \\\midrule
			
			PA.19 & Where would you place yourself on a scale from 0 to 10, on which 0 means "left" and 10 means "right"? & Ratio Scale & 13 [Left [0] to  Right[10], Don't know, No answer] \\\midrule
			
			PA.20 & How much do you agree with the following statements? Please give an answer on the scale from 0 to 10, where 0 stands for "do not agree at all" and 10 for "agree completely"". You can use the values in between to position your answer. & Group of questions & 6 questions\\
			\qquad PA.20.1 & \qquad I can understand and assess important political issues from the local level well.  & Ratio Scale & 13 [Do not agree at all [0] to Agree completely [10], Don't know, No answer]  \\ 
			\qquad PA.20.2 & \qquad I have the confidence to actively participate in a conversation about local politics & Ratio Scale & 13 [Do not agree at all [0] to Agree completely [10], Don't know, No answer]\\ 
			\qquad PA.20.3 & \qquad Local politicians care about what ordinary people think  & Ratio Scale & 13 [Do not agree at all [0] to Agree completely [10], Don't know, No answer]\\
			\qquad PA.20.4 & \qquad Local politicians strive to maintain close contact with the population  & Ratio Scale & 13 [Do not agree at all [0] to Agree completely [10], Don't know, No answer] \\
			\qquad PA.20.5 & \qquad I am confident in the way local authorities allocate the public budget & Ratio Scale  & 13 [Do not agree at all [0] to Agree completely [10], Don't know, No answer] \\
			\qquad PA.20.6 & \qquad I have confidence in the abilities of the government and politicians at the local level & Ratio Scale & 13 [Do not agree at all [0] to Agree completely [10], Don't know, No answer]\\\midrule
			
			PA.21 & On a scale from 0 (no trust) to 10 (full trust), how much do you trust the following institutions, organizations and groups? & Group of questions & 6 questions\\
			\qquad PA.21.1 & \qquad City council (government)  & Ratio Scale & 11 [no trust [0] to full trust[10]]  \\ 
			\qquad PA.21.2 & \qquad Residents' council (parliament)  & Ratio Scale & 11 [no trust [0] to full trust[10]]  \\ 
			\qquad PA.21.3 & \qquad City administration  & Ratio Scale & 11 [no trust [0] to full trust[10]] \\ 
			\qquad PA.21.4 & \qquad Political parties  & Ratio Scale & 11 [no trust [0] to full trust[10]]  \\ 
			\qquad PA.21.5 & \qquad (Traditional) media  & Ratio Scale & 11 [no trust [0] to full trust[10]] \\ 
			\qquad PA.21.6 & \qquad Social Media  & Ratio Scale & 11 [no trust [0] to full trust[10]]  \\ 
			\qquad PA.21.7 & \qquad My neighbors  & Ratio Scale & 11 [no trust [0] to full trust[10]] \\ 
			\qquad PA.21.8 & \qquad People in Aarau  & Ratio Scale & 11 [no trust [0] to full trust[10]]  \\\midrule

			PA.22 & How satisfied are you with the way local democracy works in the city of Aarau? & Ratio Scale & 9 [completely dissatisfied [1] to  completely satisfied [7], Don't know, No answer]   \\\midrule
			PA.23 & How satisfied are you with the way local democracy works in the city of Aarau? & Ratio Scale & 9 [completely dissatisfied [1] to  completely satisfied [7], Don't know, No answer]  \\\midrule
			
			PA.24 & What is the most urgent problem in your neighborhood? & Text entry & 4 [String, There are no urgent problems, Don't know, No answer] \\\midrule

		\end{tabular}
	}
\end{table}

\begin{table}[!htb]
	\caption{Survey questions for the exit phase – demographics and political opinion}\label{table:exit-demographics}
\centering
\resizebox{\textwidth}{!}{%
	\begin{tabular}{p{0.15\textwidth}p{0.62\textwidth}p{0.19\textwidth}p{0.78\textwidth}}
		
		\toprule
		\textbf{ID} & \textbf{Question} & \textbf{Type}& \textbf{Options} \\
		\midrule
		PB.1 & Language of survey & Single Choice & 2 [English, German] \\\midrule

		PB.2 & How strongly do you feel connected to Aarau? & single choice  & 7 [Not at all, Hardly, A little bit, Rather, Very much, Don't know, No answer]\\\midrule

	\end{tabular}
}
\end{table}

\section{Engagement and Digital Skills}

\begin{table}[!htb]
\caption{Survey questions for the entry phase – engagement and digital skills} \label{table:entry-digital}
\centering
\resizebox{\textwidth}{!}{%
\begin{tabular}{p{0.15\textwidth}p{0.8\textwidth}p{0.19\textwidth}p{0.8\textwidth}}
	
	\toprule
	
	\textbf{ID} & \textbf{Question} & \textbf{Type}& \textbf{Options} \\
	\midrule
	\multicolumn{4}{c}{{\bf Entry Phase}}\\ \hline
	EA.1 & To what degree do the following statements apply to you? & Group of questions & 7 questions \\
	\qquad EA.1.1 & \qquad I know how to protect a device against unathorized access (e.g. a PIN code or fingerprint) & Ratio scale & 7 [Completely disagree [1] {\em to} Completely agree [5], I don’t understand the question , No answer] \\ 
	\qquad EA.1.2 & \qquad I know how to protect devices against viruses &  Ratio scale & 7 [Completely disagree [1] {\em to} Completely agree [5], I don’t understand the question , No answer] \\ 
	\qquad EA.1.3 & \qquad I know how to adjust the privacy settings on a mobile phone or tablet &  Ratio scale & 7 [Completely disagree [1] {\em to} Completely agree [5], I don’t understand the question , No answer] \\ 
	\qquad EA.1.4 & \qquad I know how to identify suspicious e-mail messages that try to get my personal data &  Ratio scale & 7 [Completely disagree [1] {\em to} Completely agree [5], I don’t understand the question , No answer] \\ 
	\qquad EA.1.5 & \qquad I know how to delete the history of websites that I have visited before &  Ratio scale & 7 [Completely disagree [1] {\em to} Completely agree [5], I don’t understand the question , No answer]\\ 
	\qquad EA.1.6 & \qquad I know how to block messages in social media from someone I don’t want to hear from &  Ratio scale & 7 [Completely disagree [1] {\em to} Completely agree [5], I don’t understand the question , No answer] \\ \hline
	
	EA.2 & To what degree do the following statements apply
	to you? & Group of questions & 2 questions \\
	
	\qquad EA.2.1 & \qquad I tend to shy away from using digital technologies where possible. & Ratio scale & 7 [Completely disagree [1] {\em to} Completely agree [5], I don’t understand the question , No answer] \\

	\qquad EA.2.2 & \qquad I would feel safe voting online. & Ratio scale & 7 [Completely disagree [1] {\em to} Completely agree [5], I don’t understand the question , No answer] \\\hline 
	
	EA.3 & In general, how much trust do you have in online voting / e-voting solutions?  & Ratio scale & 6 [No trust at all, Rather no trust, Rather trust, A lot of trust, Don't know, No answer] \\\midrule

	EA.4 & A few questions about digital skills and media at the local level. & Group of Questions & 3 questions\\
	\qquad EA.4.1 & \qquad My preferences changed as I am now more informed  & Ratio Scale & 7 [Completely disagree [1] {\em to} Completely agree [5], I don’t understand the question , No answer] \\    
	\qquad EA.4.2 & \qquad I know how I can find answers to my questions on the Internet.  & Ratio Scale & 7 [Completely disagree [1] {\em to} Completely agree [5], I don’t understand the question , No answer] \\    
	\qquad EA.4.3 & \qquad I know where or from whom I can get help to improve my digital skills.  & Ratio Scale & 7 [Completely disagree [1] {\em to} Completely agree [5], I don’t understand the question , No answer] \\\midrule

	EA.5 & How often do you interact with the following persons? & Group of questions & 4 questions\\
	\qquad EA.5.1 & \qquad Members of the City Council  & Ratio Scale & 7 [Daily, Weekly, Quarterly, Annually, Never, Don't know, No answer] \\ 
	\qquad EA.5.2 & \qquad Members of Residents' Council & Ratio Scale & 7 [Daily, Weekly, Quarterly, Annually, Never, Don't know, No answer]\\ 
	\qquad EA.5.3 & \qquad Members of the city administration  & Ratio Scale & 7 [Daily, Weekly, Quarterly, Annually, Never, Don't know, No answer]\\
	\qquad EA.5.4 & \qquad Other inhabitants of Aarau  & Ratio Scale & 7 [Daily, Weekly, Quarterly, Annually, Never, Don't know, No answer] \\\midrule
	
	EA.6 & How strongly do you feel connected to Aarau? & Ratio Scale & 7 [Not at all, Hardly, A little bit, Rather, Very much, Don't know, No answer] \\\midrule
	
	EA.7 & For each of these three listed groups, please indicate whether you think they should have the right to vote in Aarau (for local issues). & Group of questions & 3 questions\\
	\qquad EA.7.1 & \qquad City council (government)  & Ratio Scale & 4 [No, Yes, Don't know, No answer] \\ 
	\qquad EA.7.2 & \qquad Residents' council (parliament)  & Ratio Scale & 4 [No, Yes, Don't know, No answer] \\ 
	\qquad EA.7.3 & \qquad City administration  & Ratio Scale & 4 [No, Yes, Don't know, No answer]\\ \midrule 
	
\end{tabular}
}
\end{table}

\begin{table}[!htb]
\caption{Survey questions for the exit phase – engagement and digital skills} \label{table:exit-digital}
\centering
\resizebox{\textwidth}{!}{%
\begin{tabular}{p{0.15\textwidth}p{0.8\textwidth}p{0.19\textwidth}p{0.8\textwidth}}

\toprule

\textbf{ID} & \textbf{Question} & \textbf{Type}& \textbf{Options} \\
\midrule

EB.1 & Which media do you consume? & group of questions & 12 questions\\
\qquad EB.1.1 & \qquad Daily national newspaper - Aargauer Zeitung & single choice & 6 [Never, 1-2 times, 3-4 times, 5 and more times, Don't know, No answer]\\
\qquad EB.1.2 & \qquad Online News - Aarauer Nachrichten & single choice & 6 [Never, 1-2 times, 3-4 times, 5 and more times, Don't know, No answer]\\
\qquad EB.1.3 & \qquad Daily newspaper - Der Landanzeiger & single choice & 6 [Never, 1-2 times, 3-4 times, 5 and more times, Don't know, No answer]\\
\qquad EB.1.4 & \qquad Radio - Argovia & single choice & 6 [Never, 1-2 times, 3-4 times, 5 and more times, Don't know, No answer]\\
\qquad EB.1.5 & \qquad Radio - Kanal K & single choice & 6 [Never, 1-2 times, 3-4 times, 5 and more times, Don't know, No answer]\\
\qquad EB.1.6 & \qquad Regional journal - Aargau Solothurn (Radio SRF 1) & single choice & 6 [Never, 1-2 times, 3-4 times, 5 and more times, Don't know, No answer]\\
\qquad EB.1.7 & \qquad Online News - Tele M1 & single choice & 6 [Never, 1-2 times, 3-4 times, 5 and more times, Don't know, No answer]\\
\qquad EB.1.8 & \qquad Online News - aarau.ch & single choice & 6 [Never, 1-2 times, 3-4 times, 5 and more times, Don't know, No answer]\\
\qquad EB.1.9 & \qquad Poster & single choice & 6 [Never, 1-2 times, 3-4 times, 5 and more times, Don't know, No answer]\\
\qquad EB.1.10 & \qquad Local neighborhood newspaper & single choice & 6 [Never, 1-2 times, 3-4 times, 5 and more times, Don't know, No answer]\\
\qquad EB.1.11 & \qquad Webpage of the Stadtidee & single choice & 6 [Never, 1-2 times, 3-4 times, 5 and more times, Don't know, No answer]\\
\qquad EB.1.12 & \qquad Instagram & single choice & 6 [Never, 1-2 times, 3-4 times, 5 and more times, Don't know, No answer]\\ \midrule 

EB.2 & Below you will find some statements on discussions on the Stadtidee and the voting. Please indicate on a scale of 0 to 10 how strongly you agree with these statements. 0 stands for "not at all" and 10 for "completely". You can use the values in between to grade your answer. & group of questions & 4 questions\\
\qquad EB.2.1 & \qquad I have informed many of my friends about the Stadtidee. & ratio scale & 10 [do not agree at all, fully agree]\\
\qquad EB.2.2 & \qquad I have tried to motivate many to participate in the Stadtidee & ratio scale & 10 [do not agree at all, fully agree]\\
\qquad EB.2.3 & \qquad I was often asked questions about the Stadtidee & ratio scale & 10 [do not agree at all, fully agree]\\
\qquad EB.2.4 & \qquad In political discussions about the Stadtidee, I have always been the most vocal & ratio scale & 10 [do not agree at all, fully agree]\\ \midrule 
\qquad EB.2.5 & \qquad The result of the vote was as I expected & ratio scale & 10 [do not agree at all, fully agree]\\ \midrule 
\qquad EB.2.6 & \qquad The number of projects put to the vote was overwhelming & ratio scale & 10 [do not agree at all, fully agree]\\ \midrule 
\qquad EB.2.7 & \qquad The quality of information about the projects was very poor & ratio scale & 10 [do not agree at all, fully agree]\\ \midrule 

EB.3 & Which Stadtidee events did you attend? & group of questions & 10 questions\\
\qquad EB.3.1 & \qquad Stand action at the weekly market (Saturday 18.2.2023) & single choice & 4 [I participated, I did not participate, Don't know, No answer]\\
\qquad EB.3.2 & \qquad Kick-off event for the city idea as part of the Digital Days Aarau (Friday 16.9.2022) & single choice & 4 [I participated, I did not participate, Don't know, No answer]\\
\qquad EB.3.3 & \qquad Brainstorming Aarau North (Saturday 4.3.2023) & single choice & 4 [I participated, I did not participate, Don't know, No answer]\\
\qquad EB.3.4 & \qquad Info evening with forära (Monday 6.3.2023) & single choice & 4 [I participated, I did not participate, Don't know, No answer]\\
\qquad EB.3.5 & \qquad Cargobike tour (Saturday 18.3. 2023) & single choice & 4 [I participated, I did not participate, Don't know, No answer]\\
\qquad EB.3.6 & \qquad Idea submission assistance at city office (Wednesday 22.2. 2023) & single choice & 4 [I participated, I did not participate, Don't know, No answer]\\
\qquad EB.3.7 & \qquad Idea submission assistance at city office (Wednesday 1.3. 2023) & single choice & 4 [I participated, I did not participate, Don't know, No answer]\\  
\qquad EB.3.8 & \qquad Idea submission assistance at city office (Wednesday 8.3. 2023) & single choice & 4 [I participated, I did not participate, Don't know, No answer]\\		
\qquad EB.3.9 & \qquad Idea submission assistance at city office (Wednesday 15.3. 2023) & single choice & 4 [I participated, I did not participate, Don't know, No answer]\\		
\qquad EB.3.10 & \qquad Ideas Exchange (Thursday 27.4.023) & single choice & 4 [I participated, I did not participate, Don't know, No answer]\\ \midrule		

EB.4 & Below are some statements on the exchange of information in the run-up to the vote on the Stadtidee. We are interested in how far you agree with the statements. & group of questions & 2 questions\\
\qquad EB.4.1 & \qquad I had enough opportunities for information exchange & ratio scale & 13 [Do not agree at all [0] to Agree completely [10], Don't know, No answer] \\
\qquad EB.4.2 & \qquad I had enough information about the proposed projects & ratio scale & 13 [Do not agree at all [0] to Agree completely [10], Don't know, No answer] \\

\qquad EB.4.3 & \qquad I was informed well about the voting platform & ratio scale & 13 [Do not agree at all [0] to Agree completely [10], Don't know, No answer] \\
\qquad EB.4.4 & \qquad I had enough information about how the final
vote would be calculated & ratio scale & 13 [Do not agree at all [0] to Agree completely [10], Don't know, No answer] \\

\qquad EB.4.5 & \qquad I did not feel like I had enough information
about the voting process & ratio scale & 13 [Do not agree at all [0] to Agree completely [10], Don't know, No answer] \\
EB.5 & Have you (co-)submitted a project for Stadtidee? & single choice  & 4 [Yes, No, Don't know, No answer]\\\midrule

EB.6 & What were your reasons to participate in the Stadtidee vote? You may tick more than one answer. & multiple choice  & 11 [Support for one or more projects, Interest in a new form of participation, Civic duty, To have my say on how the local budget is spent, To know what Stadtidee is about, To experience the online voting platform, Someone encouraged me, Many others have also participated, Other reason (please state), Don't know, No answer]\\\midrule   

EB.7 & What reasons led you not to take part in the Stadtidee idea? You may tick more than one answer. & multiple choice  & 14 [I did not know anything about it, Low interest in local politics, Constraints (illness, accident, absence), No time (family commitments, appointments), I forgot to participate, The projects did not appeal to me, No trust in the voting process, My vote would not have much impact on the outcome, It was too complicated for me, Nobody asked or motivated me to do so, I had a technical problem with the platform that prevented me from voting, Other reason (please state), Don't know, No answer]\\\midrule   

EB.8 & Have you gained new social contacts during and in connection with the Stadtidee? & single choice  & 5 [Yes - few new contacts, Yes - many new contacts, No, Don't know, No answer]\\\midrule  
\end{tabular}
}
\end{table}

\clearpage	
\section{City Idea Voting}

\begin{table}[!htb]
\caption{Survey questions for the entry phase – City Idea voting} \label{table:entry-city idea}
\centering
\resizebox{\textwidth}{!}{%
\begin{tabular}{p{0.15\textwidth}p{0.8\textwidth}p{0.19\textwidth}p{0.8\textwidth}}

\toprule

\textbf{ID} & \textbf{Question} & \textbf{Type}& \textbf{Options} \\
\midrule

VA.1 & In this spring, residents of Aarau will be able to submit their own project ideas to implement in their neighbourhood or the city. They will then be able to vote among the proposed projects to select the ones to implement with a maximum total budget of CHF 50,000. In this context: Please rate the following statements from not at all true (1) to completely true (5). & Group of questions & 6 Questions \\

\qquad VA.1.1 & \qquad I very much welcome the fact that the population can propose and choose their own projects. & Ratio Scale & 7 [Not true at all, Completely true, Don't know, No answer] \\ 
\qquad VA.1.2 & \qquad I will propose a project. & Ratio Scale & 7 [Not true at all [1] to Completely true [5], Don't know, No answer] \\ 
\qquad VA.1.3 & \qquad I will vote. & Ratio Scale & 7 [Not true at all [1] to Completely true [5], Don't know, No answer] \\ 
\qquad VA.1.4 & \qquad I think it would be better if the local authorities distributed the money instead of putting projects to vote. & Ratio Scale & 7 [Not true at all [1] to Completely true [5], Don't know, No answer] \\ 
\qquad VA.1.5 & \qquad The budget is too low. A higher budget would motivate me to propose a project & Ratio Scale & 7 [Not true at all [1] to Completely true [5], Don't know, No answer] \\ 
\qquad VA.1.6 & \qquad The budget is too low. A higher budget would motivate me to participate in voting & Ratio Scale & 7 [Not true at all [1] to Completely true [5], Don't know, No answer] \\\midrule

VA.2 & Eliminating a proposed project from the voting process should happen as follows: Never (VA.2.1), If it is not feasible (VA.2.2), If it is not popular(VA.2.3), If it does not have broad impact on the city (VA.2.4), if it is not environmentally friendly (VA.2.5), If there are other similar ideas that are more promising (VA.2.6), If it cannot be maintained in the long-term (VA.2.7), Don't know (VA.2.8), No answer (VA.2.9) & Multiple choice & No [0] and Yes [1]\\\midrule
\end{tabular}
}
\end{table}

\begin{table}[!htb]
\caption{Survey questions for the exit phase – city idea voting} \label{table:exit-city idea}
\centering
\resizebox{\textwidth}{!}{%
\begin{tabular}{p{0.15\textwidth}p{0.83\textwidth}p{0.19\textwidth}p{0.83\textwidth}}

\toprule

\textbf{ID} & \textbf{Question} & \textbf{Type}& \textbf{Options} \\
\midrule

VB.1 & Did you vote in the participatory budgeting vote ran in Aarau between June 12th to 25th 2023? & Single Choice & 3 [Yes, No, Don't Know/Can't Remember] \\\midrule
VB.2 & You voted by assigning a total of 10 points to at least 3 projects. We are interested in how you voted. Please repeat the voting by assigning ten points to the projects.  & multiple choice & 33 [Project 1, Project 2.... Project N]\\\midrule
VB.3 & Please provide the approximate cost of the projects you have selected. & multiple choice & 33 [Cost of Project 1, Cost of Project 2.... Cost of Project N]\\\midrule
VB.4 & Mark the projects that have won. & multiple choice & 33 Options [Project 1, Project 2.... Project N]\\\midrule
VB.5 & How did you vote? & single choice & 6 [Via smartphone, Using a computer or a laptop, Using a tablet, Someone helped me, I do not remember / don't know, No answer.]\\\midrule

VB.6 & How well informed were you about the projects that could be voted on in the Stadtidee? (You can select multiple) & multiple choice & 7 Options [I was not aware of the projects / not informed, I only read the project titles when I voted, I read the descriptions of the projects before I voted, I learned about the project(s) from others, I attended an event about the projects, Don't know, No answer.]\\\midrule   

VB.7 & When did you decide how you would vote?) & single choice & 5 Options [More than two weeks prior to the vote, Within the two weeks before to the vote, In the last moment (during voting), Don't know, No answer]\\\midrule 
\end{tabular}
}
\end{table}

\clearpage

\section{Voting Outcome Preference}
\begin{table}[!htb]
\caption{Survey questions for the entry phase – voting outcome preference} \label{table:entry-voting outcome}
\centering
\resizebox{\textwidth}{!}{%
\begin{tabular}{p{0.15\textwidth}p{0.83\textwidth}p{0.19\textwidth}p{0.83\textwidth}}

\toprule

\textbf{ID} & \textbf{Question} & \textbf{Type}& \textbf{Options} \\
\midrule

OA.1 & On a scale of 1 to 5, how important do you think these criteria are for the selection of projects to implement at a local level? (such as measures for climate adaptation or economic promotion)? & Group of questions & 9 questions\\
\qquad OA.1.1 & \qquad Cost efficiency & Ratio Scale & 7 [Not important at all [1] to Very important [5], Don't know, No answer] \\
\qquad OA.1.2 & \qquad Viable to maintain financially in the long-term  & Ratio Scale & 7 [Not important at all [1] to Very important [5], Don't know, No answer] \\
\qquad OA.1.3 & \qquad Environmental impact & Ratio Scale & 7 [Not important at all [1] to Very important [5], Don't know, No answer] \\
\qquad OA.1.4 & \qquad Promoting welfare and reducing inequality & Ratio Scale & 7 [Not important at all [1] to Very important [5], Don't know, No answer] \\
\qquad OA.1.5 & \qquad Benefit for neighborhood & Ratio Scale & 7 [Not important at all [1] to Very important [5], Don't know, No answer] \\
\qquad OA.1.6 & \qquad Benefit for city & Ratio Scale & 7 [Not important at all [1] to Very important [5], Don't know, No answer] \\
\qquad OA.1.7 & \qquad Benefit for myself & Ratio Scale & 7 [Not important at all [1] to Very important [5], Don't know, No answer] \\
\qquad OA.1.8 & \qquad Fair distribution of budget & Ratio Scale & 7 [Not important at all [1] to Very important [5], Don't know, No answer] \\
\qquad OA.1.9 & \qquad Others (please specify) & Text & 1 [String] \\\midrule

OA.2 & Please determine your preferences within the provided range, choosing within the left and the right option based on your preferences. & Group of questions &  7 questions \\
\qquad OA.2.1 & \qquad The money benefits children vs elderly & Ratio Scale & 5 [Fully agree with the first part of the question [1] to Fully agree with the second part of the question [5]] \\
\qquad OA.2.2 & \qquad The money benefits everyone vs myself & Ratio Scale & 5 [Fully agree with the first part of the question [1] to Fully agree with the second part of the question [5]] \\
\qquad OA.2.3 & \qquad The project is cost-efficient vs is environmental-friendly / ecologically & Ratio Scale & 5 [Fully agree with the first part of the question [1] to Fully agree with the second part of the question [5]] \\
\qquad OA.2.4 & \qquad The project strengthens equal opportunities vs is environmentally friendly/ecological & Ratio Scale & 5 [Fully agree with the first part of the question [1] to Fully agree with the second part of the question [5]]\\
\qquad OA.2.5 & \qquad The project strengthens equal opportunities vs is cost-efficient & Ratio Scale & 5 [Fully agree with the first part of the question [1] to Fully agree with the second part of the question [5]] \\
\qquad OA.2.6 & \qquad Who decides is the population vs the residents' council & Ratio Scale & 5 [Fully agree with the first part of the question [1] to Fully agree with the second part of the question [5]] \\
\qquad OA.2.7 & \qquad The money benefits local area vs citywide & Ratio Scale & 5 [Fully agree with the first part of the question [1] to Fully agree with the second part of the question [5]] \\\midrule

OA.3 & You now see nine thematic areas in which urban projects can be realised. Please select the ones you support.  The nine areas are Education (OA.3.1), Urban greenery (e.g. parks, greenery) (OA.3.2), Public space (e.g. squares) (OA.3.3), Welfare (for people living below the poverty line) (OA.3.4), Culture (OA.3.5), Environmental protection (OA.3.6), Public transit and roads (OA.3.7), Sport (OA.3.8), and Health (OA.3.9) & Multiple choice &  No [0] and Yes [1]\\\midrule

OA.4 & On a scale of 1 to 5, how important is it to you that the following group benefits from urban projects? & Group of questions & 7 questions \\
\qquad OA.4.1 & \qquad Families with children & Ratio Scale & 7 [Not important at all [1] to Very important [5], Don't know, No answer] \\
\qquad OA.4.2 & \qquad Children & Ratio Scale & 7 [Not important at all [1] to Very important [5], Don't know, No answer] \\
\qquad OA.4.3 & \qquad Youth & Ratio Scale & 7 [Not important at all [1] to Very important [5], Don't know, No answer]] \\
\qquad OA.4.4 & \qquad Adults & Ratio Scale & 7 [Not important at all [1] to Very important [5], Don't know, No answer] \\
\qquad OA.4.5 & \qquad People with disabilities & Ratio Scale & 7 [Not important at all [1] to Very important [5], Don't know, No answer] \\
\qquad OA.4.6 & \qquad Elderly & Ratio Scale & 7 [Not important at all [1] to Very important [5], Don't know, No answer] \\
\qquad OA.4.7 & \qquad Poor people & Ratio Scale & 7 [Not important at all [1] to Very important [5], Don't know, No answer] \\\midrule
\end{tabular}
}
\end{table}	

\begin{table}[!htb]
\caption{Survey questions for the exit phase – voting outcome preference} \label{table:exit-voting outcome}
\centering
\resizebox{\textwidth}{!}{%
\begin{tabular}{p{0.15\textwidth}p{0.8\textwidth}p{0.19\textwidth}p{0.8\textwidth}}
\toprule

\textbf{ID} & \textbf{Question} & \textbf{Type}& \textbf{Options} \\
\midrule

OB.1 & What's your impression of the Stadtidee voting result? Rate the following statements on a scale from 0 (do not agree at all) to 10 (fully agree). & group of questions & 9 questions\\
\qquad OB.1.1 & \qquad I am satisfied with the outcome & ratio scale & 11 [do not agree at all [0] to fully agree [10]]\\
\qquad OB.1.2 & \qquad I accept the outcome & ratio scale & 11 [do not agree at all [0] to fully agree [10]]\\
\qquad OB.1.3 & \qquad I was able to influence the outcome & ratio scale & 11 [do not agree at all [0] to fully agree [10]]\\
\qquad OB.1.4 & \qquad The winning projects are too similar & ratio scale & 11 [do not agree at all [0] to fully agree [10]]\\
\qquad OB.1.5 & \qquad The budget should have been spent on fewer projects that are more expensive & ratio scale & 11 [do not agree at all [0] to fully agree [10]]\\
\qquad OB.1.6 & \qquad The budget should have been spent on more projects that are cheaper & ratio scale & 11 [do not agree at all [0] to fully agree [10]]\\
\qquad OB.1.7 & \qquad I feel the outcome of the Stadtidee votes accurately represents the will of Aarau citizens & ratio scale & 11 [do not agree at all [0] to fully agree [10]]\\
\qquad OB.1.8 & \qquad I think it's good that people without the Swiss nationality were also allowed to vote & ratio scale & 11 [do not agree at all [0] to fully agree [10]]\\
\qquad OB.1.9 & \qquad I think it is good that minors were also allowed to vote & ratio scale & 11 [do not agree at all [0] to fully agree [10]]\\\midrule

OB.2 & When considering the Stadtidee voting outcome, I would have preferred that the selected projects... & single choice & 6 [Meet the needs of others, even if my preferred projects are not selected, Meet my needs, even if the preferred projects of others are not selected, Don't Know, No Answer.]\\\midrule

OB.3 & Imagine you can distribute 50,000 CHF for city projects (in general, regardless of the Stadtidee). Which of these options would you choose?   & single choice  & 7 [50 very small projects each costing 1,000 CH, 20 smaller projects each costing 2,500 CHF, 10 medium-sized projects each costing 5,000 CHF, 5 larger projects each costing 10,000 CHF, 1 large project costing 50,000 CHF, Don't know, No answer]\\\midrule  

OB.4 & If you were asked to vote again for the Stadtidee, how would you do so? To vote, please distribute 10 points across projects (selecting a minimum of three projects).  & multiple choice & 33 [Project 1, Project 2.... Project N]\\\midrule

OB.5 & How would you characterize the preferences you expressed now compared to the original preferences you voted for? & group of questions & 4 questions\\
\qquad OB.5.1 & \qquad My preferences changed as I am now more informed & ratio scale & 13 [Not at all true [0] to Completely true [10], Don't know, No answer] \\    
\qquad OB.5.2 & \qquad My preferences did not change; I assigned exactly or approximately the same scores to the same projects & ratio scale & 13 [Not at all true [0] to Completely true [10], Don't know, No answer] \\
\qquad OB.5.3 & \qquad I did not accept one or more winning projects and gave fewer or no points & ratio scale & 13 [Not at all true [0] to Completely true [10], Don't know, No answer] \\    
\qquad OB.5.4 & \qquad I did not accept one or more losing projects and gave them a higher score & ratio scale & 13 [Not at all true [0] to Completely true [10], Don't know, No answer] \\\midrule

OB.6 & Do you have any feedback about the Stadtidee vote that was not covered within this survey? & open ended  & \\\midrule

OB.7 & How important are the following aspects of the Stadtidee vote for you? Please rank the following aspects:  & ranking  & 8 [A fair share of the budget, Social connections with other citizens, A direct say in what happens in the city, Creative ideas for projects, Cost-effective use of the budget, Learning about new forms of participation and decision-making, Don't know, No answer]\\\midrule

OB.8 & Below are some statements on the exchange of information in the run-up to the vote on the Stadtidee. We are interested in how far you agree with the statements & group of questions & 4 questions\\
\qquad OB.8.1 & \qquad I can understand and assess important political
issues on the local level well & ratio scale  & 13 [Do not agree at all [0]  to Completely agree [10], Don't know, No answer]\\
\qquad OB.8.2 & \qquad I have confidence in the abilities of the government
and politicians on the local level & ratio scale  & 13 [Do not agree at all [0]  to Completely agree [10], Don't know, No answer]\\
\qquad OB.8.3 & \qquad I have confidence in other people & ratio scale  & 13 [Do not agree at all [0]  to Completely agree [10], Don't know, No answer]\\
\qquad OB.8.4 & \qquad The Stadtidee budget of 50'000 CHF was sufficient & ratio scale  & 13 [Do not agree at all [0]  to Completely agree [10], Don't know, No answer] 
\\\midrule
\end{tabular}
}
\end{table}

\clearpage
\begin{figure}[!htb]
\centering
\includegraphics[width=0.5\textwidth]{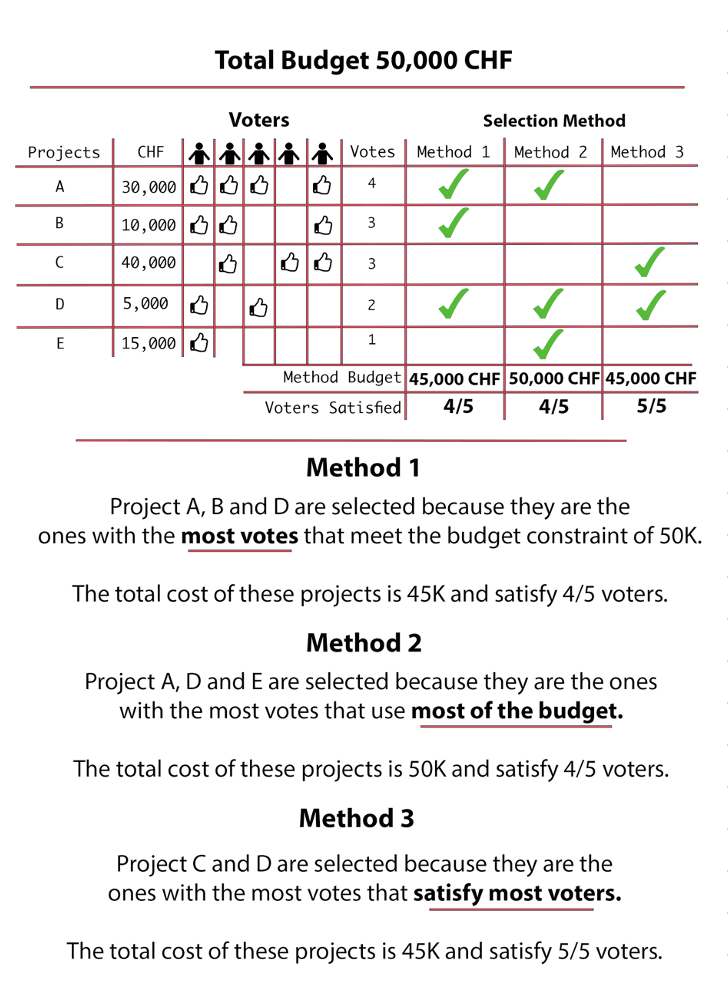} 
\caption{A hypothetical voting scenario with a budget of 50,000 CHF to distribute among 5 projects, which receive approvals by 5 voters. Three illustrative voting outcomes are calculated with three ballot aggregation methods coming with an explanation. This scenario is covered within Questions~MA.1 and~MA.5 in Table~\ref{table:entry-voting input}.}
\label{fig:ballot-aggregation}
\end{figure}
\begin{figure}[!htb]
\centering
\includegraphics[width=0.9\textwidth]{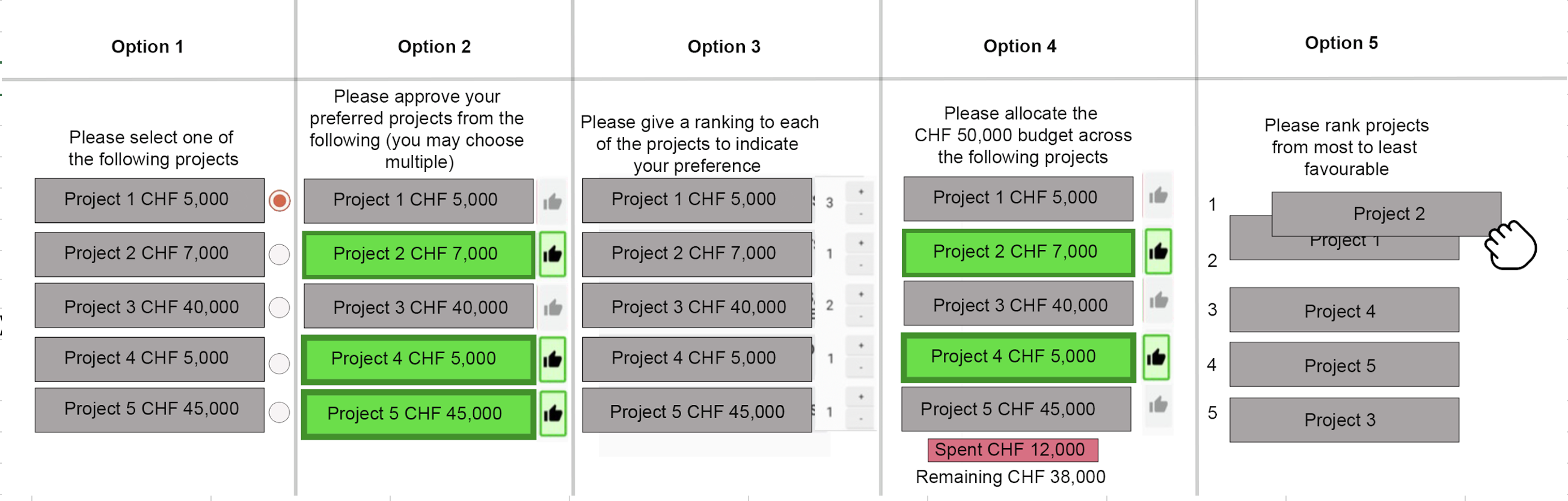}
\caption{\textbf{Choice with 5 ballot formats}: single choice, approval, score, Knapsack and ranked voting. See Questions~MA.6-MA.9 in Table~\ref{table:entry-voting input}.}
\label{fig:ballot-formats}
\end{figure}

\section{Voting Input and Aggregation Methods}

\begin{table}[!htb]
\caption{Survey questions for the entry phase – voting input and aggregation method} \label{table:entry-voting input}
\centering
\resizebox{\textwidth}{!}{%
\begin{tabular}{p{0.15\textwidth}p{0.8\textwidth}p{0.19\textwidth}p{0.8\textwidth}}

\toprule

\textbf{ID} & \textbf{Question} & \textbf{Type}& \textbf{Options} \\
\midrule

Common & \multicolumn{3}{p{1.3\textwidth}}{The city has made a budget of CHF 50,000 available. The population can propose projects and vote on them. In the following scenario, 5 projects were proposed that cost different amounts. Five people vote by indicating for each project whether they want to support it or not. You can see three selection methods of deciding on the basis of the votes cast. Please familiarize yourself with the scenario. Afterwards we will ask you a few questions. Image shown Fig.~\ref{fig:ballot-aggregation}.}  \\\midrule

MA.1 & Which method to you prefer for the selection of the projects? Please rank them from 1 to 3. With the options Method 1: most votes, Method 2: most of the budget, Method 3: satisfy most voters & Multiple choice & 5 [Most preferred, Second most preferred, Third most preferred, Don't know, No answer] \\\midrule

MA.2 & Can you briefly describe the reason / reasons for your preference? & Text & String \\\midrule

MA.3 & How fair do you perceive these three methods ? Please rank them from 1 to 3. With the options Method 1: most votes, Method 2: most of the budget, Method 3: satisfy most voters. & Multiple choice & 5 [Fairest, Second fairest, Third fairest, Don't know, No answer] \\\midrule

MA.4 & How easy are the three methods to understand/comprehend? Please rank them from 1 to 3. With the options Method 1: most votes, Method 2: most of the budget, Method 3: satisfy most voters & Multiple choice & 5 [Easiest method, Middle method, Most complex method, Don't know, No answer] \\\midrule

MA.5a & Suppose you voted for two projects and none of them was selected. To what extent does the method make you accept the outcome as legitimate? Please rank the methods from 1 to 3. With the options Method 1: most votes, Method 2: most of the budget, Method 3: satisfy most voters &  Multiple choice & 5 [Method that makes me accept the decision the most, Second most preferred, Third most preferred, Don't know, No answer] \\\midrule

MA.5b & Suppose you have voted for two projects and one of them has been selected. To what extent does the method make you accept the outcome as legitimate? Please rank the methods from 1 to 3. With the options Method 1: most votes, Method 2: most of the budget, Method 3: satisfy most voters & Multiple choice & 5. [Method that makes me accept the decision the most, Second most preferred, Third most preferred, Don't know, No answer] \\\midrule

MA.5c & Suppose you have voted for two projects and two of them have been selected. To what extent does the method make you accept the outcome as legitimate? Please rank the methods from 1 to 3. With the options Method 1: most votes, Method 2: most of the budget, Method 3: satisfy most voters &  Multiple choice & 5 [Method that makes me accept the decision the most, Second most preferred, Third most preferred, Don't know, No answer] \\\midrule

MA.6 & Voting Method 1: Please select one project that you favour. Example of voting used shown in Fig.~\ref{fig:ballot-formats}. & Single Choice &  3 [Selection, Don't know, No answer] \\\midrule

MA.7 & Voting Method 2: Please select the projects you approve (you can select multiple). Example of voting used shown in Fig.~\ref{fig:ballot-formats}. & Approval & 4 [Approve, Reject, Don't know, No answer] \\\midrule

MA.8 & Voting Method 3: Please assign a score to each project (1,2,3,4,5) based upon your preferences for the projects.  1 means lowest preference and 5 means highest. Example of voting used shown in Fig.~\ref{fig:ballot-formats}. & Score & 8 [No support at all[0] to Very Strong support[6], Don't know, No answer] \\\midrule

MA.9 & Voting Method 4: You have been given a budget of CHF 50,000, please choose your most preferred projects with a total cost that meets the budget. Example of voting used shown in Fig.~\ref{fig:ballot-formats}. & Multiple choice & 5 [Selection 1, Selection 2, Selection 3, Don't know, No answer] \\\midrule

MA.10 & Please rank, from most to least, which voting
method captures your opinion to what extent.  & Ranking & 7 [least accurately [1] to most accurately [5], Don't know, No answer]

\\\midrule
\end{tabular}
}
\end{table}

\begin{table}[!htb]
\caption{Survey questions for the exit phase – voting input and aggregation method} \label{table:exit-voting input}
\centering
\resizebox{\textwidth}{!}{%
\begin{tabular}{p{0.15\textwidth}p{0.8\textwidth}p{0.19\textwidth}p{0.8\textwidth}}

\toprule

\textbf{ID} & \textbf{Question} & \textbf{Type}& \textbf{Options} \\
\midrule

MB.1 & For voting you were required to assign 10 points to at least 3 projects. Assess the following statements on a scale from 0 (do not agree at all) to 10 (fully agree). & group of questions & 6 questions\\
\qquad MB.1.1 & \qquad The voting method was effective to capture my preferences & ratio scale & 11 [do not agree at all [0] to fully agree  [10]]\\
\qquad MB.1.2 & \qquad The method was complex & ratio scale & 11 [do not agree at all [0] to fully agree  [10]]\\
\qquad MB.1.3 & \qquad Less than 10 points to distribute would be more effective & ratio scale & 11 [do not agree at all [0] to fully agree  [10]]\\
\qquad MB.1.4 & \qquad More than 10 points to distribute would be more effective & ratio scale & 11 [do not agree at all [0] to fully agree  [10]]\\
\qquad MB.1.5 & \qquad Limiting my choices to at least 3 projects does not match my preferences & ratio scale & 11 [do not agree at all [0] to fully agree  [10]]\\
\qquad MB.1.6 & \qquad Knowing I had to distribute my votes to at least three projects encouraged me to consider a variety of different projects. & ratio scale & 11 [do not agree at all [0] to fully agree  [10]]\\\midrule

MB.2 & Would you prefer any of the following alternative methods to vote instead? & multiple choice & 8 [Selecting projects that I approve (without scoring), Selecting projects that I approve and ordering them according to my preferences, Selecting a number of projects that do not surpass the total budget, No alternative method, the one that was applied fits, Don't know, No answer]\\\midrule

MB.3 & We used a new calculation method to determine the voting result.  It is called the "methods of equal shares." Perhaps you can still remember. Please indicate on a scale of zero to ten, how much you agree with the following statements. & group of questions & 8 questions\\
\qquad MB.3.1 & \qquad The choice of the calculation method is important for me. & ratio scale & 11 [do not agree at all [0] to fully agree  [10]]\\
\qquad MB.3.2 & \qquad I understand why the method of equal shares was chosen & ratio scale & 11 [do not agree at all [0] to fully agree  [10]]\\
\qquad MB.3.3 & \qquad I understand how the method of equal shares calculated the winner projects & ratio scale & 11 [do not agree at all [0] to fully agree  [10]]\\
\qquad MB.3.4 & \qquad I would like the method of equal shares to be used in a future participatory budgeting voting & ratio scale & 11 [do not agree at all [0] to fully agree  [10]]\\
\qquad MB.3.5 & \qquad The method of equal shares benefits people like me & ratio scale & 11 [do not agree at all [0] to fully agree  [10]]\\
\qquad MB.3.6 & \qquad I understand why some of the projects with more votes are not funded. & ratio scale & 11 [do not agree at all [0] to fully agree  [10]]\\
\qquad MB.3.7 & \qquad I think it's fair that some projects with more votes are not funded. & ratio scale & 11 [do not agree at all [0] to fully agree  [10]]\\
\qquad MB.3.8 & \qquad I think the projects with the most votes should be chosen, regardless of cost & ratio scale & 11 [do not agree at all [0] to fully agree  [10]]\\\midrule

MB.4 & Which outcome do you prefer? Image shown Fig.~\ref{fig:q2224graphic} & single choice  & 5 [Outcome by method 1, Outcome by method 2, Outcome by method 3, Don't know, No answer]\\\midrule    

MB.5 & Can you briefly describe the reason(s) for your preference? & open ended  & String\\\midrule  

MB.6 & Which outcome do you perceive as the most fair? Image shown Fig.~\ref{fig:q2224graphic}. & single choice & 5 [Outcome by method 1, Outcome by method 2, Outcome by method 3, Don't know, No answer]\\\midrule    

MB.7 & Which method do you prefer? Image shown Fig.~\ref{fig:q2224graphic}. & single choice & 5 [Outcome by method 1, Outcome by method 2, Outcome by method 3, Don't know, No answer]\\\midrule    

MB.8 & Can you briefly describe the reason(s) for your preference? & open ended  & String\\\midrule  

MB.9 & Which outcome do you perceive as the most fair? Image shown Fig.~\ref{fig:q2224graphic}. & single choice & 5 [Outcome by method 1, Outcome by method 2, Outcome by method 3, Don't know, No answer]\\\midrule  

MB.10 & Which method is the easier to understand/comprehend? Image shown Fig.~\ref{fig:q2224graphic}. & sigle choice & 5 [Outcome by method 1, Outcome by method 2, Outcome by method 3, Don't know, No answer]\\\midrule

MB.11 & Which way of budget allocation (traditionally, for example, via parliament or, as in the case of the city idea, via the population) would most likely lead to the outcomes you want. Please indicate that for the following keywords. & group of questions & 9 questions\\
\qquad MB.11.1 & \qquad A cost-effective allocation of funding & ratio scale & 13 [Traditional budget allocation (e.g. decision by Einwohnerrat or Stadtrat) [0] to Stadtidee (citizen vote) [10], Don't know, No answer]\\
\qquad MB.11.2 & \qquad Resilient to corruption & ratio scale & 13 [Traditional budget allocation (e.g. decision by Einwohnerrat or Stadtrat) [0] to Stadtidee (citizen vote) [10], Don't know, No answer]\\
\qquad MB.11.3 & \qquad A fair distribution of funding & ratio scale & 13 [Traditional budget allocation (e.g. decision by Einwohnerrat or Stadtrat) [0] to Stadtidee (citizen vote) [10], Don't know, No answer]\\
\qquad MB.11.4 & \qquad An inclusive distribution of funding that would benefit minority groups & ratio scale & 13 [Traditional budget allocation (e.g. decision by Einwohnerrat or Stadtrat) [0] to Stadtidee (citizen vote) [10], Don't know, No answer]\\    
\qquad MB.11.5 & \qquad A legitimate distribution of funding & ratio scale & 13 [Traditional budget allocation (e.g. decision by Einwohnerrat or Stadtrat) [0] to Stadtidee (citizen vote) [10], Don't know, No answer]\\
\qquad MB.11.6 & \qquad Democratic & ratio scale & 13 [Traditional budget allocation (e.g. decision by Einwohnerrat or Stadtrat) [0] to Stadtidee (citizen vote) [10], Don't know, No answer]\\
\qquad MB.11.7 & \qquad Transparant budget distribution & ratio scale & 13 [Traditional budget allocation (e.g. decision by Einwohnerrat or Stadtrat) [0] to Stadtidee (citizen vote) [10], Don't know, No answer]]\\
\qquad MB.11.8 & \qquad Trustworthy to distribute funds & ratio scale & 13 [Traditional budget allocation (e.g. decision by Einwohnerrat or Stadtrat) [0] to Stadtidee (citizen vote) [10], Don't know, No answer]\\
\qquad MB.11.9 & \qquad Traditional budget allocation (e.g. decision by Einwohnerrat or Stadtrat) & ratio scale & 13 [Traditional budget allocation (e.g. decision by Einwohnerrat or Stadtrat) [0] to Stadtidee (citizen vote) [10], Don't know, No answer]\\\midrule  

MB.12 & Take a look at the figure below and let us know how much you agree with the following statements. Image shown Fig.~\ref{fig:q31graphic}. & group of questions & 5 questions\\
\qquad MB.12.1 & \qquad I am familiar with this figure & ratio scale & 6 [Totally disagree, Rather disagree, Rather agree, Fully agree, Don't know, No answer]\\\midrule
\qquad MB.12.2 & \qquad I understand this figure & ratio scale  & 6 [Totally disagree, Rather disagree, Rather agree, Fully agree, Don't know, No answer]\\\midrule
\qquad MB.12.3 & \qquad I understand why some of the projects with more votes are not funded & ratio scale & 6 [Totally disagree, Rather disagree, Rather agree, Fully agree, Don't know, No answer]\\\midrule
\qquad MB.12.4 & \qquad I think it is fair that an expensive project needs more support & ratio scale & 6 [Totally disagree, Rather disagree, Rather agree, Fully agree, Don't know, No answer]\\\midrule
\qquad MB.12.5 & \qquad This figure is too complicated & ratio scale & 6 [Totally disagree, Rather disagree, Rather agree, Fully agree, Don't know, No answer]\\\midrule

\end{tabular}
}
\end{table}

\begin{figure}[!htb]
\centering
\includegraphics[width=1.0\textwidth]{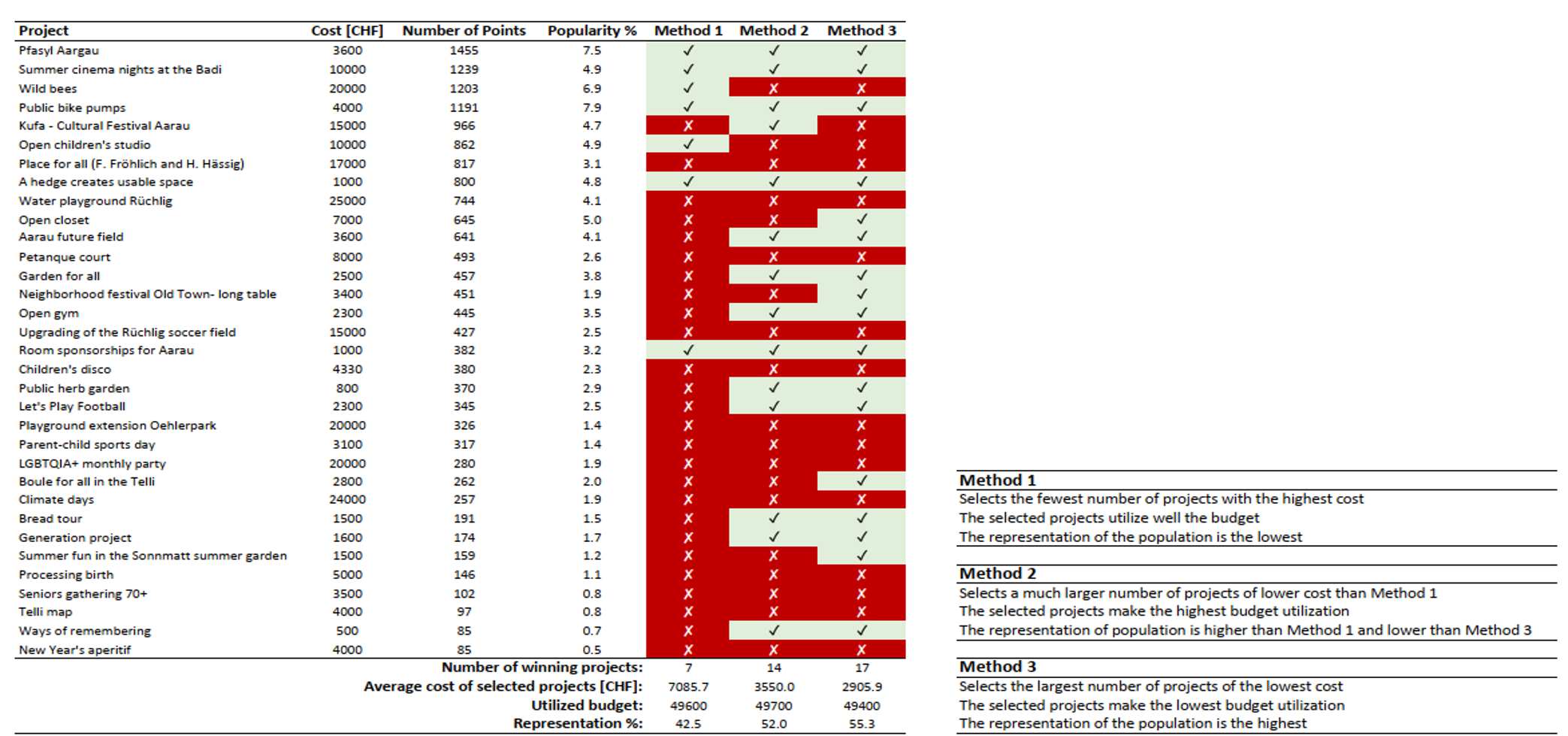} 
\caption{Voting outcomes with different ballot aggregation methods and their explanation. Method 1: utilitarian greedy. Method 2: Equal shares with greedy completion. 3. Equal shares with Add1U. Image accompanied Questions MB.4 to MB.10 in Table~\ref{table:exit-voting input}.}
\label{fig:q2224graphic}
\end{figure}

\begin{figure}[h]
\centering
\includegraphics[width=0.7\textwidth]{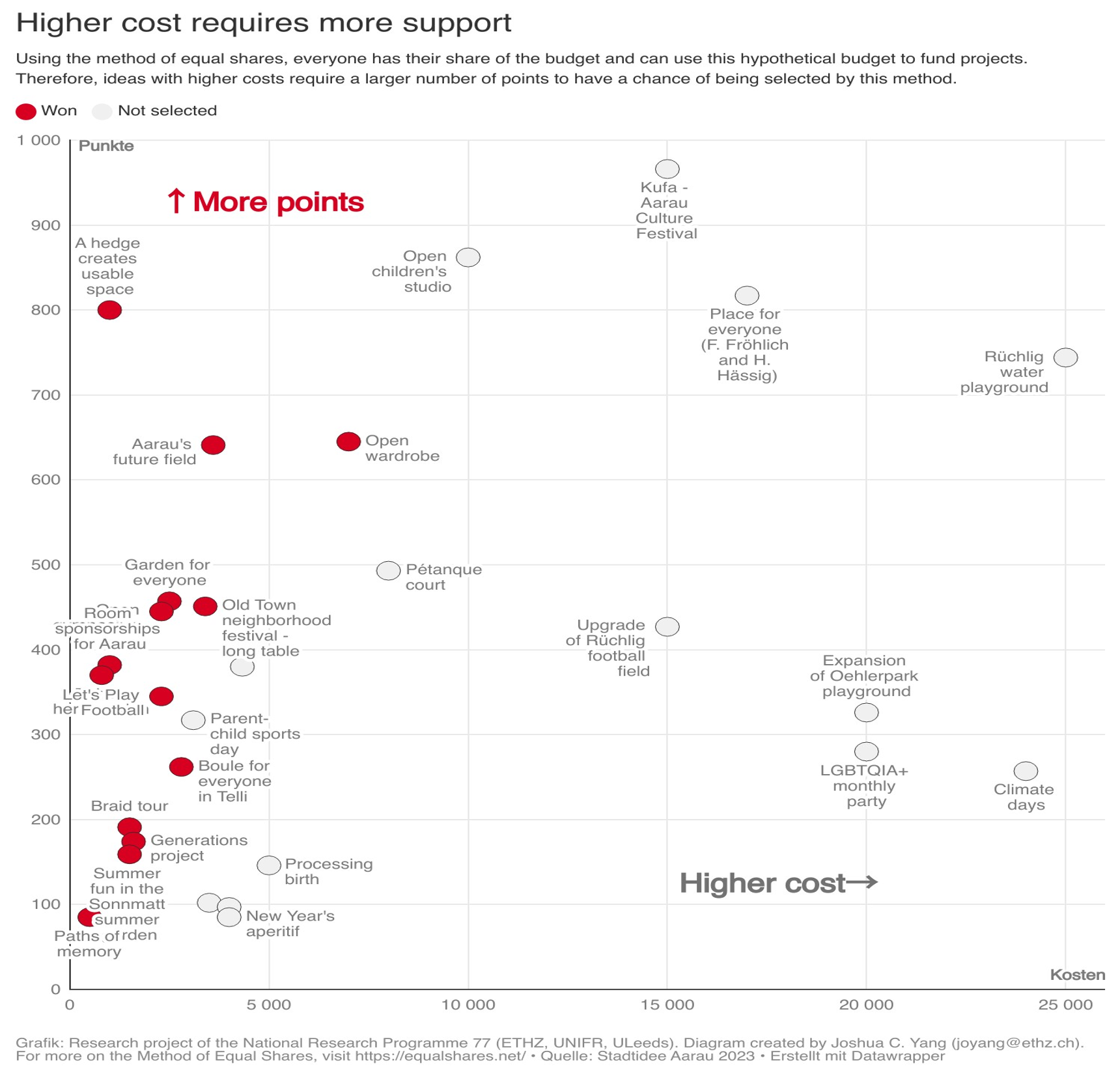} 
\caption{Cost based analysis of the voting aggregation. Question MB.12 in Table~\ref{table:exit-voting input}}
\label{fig:q31graphic}
\end{figure}

\clearpage	

\section{The Influence of Ballot Aggregation on Voter Representation}\label{sec:voter-representation}

\begin{figure}[!htb]
\centering
\includegraphics[width=0.5\textwidth]{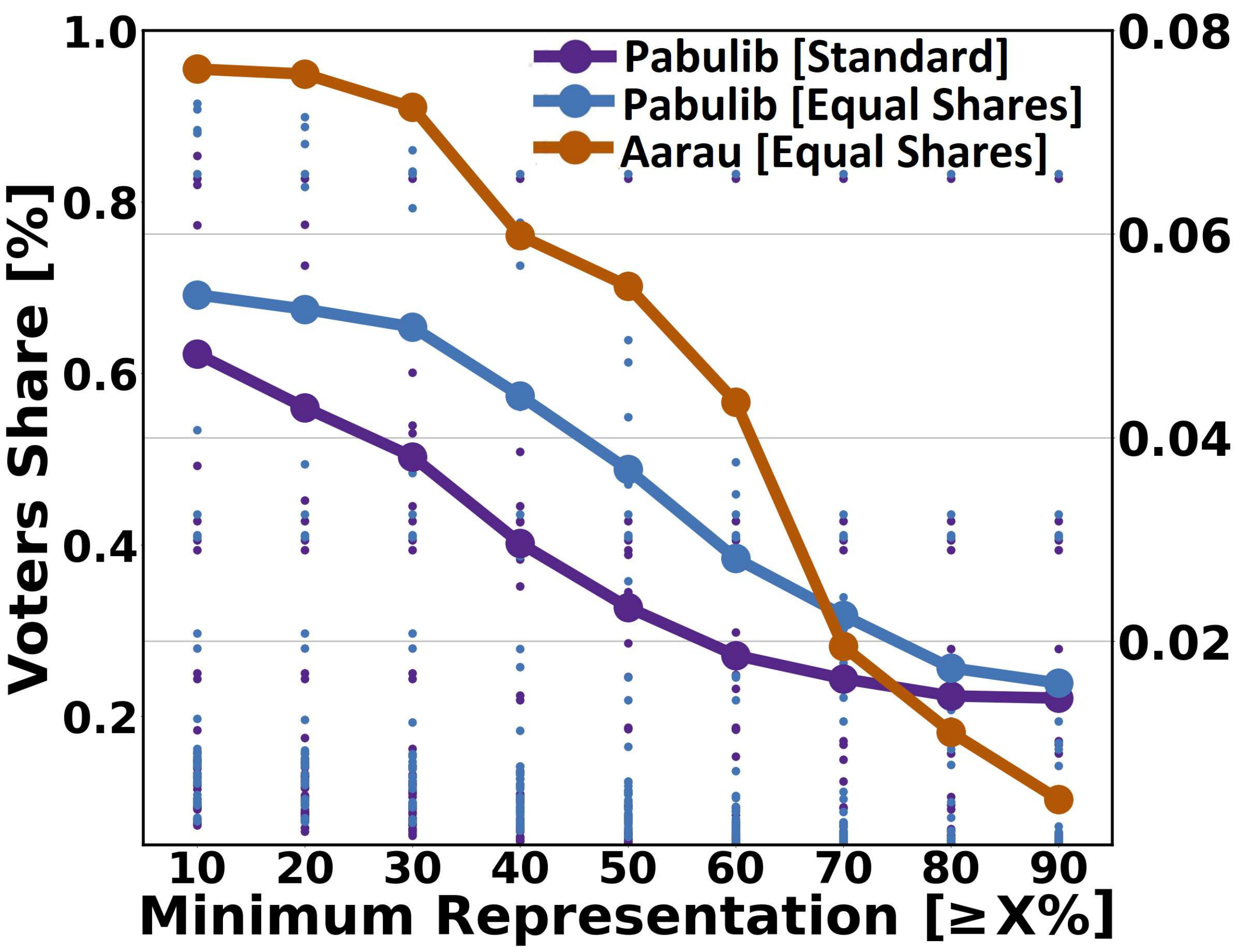}
\caption{\textbf{22.6\% more voters on average get any minimum representation level using the Method of Equal Shares as compared to earlier participatory budgeting processes based on the standard utilitarian greedy method.} Compared to Figure~\ref{fig:representation} in the main paper, this figure calculates the mean representation across all voters and elections.}
\label{fig:pabulib-representation-voters}
\end{figure}

\cparagraph{Calculating the minimum voters representation} Figure~\ref{fig:pabulib-representation-voters} complements Figure~\ref{fig:representation} in the main paper, averaging the representation of the voters over all election (instead of averaging per election and then among elections). The results and conclusions are similar to what is presented in the main paper. 

\cparagraph{Matching participatory budgeting elections for comparison} Three matching criteria are selected to identify past elections with high similarity with the Aarau participatory budgeting campaign to compare with rigorously. These criteria include: (i) \emph{Number of proposed projects}, (ii) \emph{relative budget allocation}, which is the total available budget divided by the cost of all proposed projects, and (iii) \emph{project budget share}, which is the mean budget cost of the proposed projects divided by the total available budget. These three criteria are chosen to eliminate irrelevant elections and control for these factors that can strongly influence the representation calculations, i.e in elections with more proposed projects and budget, there are more ways for voters to end with higher representation. The matching is done based on quartiles. The participatory budgeting process in Aarau belongs to the top quartile (Q4) in terms of having a high number of proposed projects. It also belongs to the lowest quartile (Q1) in terms of very low relative budget allocation and project budget share. In other words, the case of Aarau represents a scenario where many low-cost projects compete for a small budget, which is a good crash test for assessing voters representation. Table~\ref{tab:representation-gain} shows the representation gain of Aarau based on Equal Shares vs. the past election of Pabulib with the original standard method and a hypothetical application of Equal Shares. Both comparisons are shows for (i) each combination of criteria (4 columns) and (ii) 3 ballot formats (approval, cumulative and jointly both).

\begin{table*}[!htb]
\centering
\caption{Representation gain between the participatory budgeting process of Aarau based on the Method of Equal Shares vs. past elections of Pabulib that used the standard majoritarian method and a hypothetic application of Equal Shares. The representation gain is shown for four combination of the three matching criteria and two ballot formats, including their joint set.}
\label{tab:representation-gain}
\resizebox{\textwidth}{!}{
\begin{tabular}{l l l l l l l l l l}
\hline
\textbf{Similarity Criteria} & \textbf{Quartile} 
& \multicolumn{4}{c}{\makecell{\textbf{Aarau (Equal Shares) vs} \\ \textbf{Pabulib (Standard)}}} 
& \multicolumn{4}{c}{\makecell{\textbf{Aarau (Equal Shares) vs} \\ \textbf{Pabulib (Equal Shares)}}} \\
\hline

Number of Projects & Q4 (high)
& $\checkmark$ & $\checkmark$ & $\checkmark$ & $\checkmark$
& $\checkmark$ & $\checkmark$ & $\checkmark$ & $\checkmark$ \\

Relative Budget Allocation & Q1 (low)
& $\times$ & $\checkmark$ & $\times$ & $\checkmark$
& $\times$ & $\checkmark$ & $\times$ & $\checkmark$ \\

Project Budget Share & Q1 (low)
& $\times$ & $\times$ & $\checkmark$ & $\checkmark$
& $\times$ & $\times$ & $\checkmark$ & $\checkmark$ \\\hline
Number of Elections & & 182 & 77 & 126 & 38 &182 & 77 & 126 & 38  \\
\hline
\textbf{Ballot formats} & & & & & & & & & \\ \hline

Approval   &  
& 8.4\%  & 16.3\% & 0.4\% & 22.5\%
& -0.14\% & 5.2\%  & -4.1\% & 14.1\% \\

Cumulative &  
& 10.1\% & 18.1\% & 9.6\% & 21.8\%
& -4.0\% & 6.7\%  & -6.7\% & 5.6\% \\\hline

\textbf{Joint}      &  
& \textbf{9.0}\%  & \textbf{17.8}\% & \textbf{9.0}\% & \textbf{21.9}\%
& \textbf{-3.5}\% & \textbf{6.5}\%  & \textbf{-6.5}\% & \textbf{7.1}\% \\

\hline
\end{tabular}
}
\end{table*}

\cparagraph{Evaluation of representation gain} The results in Table~\ref{tab:representation-gain} show that the approach of Aarau comes with the highest representation gain (21.9\%) when compared to election of similar nature (all used criteria). This confirms that this approach generalizes well as it can be used as a blueprint to get the maximum benefits for citizens. 


\section{Choice Analysis of Ballot Aggregation Methods}\label{sec:choice-ballot-aggregation}

\begin{figure}[!htb]
\centering
\includegraphics[width=1.0\textwidth]{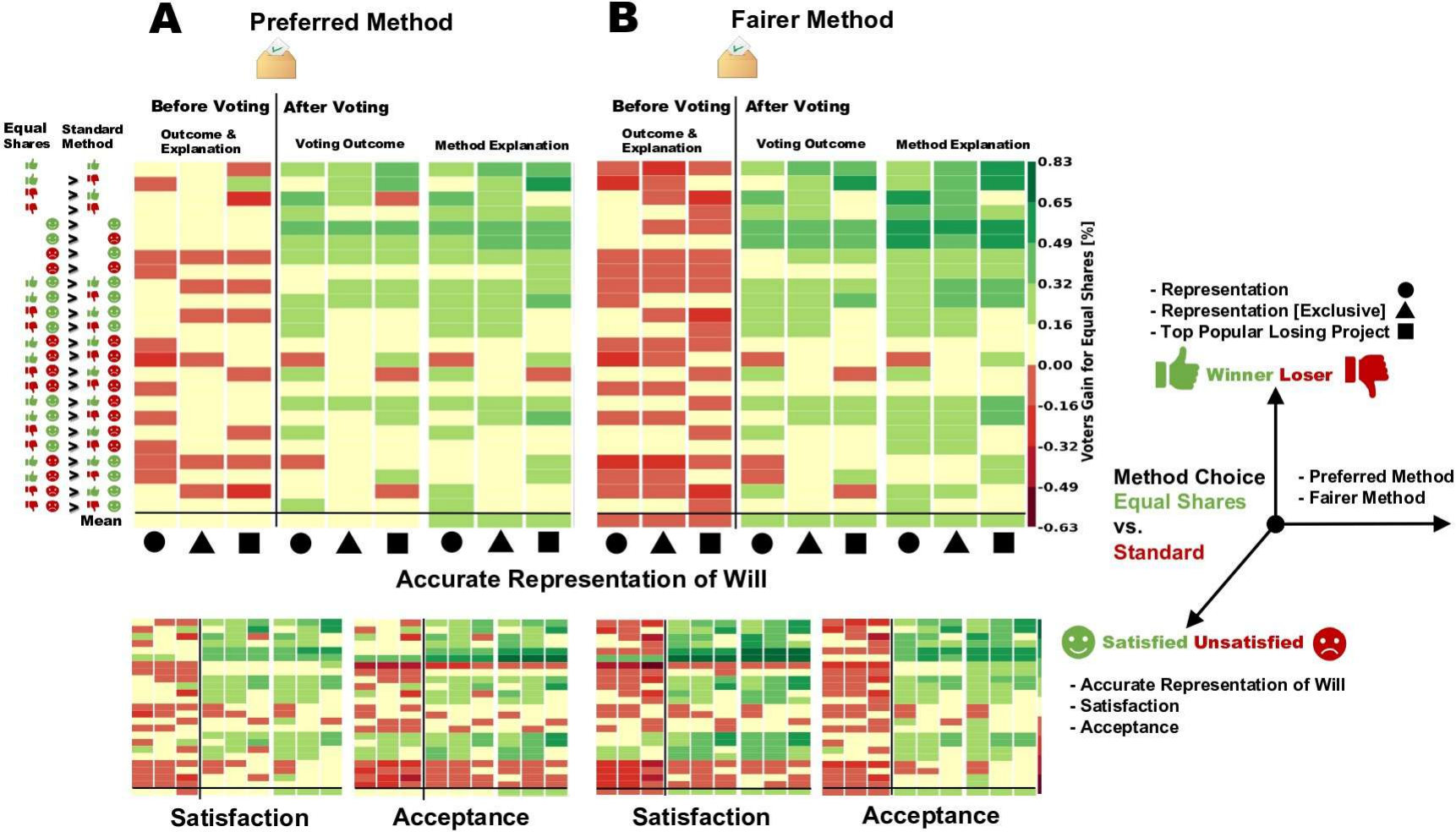}
\caption{\textbf{Equal shares is found more preferred and fairer after voting than the standard ballot aggregation method. Voting outcomes themselves are sufficient for individuals to determine their choice of ballot aggregation method. The explanation of the method does not significantly contribute to shifting the choice to equal shares. There is no clear preference over the two methods before voting, while the greedy method is found fairer}. The legitimacy of equal shares is assessed over 24 criteria that determine the (A) preference and (B) fairness of equal shares over the standard method for all 24 combinations of winners/losers and satisfied/unsatisfied voters. Winners/losers are determined in three different ways: 1st and 4th quartile of representation  for the winning projects that are (i) non-exclusive and (ii) exclusive between the two ballot aggregation methods. (iii) The non-supporters and supporters of the top popular losing project. Satisfied/unsatisfied are determined by three post-voting survey questions that assess: (i) satisfaction (OB1.1), (ii) acceptance of outcome (OB1.2) and (iii) accurate representation of will (OB1.7). The colors determine the difference between the voters share between equal shares and the standard method. The `*' determines whether the differences between the different stages in statistically significant (p<0.05).}
\label{fig:legitimacy_fairness_preference_heat}
\end{figure}

Voters' perceptions regarding the preference and fairness of the equal shares and standard methods are illustrated in Figure~\ref{fig:legitimacy_fairness_preference_heat}. Pre-voting perceptions of preference and fairness were obtained from the entry survey questions MA.1 and MA.3 in Table~\ref{table:entry-voting input}, respectively. Post-voting perceptions were captured through exit survey questions (Table~\ref{table:exit-voting input}) -  MB.4 (based only on the outcome) and MB.7 (based on outcome and explanations) for preference, and MB.6 (based only on the outcome) and MB.9 (based on outcome and explanations) for fairness.

\begin{figure}[!htb]
\centering
\includegraphics[width=0.98\textwidth]{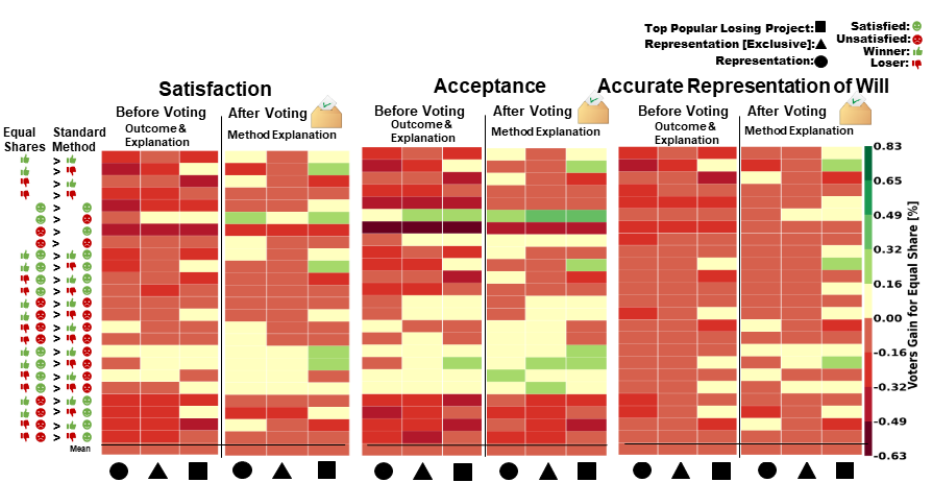}
\caption{\textbf{Overall, equal shares is perceived simpler after the explanations, however voters who preferred the standard method do not find  equal shares simple even after explanations. Voters who are satisfied and prefer equal shares finds the method even more  simpler after the explanations.} The legitimacy of equal shares is assessed over 24 criteria that determine the (A) preference and (B) fairness of equal shares over the standard method for all 24 combinations of winners/losers and satisfied/unsatisfied voters. Winners/losers are determined in three different ways: 1st and 4th quartile of representation score for the winning projects that are (i) non-exclusive and (ii) exclusive between the two ballot aggregation methods. (iii) The non-supporters and supporters of the top popular losing project. Satisfied/unsatisfied voters are determined by three post-voting survey questions that assess: (i) satisfaction (OB1.1), (ii) acceptance of outcome (OB1.2) and (iii) accurate representation of will (OB1.7). The colors determine the difference of the voters share between equal shares and the standard method.}
\label{fig:convenience}
\end{figure}

Further,  perceptions of voters regarding the simplicity of the equal shares and the standard methods are analyzed in Figure~\ref{fig:convenience}. Voters were asked about the perceived simplicity of these aggregation methods before voting, as part of the pre-voting survey (Question MA.4 in Table~\ref{table:entry-voting input}), and again after the methods were explained, in the post-voting survey (Question MB.10 in Table~\ref{table:exit-voting input}). The shifts in voters' perceptions are illustrated in Figure~\ref{fig:convenience}. Notably, we observe that voters who prefer only the standard method continue to find the equal shares method similarly complex, even after receiving an explanation.

\noindent {\bf When shift to equal shares is not achieved.}
We also found some cases where more voters shifted to the standard majoritarian voting method after voting, namely (i) satisfied voters preferring the standard method over unsatisfied voters preferring equal shares; (ii) voters accepting the voting outcome and finding the standard method fairer over voters who do not accept the voting outcome and find equal shares fairer; (iii) unsatisfied losers (or winners), who prefer the standard method over unsatisfied winners (or losers), who choose equal shares, when winners/losers are determined by the representation and their choice for the top popular losing project (`Wild bees' paradise'). See Figure~\ref{fig:legitimacy_fairness_preference_heat} for a more disaggregated view.

\noindent {\bf Pairwise Winners.} Based on the 33 projects, 73 voters were asked to evaluate a randomly selected set of project pairs, ranging from 3 to 40 pairs per voter, and indicate their preference for each pair. For any given pair, a voter could prefer one project over the other or express no preference (i.e., a tie). In total, 1,682 project pairs were evaluated. Using the Condorcet method~\cite{navarrete2024understanding, kulakowski2022similarity} to evaluate pairwise wins, we find that the project Pfasyl Aargau has the highest number of wins and thus emerges as the Condorcet winner (Table~\ref{tab:pairwise_winners}). Interestingly, this project is also the highest-rated project in the Aarau vote, indicating that Pfasyl Aargau was consistently preferred in direct comparisons. However, the next three projects with the highest pairwise wins do not appear among the top five in the overall Aarau voting results. This  can be explained by the fact that the project pairs were selected randomly, and the voters did not evaluate all 33 projects.

\begin{table}[htbp]
\centering
\resizebox{0.8\textwidth}{!}{%
\begin{tabular}{crp{7cm}r}
\hline
\textbf{ID} & \textbf{Cost} & \textbf{Name} & \textbf{Pairwise Winners (Rank)} \\
\hline
15 & 3600 & Pfasyl Aargau & 1 \\
22 & 15000 & CufA - Cultural Festival Aarau & 2 \\
29 & 2500 & A Garden for All & 2 \\
31 & 25000 & Ruchlig water playground & 3 \\
21 & 4000 & Public bicycle pumps & 4 \\
23 & 17000 & One Place for all & 5 \\
14 & 8000 & Petanque court & 6 \\
4 & 20000 & Wild bees' paradise & 7 \\
24 & 800 & Public herb garden & 7 \\
2 & 2800 & Boule for all in Telli & 8 \\
13 & 10000 & Open children's studio & 8 \\
16 & 1000 & Sponsoring a space for Aarau & 8 \\
1 & 15000 & Upgrade Ruchlig soccer field & 9 \\
33 & 20000 & Playground extension Oehlerpark & 9 \\
3 & 1600 & Intergenerational project & 10 \\
5 & 3100 & Parent-Child Fun and Action Day & 10 \\
25 & 3600 & Aarau Future Acre & 11 \\
32 & 1000 & Usable space with a hedge & 11 \\
19 & 500 & Ways of remembering & 12 \\
27 & 4000 & New edition of the Telli Map & 12 \\
8 & 3400 & Long Table Festival & 13 \\
9 & 2300 & Let's Play Football & 13 \\
11 & 2300 & Open sports hall & 13 \\
12 & 7000 & Open closet & 13 \\
28 & 24000 & Climate days for Aarau & 13 \\
30 & 10000 & Summery cinema nights in the Badi & 13 \\
10 & 20000 & LGBTQIA+ monthly party & 14 \\
26 & 1500 & Summer fun in the Sonnmatt summer garden & 14 \\
20 & 1500 & Bread tour & 15 \\
17 & 3500 & Seniors gathering 70+ & 16 \\
7 & 4330 & Children's Disco & 17 \\
18 & 5000 & Processing birth & 18 \\
6 & 4000 & Grüezi 2024 - New Year's Party & 19 \\
\hline
\end{tabular}%
}
\caption{{\bf Pfasyl Aargau is the top scored project in the Aarau voting and also has the highest number of pairwise wins in the comparison voting phase.} Out of the 1,056 valid project pairs possible from the 33 proposed projects, voters were randomly assigned a subset of pairs to evaluate. For each assigned pair, voters indicated their preference, resulting in one project being preferred (a win) over the other, or a tie if no clear preference was expressed.}
\label{tab:pairwise_winners}
\end{table}

\section{Explaining Shifts to Ballot Aggregation Methods}\label{sec:shifts-methods}
The machine learning framework is designed using personal human traits as features, serving as independent variables, while the shift in preference for aggregation methods is treated as the dependent variable. We consider two aggregation methods, resulting in four possible scenarios: standard to standard, standard to equal shares, equal shares to standard, and equal shares to equal shares.

We develop various models to predict shifts in aggregation methods, aiming to analyze them also in the context of preference, fairness and simplicity, as well as across different survey phases, including: a) before voting to after voting (with only the outcome) and  b) before voting to after voting (with both the outcome and explanations).

We experimented with various supervised classification machine learning models, including decision trees~\cite{de2013decision}, CatBoost~\cite{hancock2020catboost}, support vector machines~\cite{mammone2009support}, and multilayer perceptrons (MLPs)\cite{wu2018development}. Given that we are working with categorical data, we found that the CatBoost classifier and the multilayer perceptron (configured with two hidden layers of 90 neurons each using ReLU activation) performed similarly and outperformed the other models. The F1-scores and accuracy for the multilayer perceptron classification are presented in Table~\ref{tab:accuracy_results}.

\begin{table}[h]
\caption{{\bf Personal human traits serve as strong predictors of shifts in aggregation methods, particularly in modeling preference changes between the pre-voting phase and post-voting phase (considering both voting outcomes and explanations), outperforming other models in explanatory performance}. Hyperparameters for the neural network-based model: two dense layers with 90 neurons each, Leaky ReLU activation function, categorical cross-entropy loss, Adam optimizer, 5 fold cross validation and the Synthetic Minority Oversampling Technique (SMOTE)~\cite{chawla2002smote} has been applied by 20\% for all classes. The model is trained for 600 epochs.}
\label{tab:accuracy_results}
\centering
\resizebox{0.8\textwidth}{!}{%
\begin{tabular}{l l c c}
\toprule
& & \textbf{F1-Score [\%]} & \textbf{Accuracy  [\%]} \\
\midrule
\multirow{3}{*}{\textbf{Preference}} 
& Before Voting to Voting Outcome & 81.98 & 83.21 \\
& Before Voting to Method Explanations & 84.56 & 85.92 \\

\midrule
\multirow{3}{*}{\textbf{Fairness}} 
& Before Voting to Voting Outcome & 80.78 & 81.32 \\
& Before Voting to Method Explanations & 83.01 & 83.44 \\

\midrule
\textbf{Simplicity} & Before Voting to Method Explanations & 74.02 & 74.51 \\
\bottomrule
\end{tabular}
}

\end{table}

\begin{table}[ht]
\caption{{\bf Trust, willingness to compromise, budget flexibility, and the perceived representation of voters play a significant role in predicting shifts in the choice of aggregation methods; both before voting and after the voting outcome is known. } Neural networks were employed to predict these shifts, treating the choice changes as dependent variables and various personal human traits as independent variables. The top two personal traits that, when removed from the feature set, resulted in the highest and second-highest losses in prediction accuracy have been identified and listed.}
\label{tab:feature-loss}

\centering
\resizebox{\textwidth}{!}{
\begin{tabular}{>{\raggedright}m{2.5cm} >{\raggedright}m{4cm} >{\raggedright}m{4.5cm} c >{\raggedright}m{4.5cm} c}
\toprule
\textbf{} & \textbf{Before Voting - After Voting} & \textbf{Top 1 Feature} & \textbf{Loss in F1-Score} & \textbf{Top 2 Feature} & \textbf{Loss in F1-Score} \\
\midrule
\textbf{Preference} 
& Before Voting to Voting Outcome 
& Supporter / Opponent of Popular Loser Project 
& 4.81 
& Compromising / Uncompromising 
& 3.79 \\
& Before Voting to Method Explanations 
& Flexible / Inflexible Budget 
& 4.17 
& Trustful / Mistrustful 
& 4.02 \\
\midrule
\textbf{Fairness} 
& Before Voting to Voting Outcome 
& Compromising / Uncompromising 
& 3.35 
& Supporter / Opponent of Family Project 
& 3.12 \\
& Before Voting to Method Explanations 
& Altruistic / Self-interested 
& 3.29 
& Flexible / Inflexible Budget 
& 3.24 \\
\midrule
\textbf{Simplicity} 
& Before Voting to Method Explanations 
& Well-represented / Under-represented 
& 1.32 
& Compromising / Uncompromising 
& 1.15 \\
\bottomrule
\end{tabular}
}

\end{table}

\begin{table}[ht]
\centering
\caption{The most relevant 10 human traits that contribute in explaining the shifts in the choice of the aggregation methods, their corresponding survey question references or data sources, and whether they were collected during the entry or exit phase.}
\label{tab:trait-survey-map}
\resizebox{\textwidth}{!}{
\begin{tabular}{ccc}
\toprule
\textbf{Trait} & \textbf{Survey Question Number / Source} & \textbf{Entry / Exit} \\
\midrule
Trustful / Mistrustful & PA.21.2 in Table~\ref{table:entry-demographics} & Entry \\
Satisfied / Unsatisfied & OB.1.1 in Table S8 & Exit \\
Supporter / Opponent of Popular Loser Project & Extracted from actual votes & --- \\
Well-represented / Under-represented & Calculated feature based on number of selected projects in the winner project & --- \\
Flexible / Inflexible with Budget & Calculated based on the average preference for low-cost projects & --- \\
Aware / Unaware of Projects & MB.13.2 in Table~\ref{table:exit-voting input} & Exit \\
Supporter / Opponent of Expressive Ballots & MB.1.1 in Table~\ref{table:exit-voting input} & Exit \\
Compromising / Uncompromising & MA.5.a in Table~\ref{table:entry-voting input} & Entry \\
Altruistic / Self-interested & OA.1.7 in Table~\ref{table:entry-voting outcome} & Entry \\
Supporter / Opponent of Family Projects & OA.4.1 in Table~\ref{table:entry-voting outcome} & Entry \\
\bottomrule
\end{tabular}
}

\end{table}

To evaluate the contribution of personal human traits in predicting shifts in the choice of aggregation methods, we conduct a feature ablation study following the approach in earlier work~\cite{hameed2022based}. This analysis allows us to measure the relative importance of each trait by observing the drop in prediction accuracy when individual features are removed. Traits that play a significant and positive role in the prediction lead to a greater loss in accuracy upon removal, compared to those with minimal contribution. Table~\ref{tab:feature-loss} presents the top two human traits whose absence results in the highest accuracy loss. These are shown separately for models aimed at understanding shifts in aggregation preferences in the context of preference, fairness, and complexity; both before and after the voting phases.

The top 10 human traits, ranked by their importance across all models based on the feature ablation study, are listed in Table~\ref{tab:trait-survey-map}. Each trait is accompanied by its corresponding survey question through which it was originally reported.

We analyze the individual contribution of the top 10 personal traits to shifts in aggregation method preferences for each voter using the SHAP (Shapley Additive Explanations) method, a local explainable AI technique~\cite{nohara2019explanation}. Support or non-support for each trait is analyzed by mapping SHAP importance values to the actual reported values for each voter, followed by a quartile-based analysis~\cite{nohara2019explanation}.

\begin{figure}[!htb]
\centering
\includegraphics[width=0.98\textwidth]{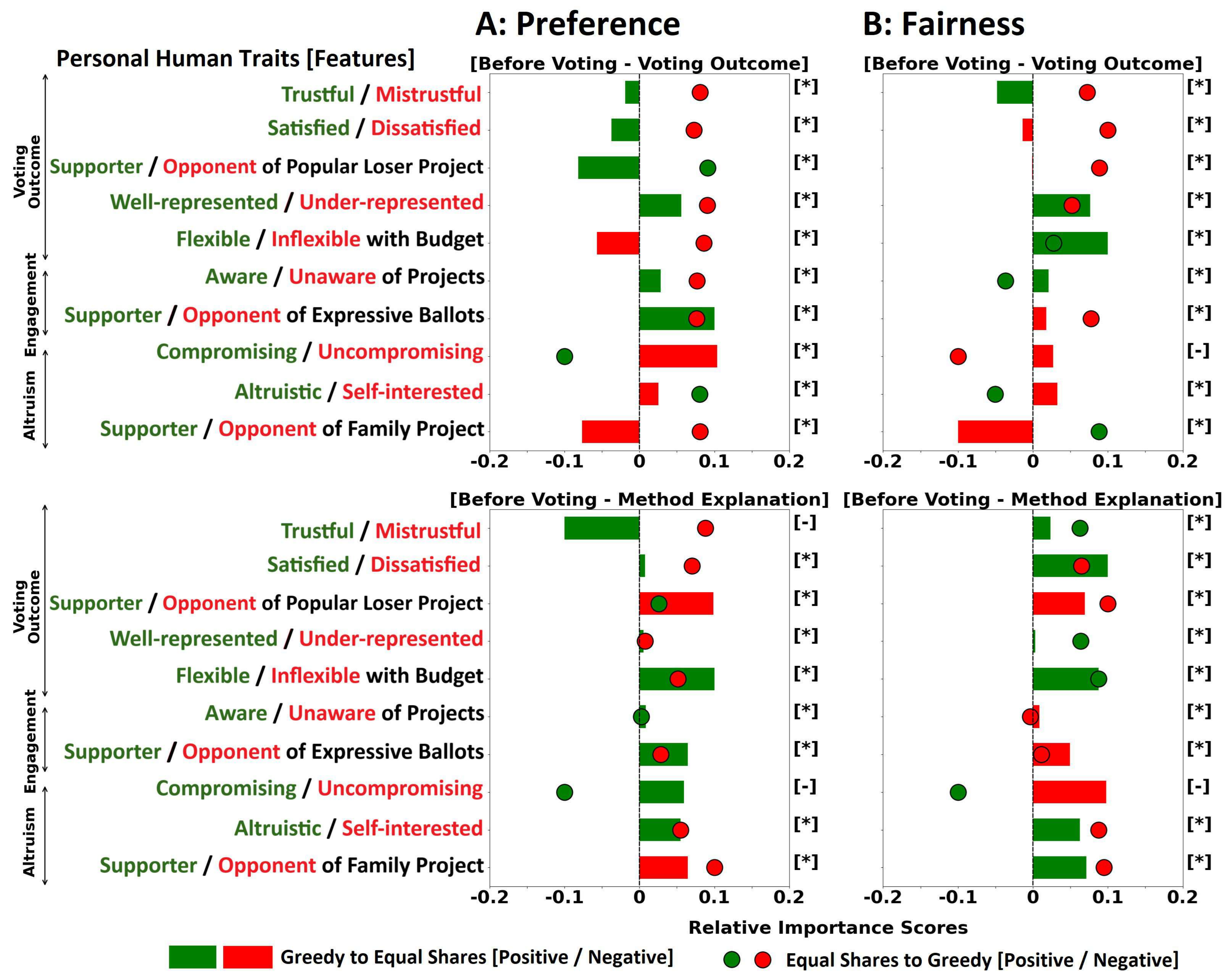}
\caption{\textbf{
The shift to equal shares as the most fair  or preferred method after voting is explained by distinguishing human traits that reflect the promotion of strong democratic values such as being altruistic, compromising and  a commitment to inclusivity, by supporting more expressive ballots or flexible budget allocations that accommodate additional family projects. Interestingly, even voters who are content with the outcome of the standard method, upon understanding how different methods work (after explanations), recognize and acknowledge the shift to equal shares as the fairest and most preferred choice. Shifting to the standard method is faced to a large extent as a backlash to these democratic values, with voters manifesting dissatisfaction, under-representation and self-interest. } Classification models (around 80\% accuracy for all 4 models, refer Table~\ref{tab:accuracy_results}), representing the preferred ballot aggregation method before and after voting (with explanation of methods). The independent variables represent voter-specific human traits that explain (A) preference and (B) perceived fairness both before and after voting, particularly in the transitions from the standard method to equal shares and vice versa. The `*' determines whether each feature is statistically significant (p<0.05) in explaining the voters' choice of method before and after voting. The analysis shown here is only for the two shift scenarios - standard to equal shares  and equal shares to standard.}
\label{fig:ES_G_G_ES}
\end{figure}

\begin{figure}[!htb]
\centering
\includegraphics[width=0.98\textwidth]{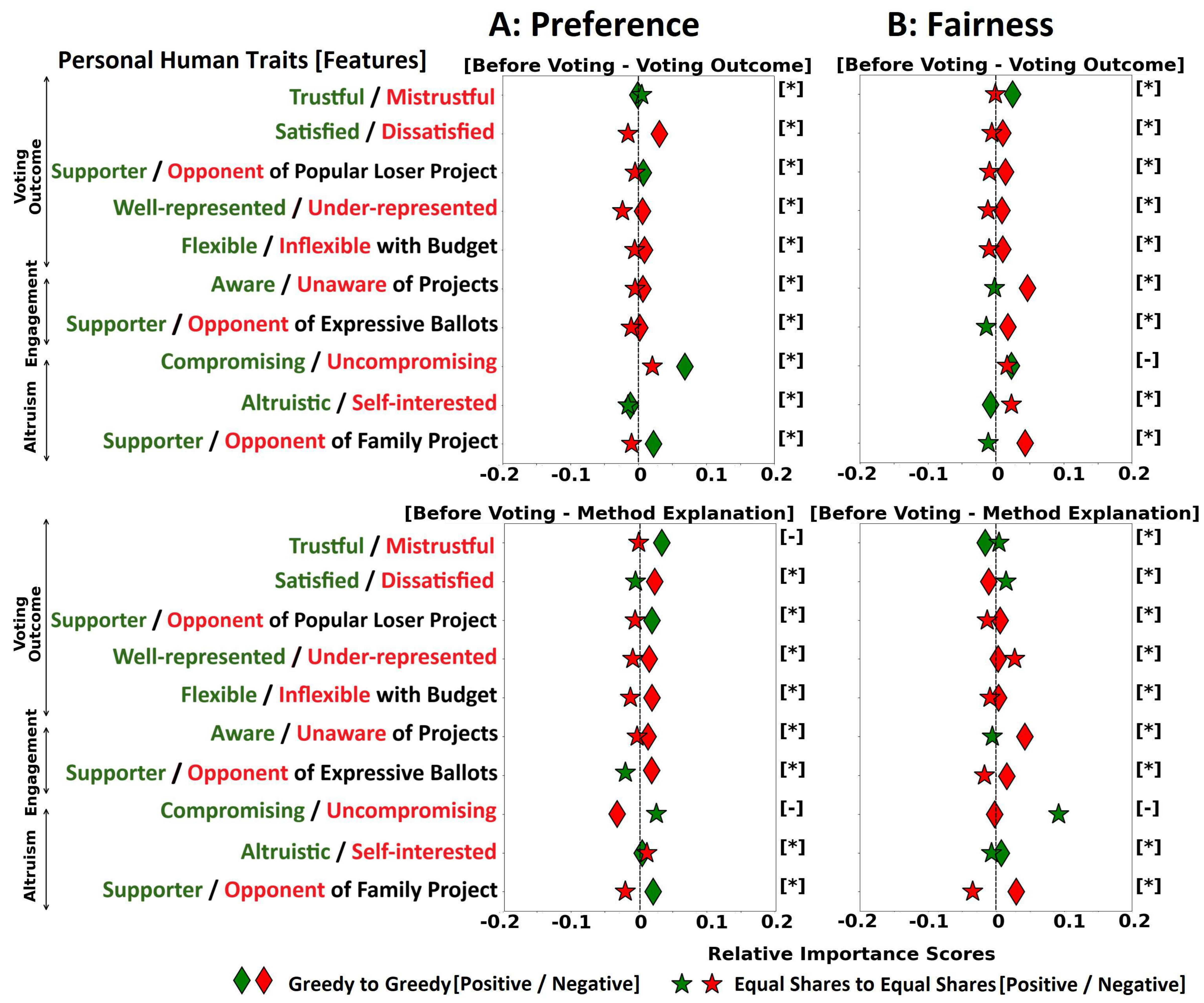}
\caption{\textbf{
The decision to remain consistent with the choice of equal shares as the fairest or most preferred voting method (after considering voting outcomes and method explanations) is explained by distinguishing human traits that promote strong democratic values, such as a willingness to compromise. On the contrary, the decision to remain consistent with the choice of the greedy method is due to voters being unaware of the projects and showing no support for expressive ballots.} Classification models (around 80\% accuracy for all 4 models, refer Table~\ref{tab:accuracy_results}), representing the preferred ballot aggregation method before and after voting (with explanation of methods). The independent variables represent voter-specific human traits that explain (A) preference and (B) perceived fairness both before and after voting, particularly in the transitions from the standard method to equal shares and vice versa. The `*' determines whether each feature is statistically significant (p<0.05) in explaining the voters' choice of method before and after voting. The analysis shown here is only for the two scenarios - standard to standard and equal shares to equal shares.}
\label{fig:ES_G}
\end{figure}

\begin{figure}[!htb]
\centering
\includegraphics[width=0.98\textwidth]{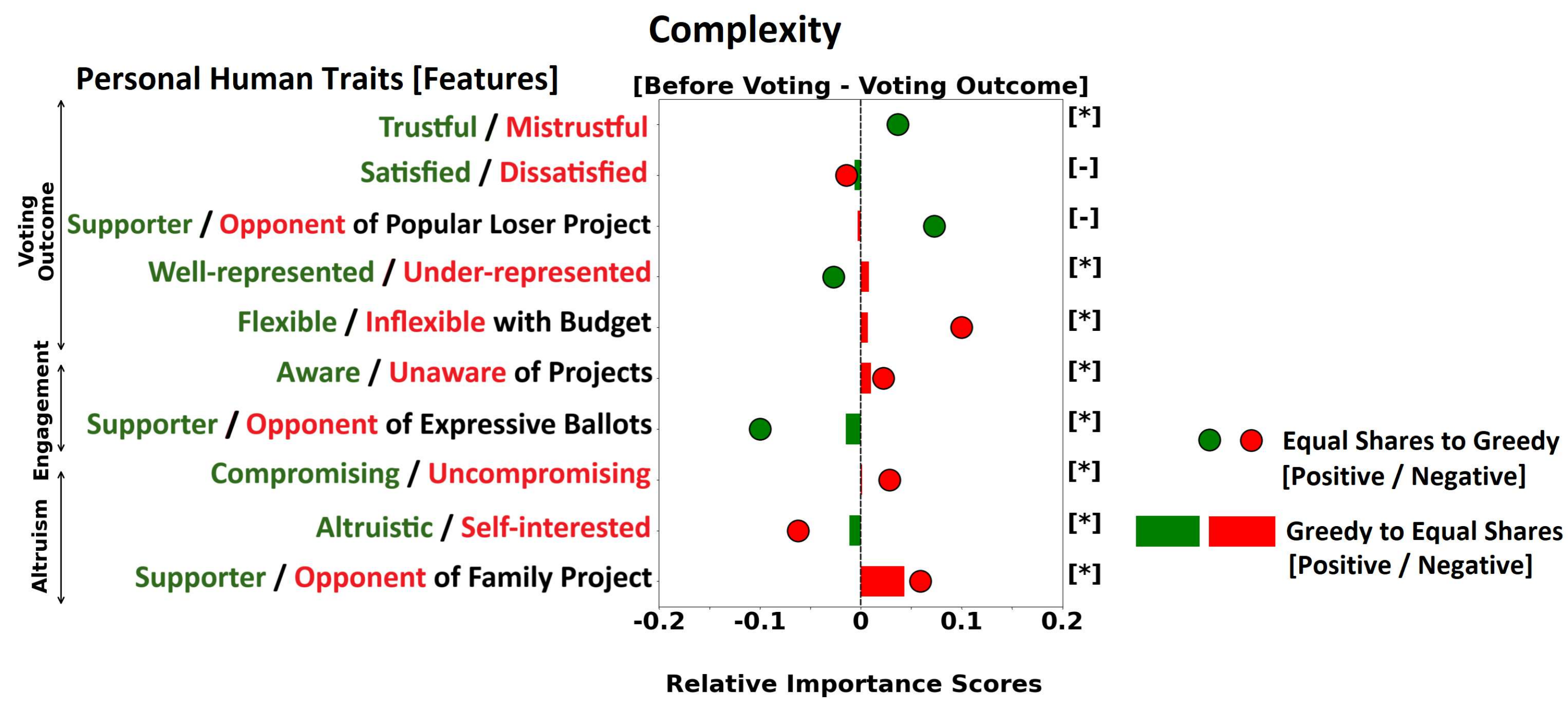}
\caption{\textbf{
The voters who are uncompromising prefer the less complex standard method compared to equal shares. On the contrary, voters who are not well-represented finds equal shares more relevant and not  complex.} Classification model (with 74.02\% accuracy, refer Table~\ref{tab:accuracy_results}), representing the preferred ballot aggregation method before and after voting (with explanation of methods). The independent variables represent voter-specific human traits that explain complexity both before and after voting, particularly in the transitions from the standard method to equal shares and vice versa. The `*' determines whether each feature is statistically significant (p<0.05) in explaining the voters' choice of method before and after voting. The analysis shown here is only for the two  scenarios - standard to equal shares and equal shares to standard.}
\label{fig:Convenience_Explain}
\end{figure}

We present the relevant personal human traits and their importance in terms of support and non-support across different phases, further segregating them based on the type of shift (class-wise). The significance of these traits in characterizing shifts from greedy to equal shares and from equal shares to greedy, particularly in the context of perceived fairness and implicit voter preference for the method, is illustrated in Figure~\ref{fig:ES_G_G_ES}. The shifts between choices of the aggregation methods in the context of the complexity of the methods have been explained in Figure~\ref{fig:Convenience_Explain}. Additionally, the explainability of non-shifts in choice of aggregation methods such as greedy to greedy or equal shares to equal shares, has been analyzed in Figure~\ref{fig:ES_G}.

\section{Communication and Explanation of Equal Shares}\label{sec:equal-shares-explanation}

\begin{figure}[h!]
\centering
\includegraphics[width=0.4\linewidth]{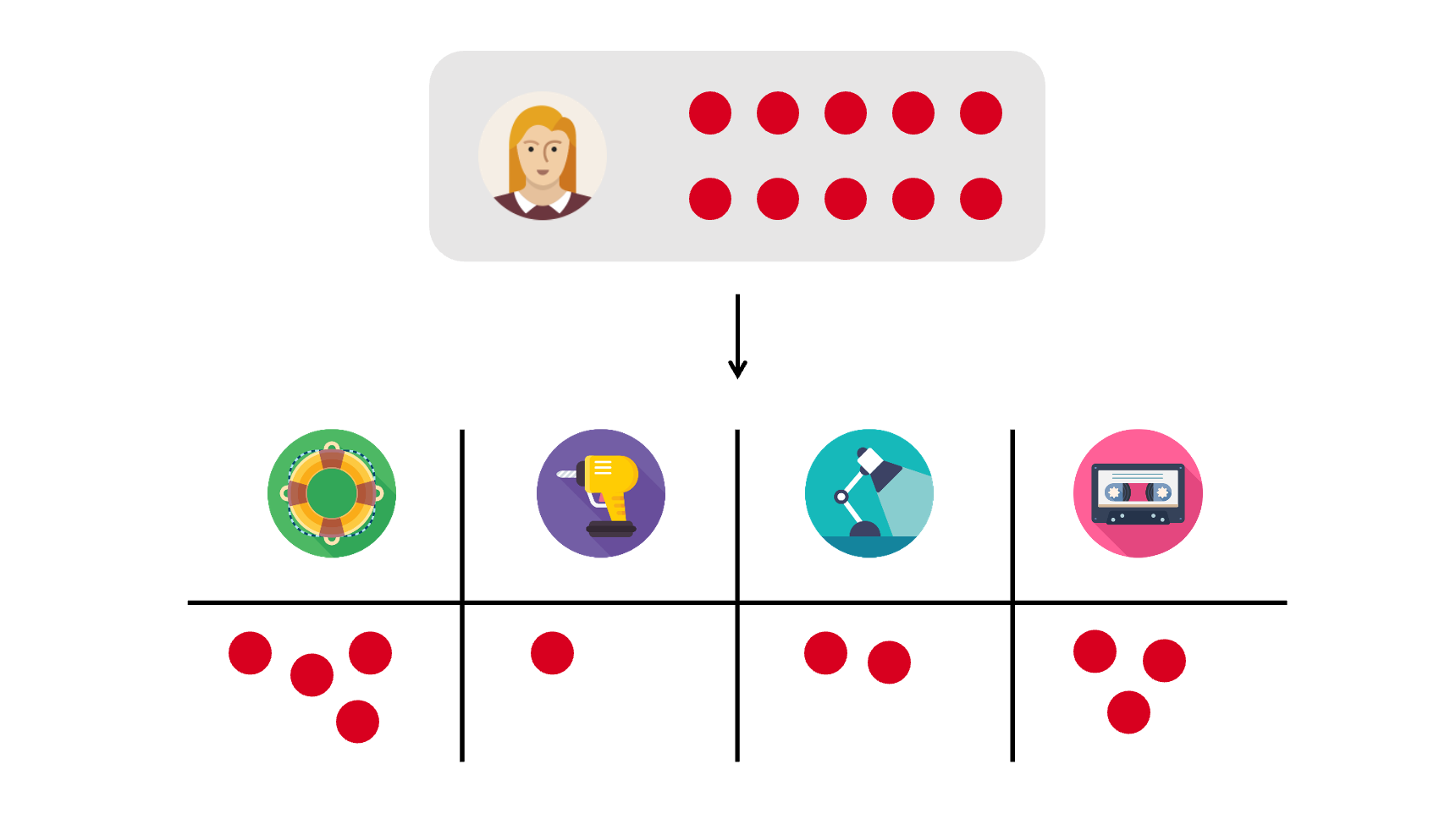}
\caption{Visualization of voting input instruction using dot stickers.}
\label{fig:cumulative}
\end{figure}

\begin{figure}[h!]
\centering
\includegraphics[width=0.9\linewidth]{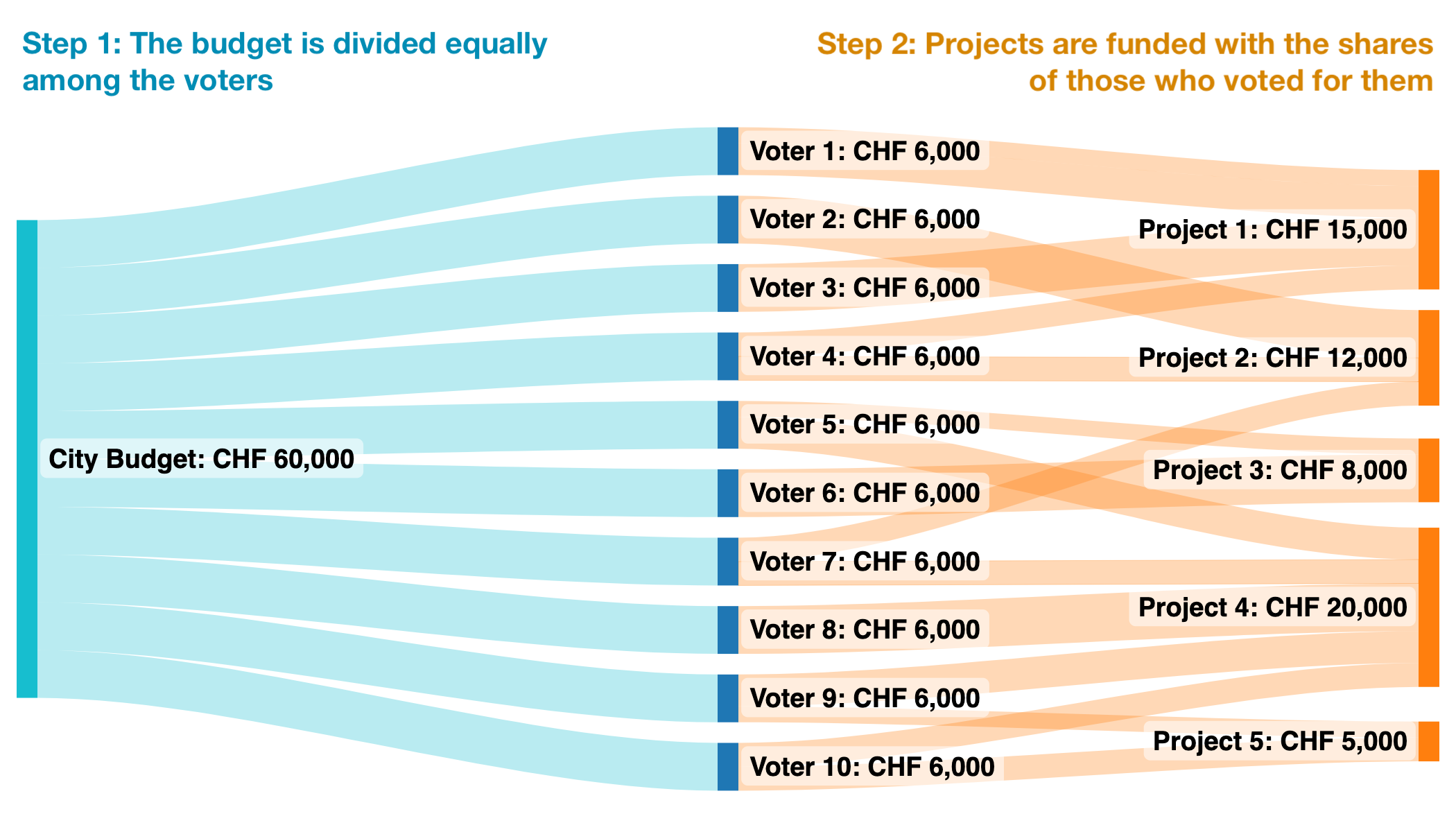}
\caption{Visualization of the underlying philosophy behind the method of equal shares.}
\label{fig:mes-flow}
\end{figure}

\begin{figure}[h!]
\centering
\begin{tabular}{ccc}
\includegraphics[width=0.3\linewidth]{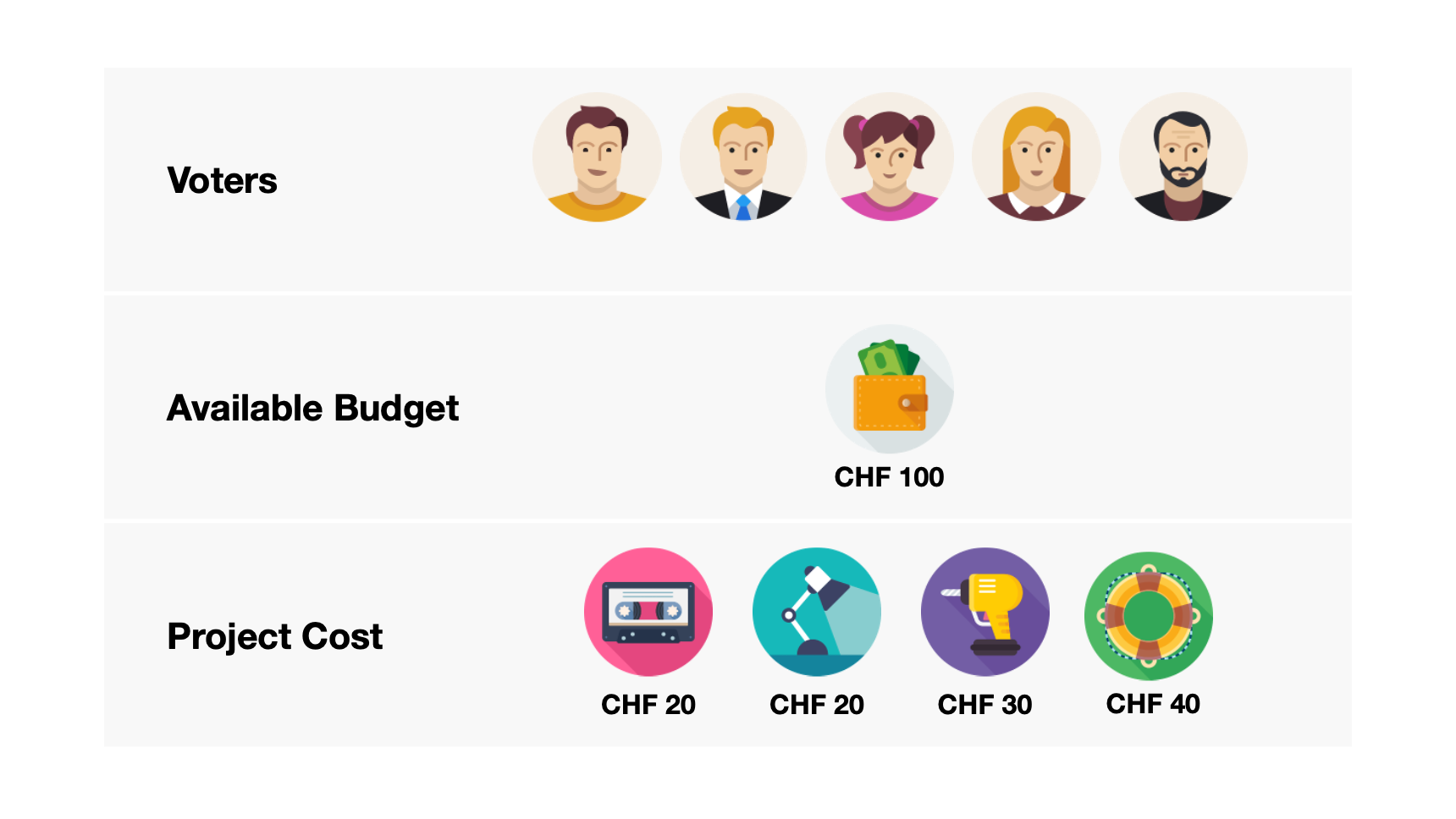} &
\includegraphics[width=0.3\linewidth]{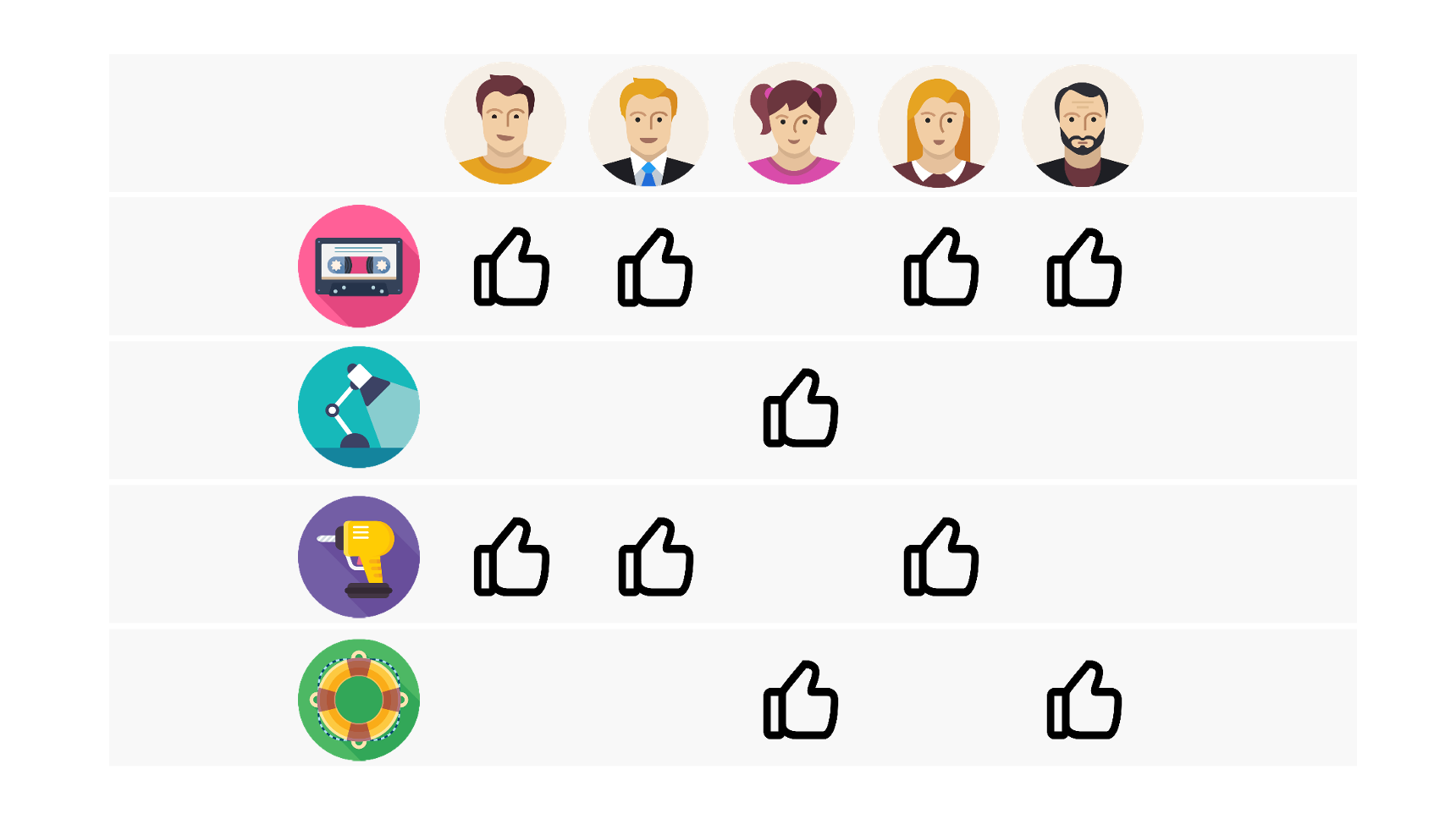} &
\includegraphics[width=0.3\linewidth]{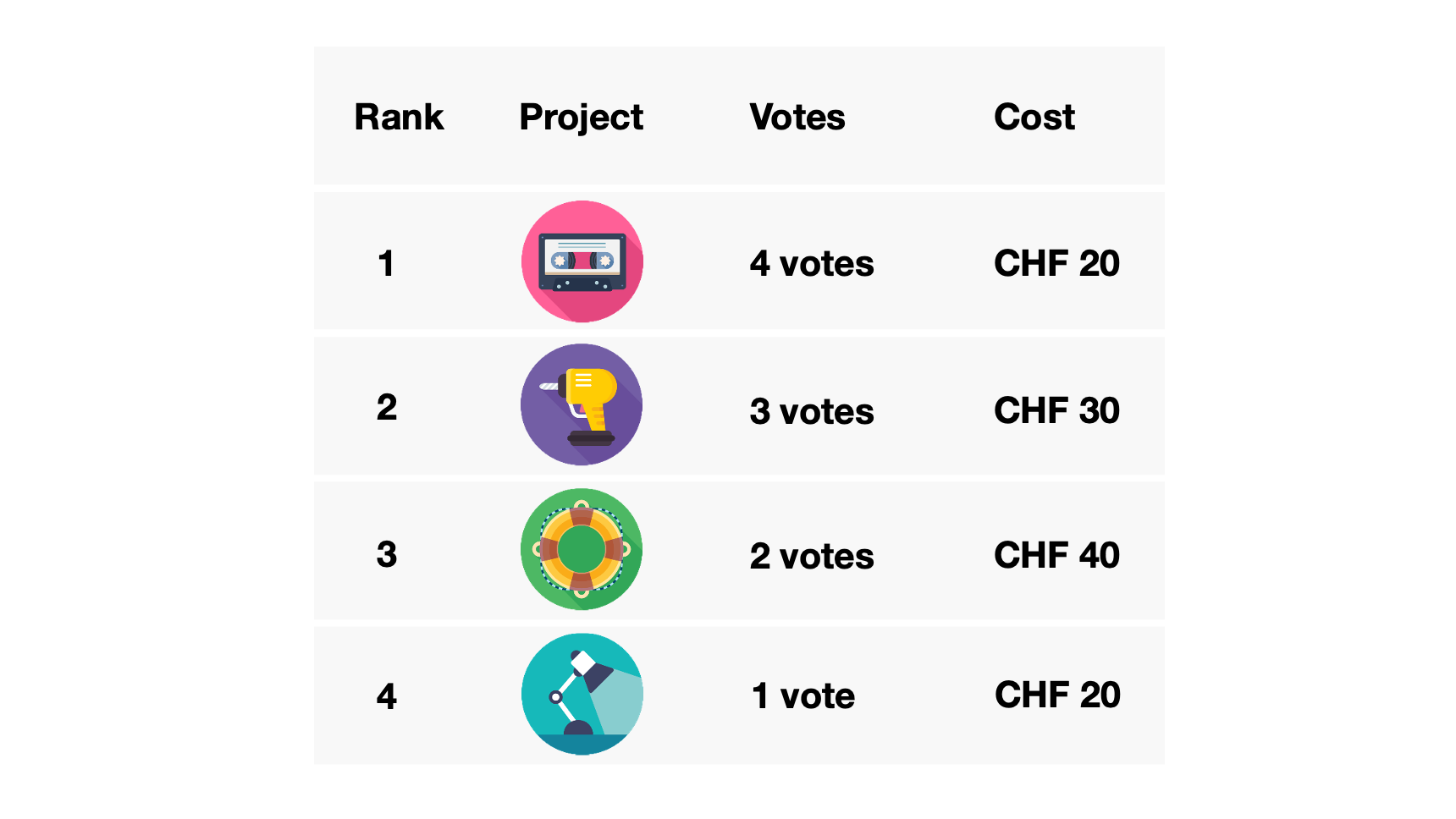} \\
\includegraphics[width=0.3\linewidth]{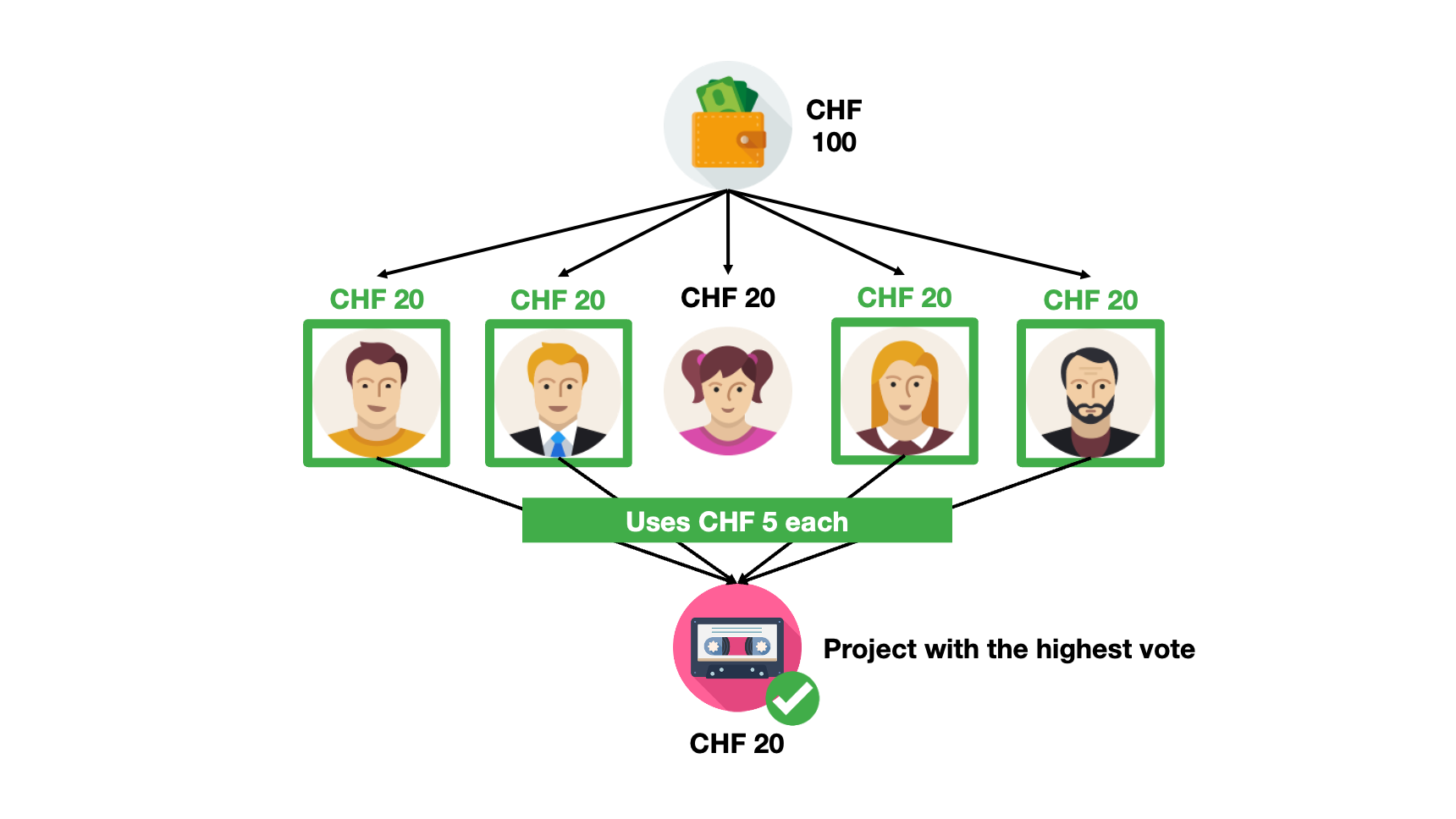} &
\includegraphics[width=0.3\linewidth]{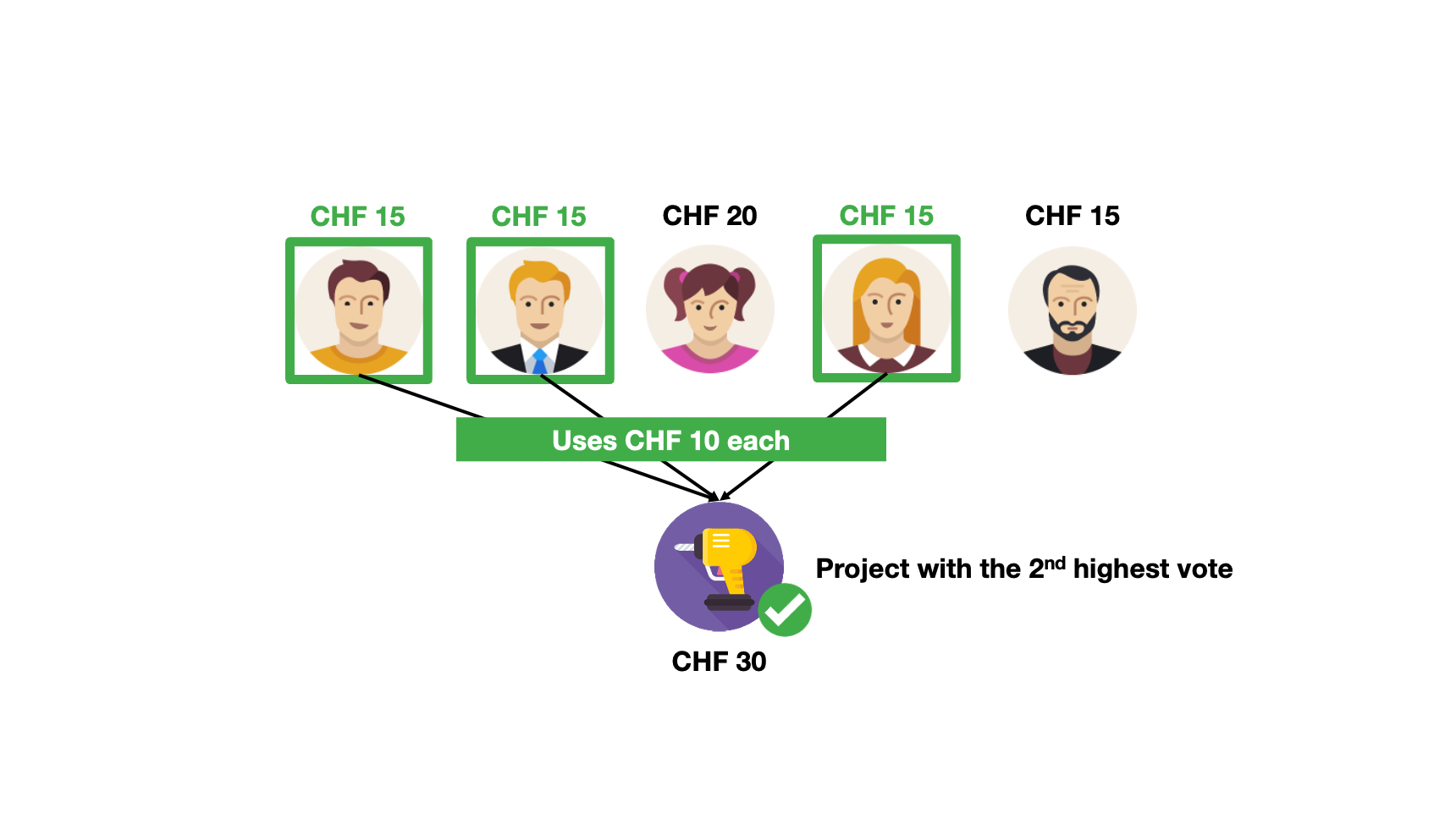} &
\includegraphics[width=0.3\linewidth]{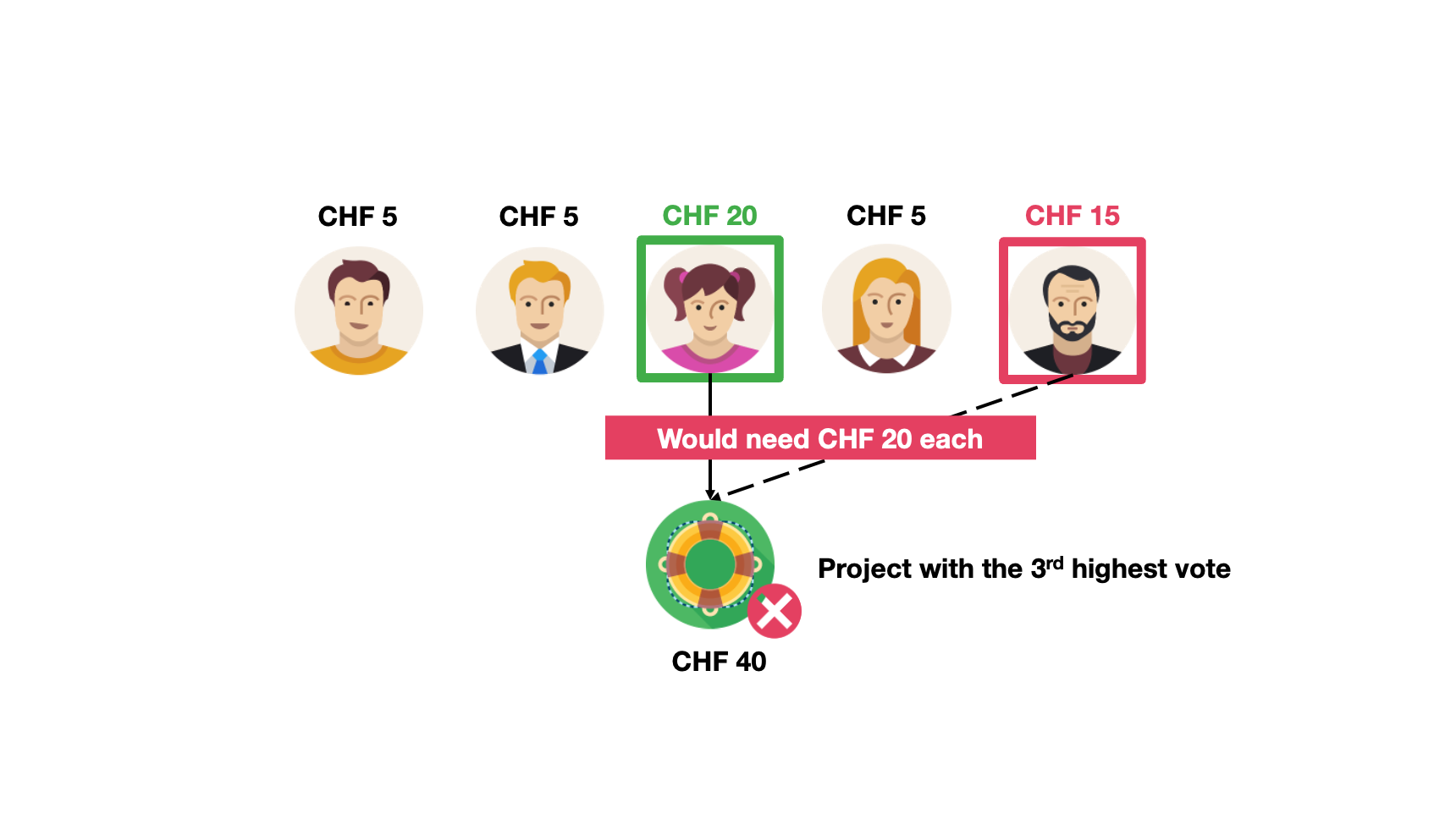} \\
\includegraphics[width=0.3\linewidth]{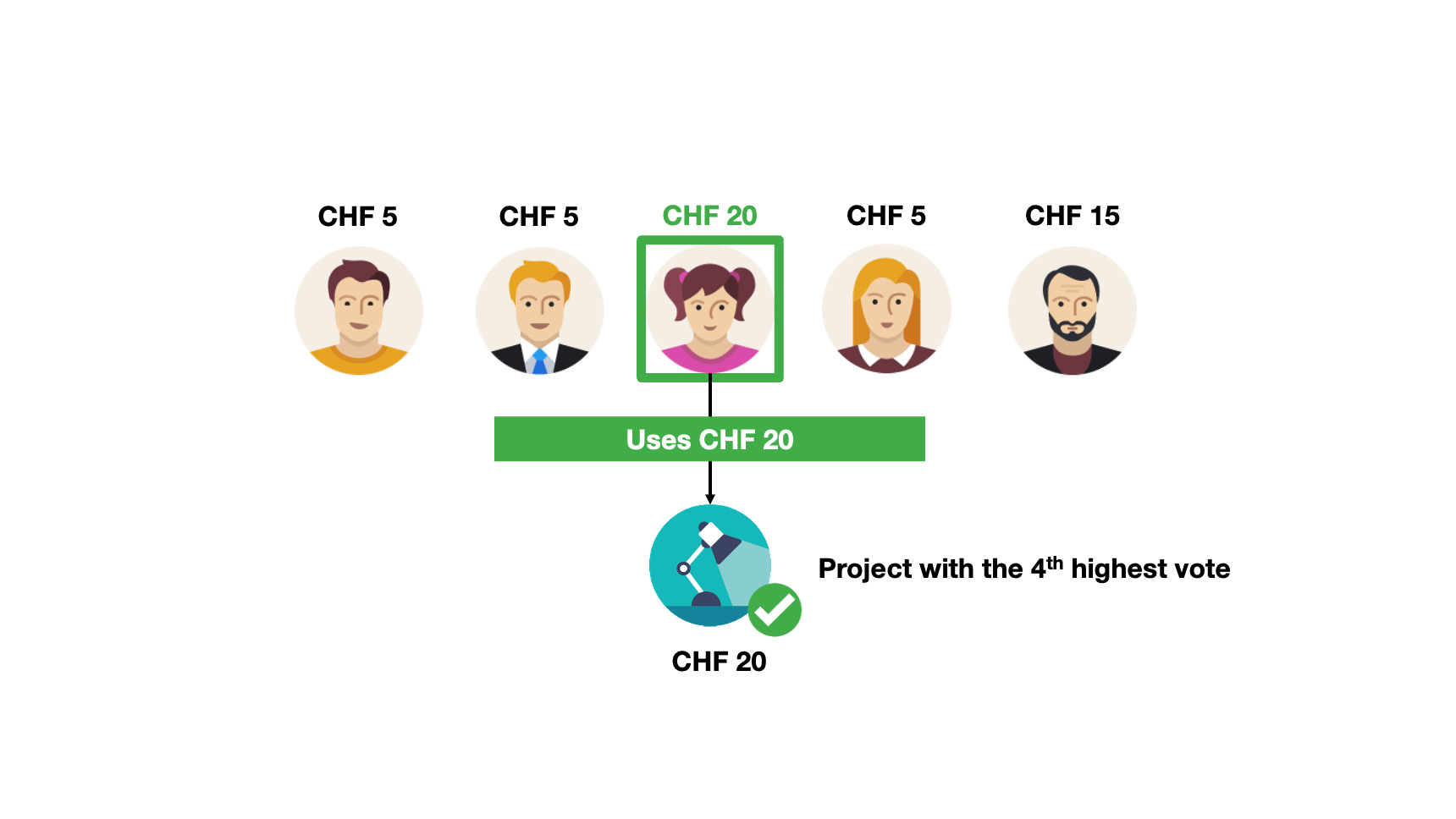} &
\includegraphics[width=0.3\linewidth]{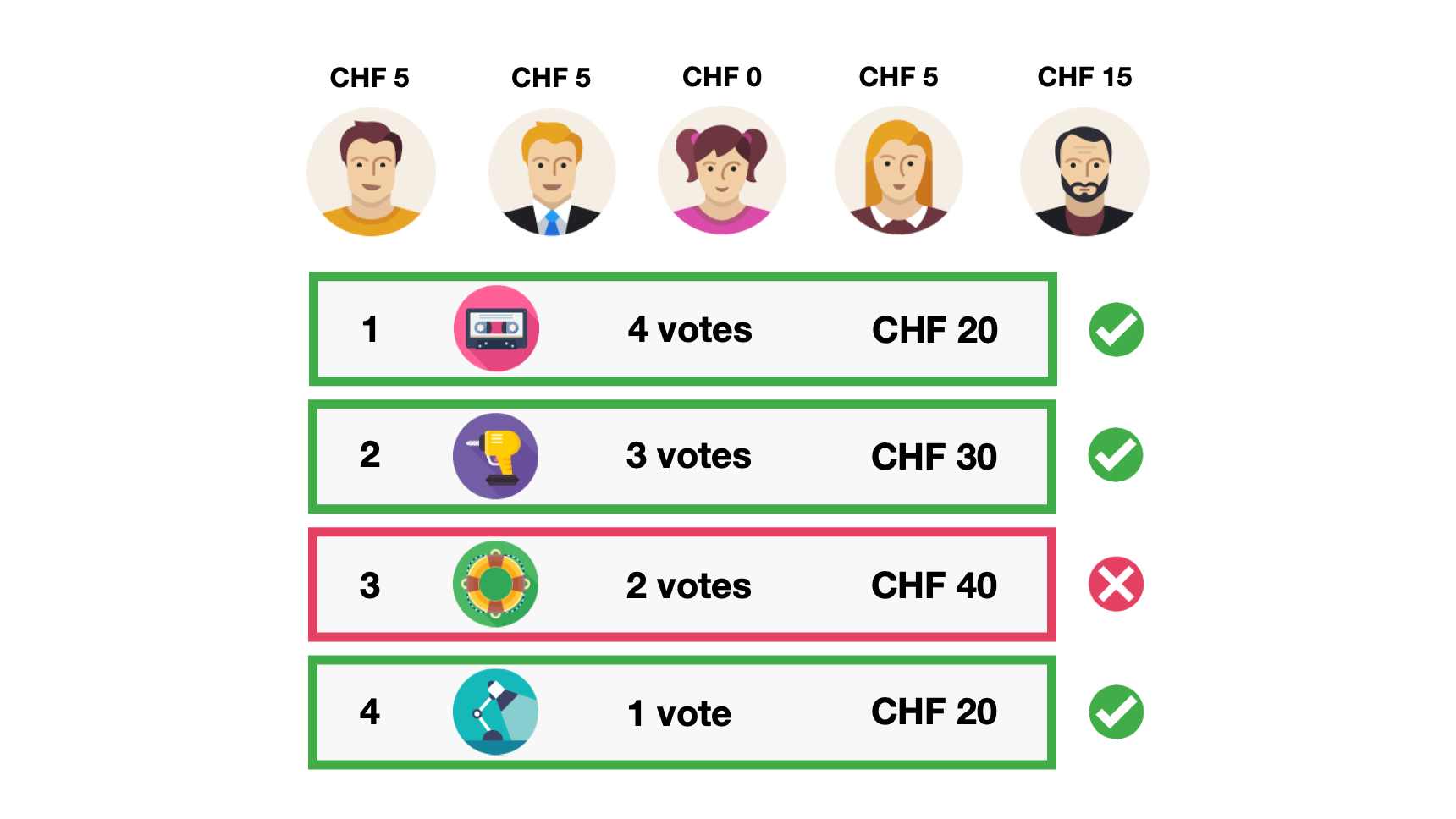} &
\includegraphics[width=0.3\linewidth]{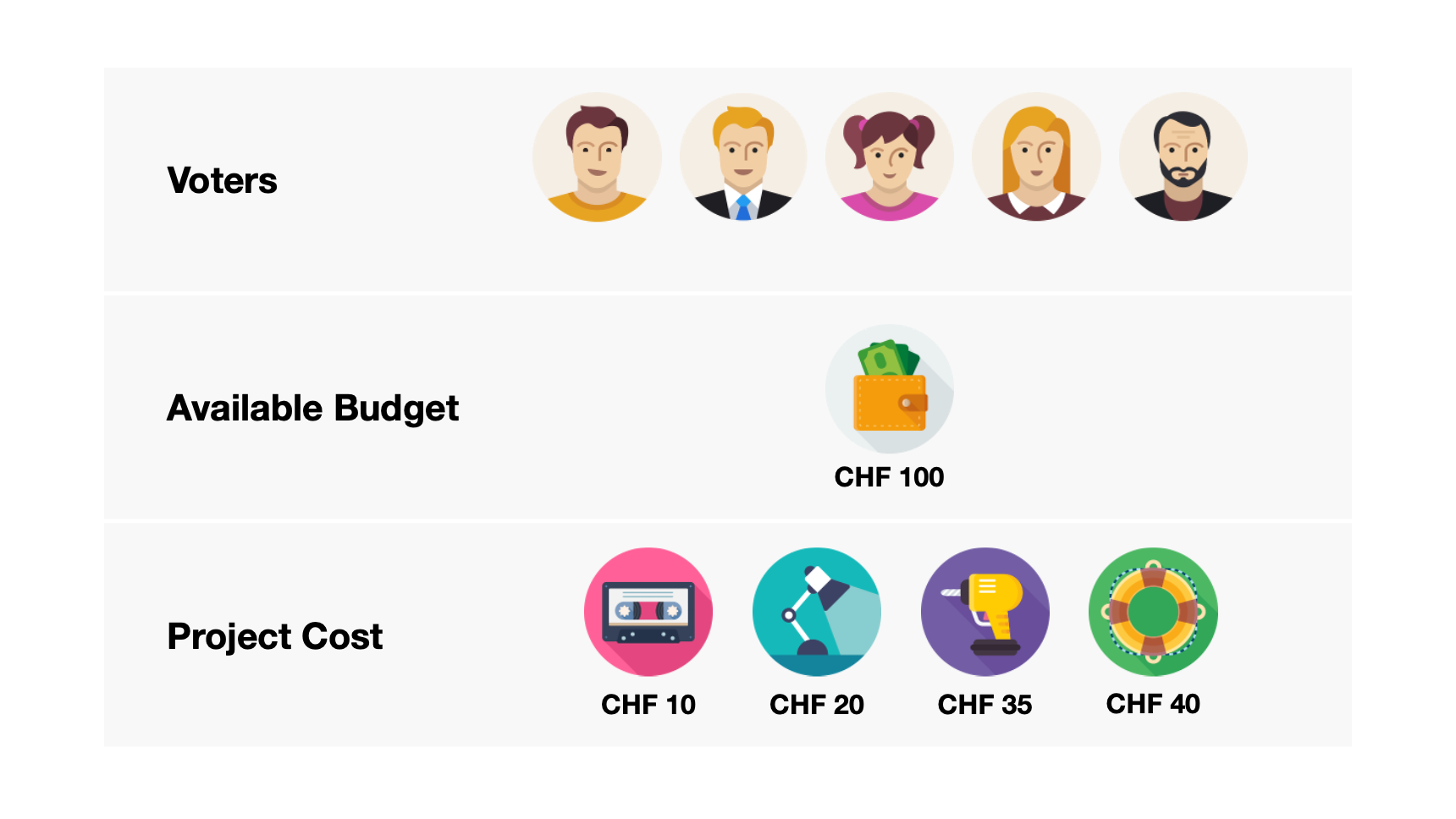} \\
\end{tabular}
\caption{Step-by-step visual representation of the method of equal shares mechanism.}
\label{fig:step}
\end{figure}

As citizens were unlikely to have encountered the Method of equal shares and cumulative voting prior to the Aarau PB vote, we applied Explainable AI strategies, focusing on transparency design and post-hoc explanations. Prior to the voting process, we conducted comprehensive pre-voting education through in-person meetings and an online platform. The emphasis was on providing straightforward graphical explanations of the voting process (e.g., Fig.~\ref{fig:cumulative}), along with simplified explanations for the method of equal shares (e.g., Fig.~\ref{fig:mes-flow}). This phase aimed at ensuring participants, especially idea owners, grasped the underlying mechanisms and principles of the voting methods. We also adopted an approach to explain the process using "explanation-by-example", as highlighted in Fig.~\ref{fig:step}. The diagram presented a simple scenario with voters, budgets, and project costs, using avatars, icons, and a concluding ranking table to simplify the workings of the mechanism.\\\\

\begin{figure}[h!]
\centering
\includegraphics[width=0.8\linewidth]{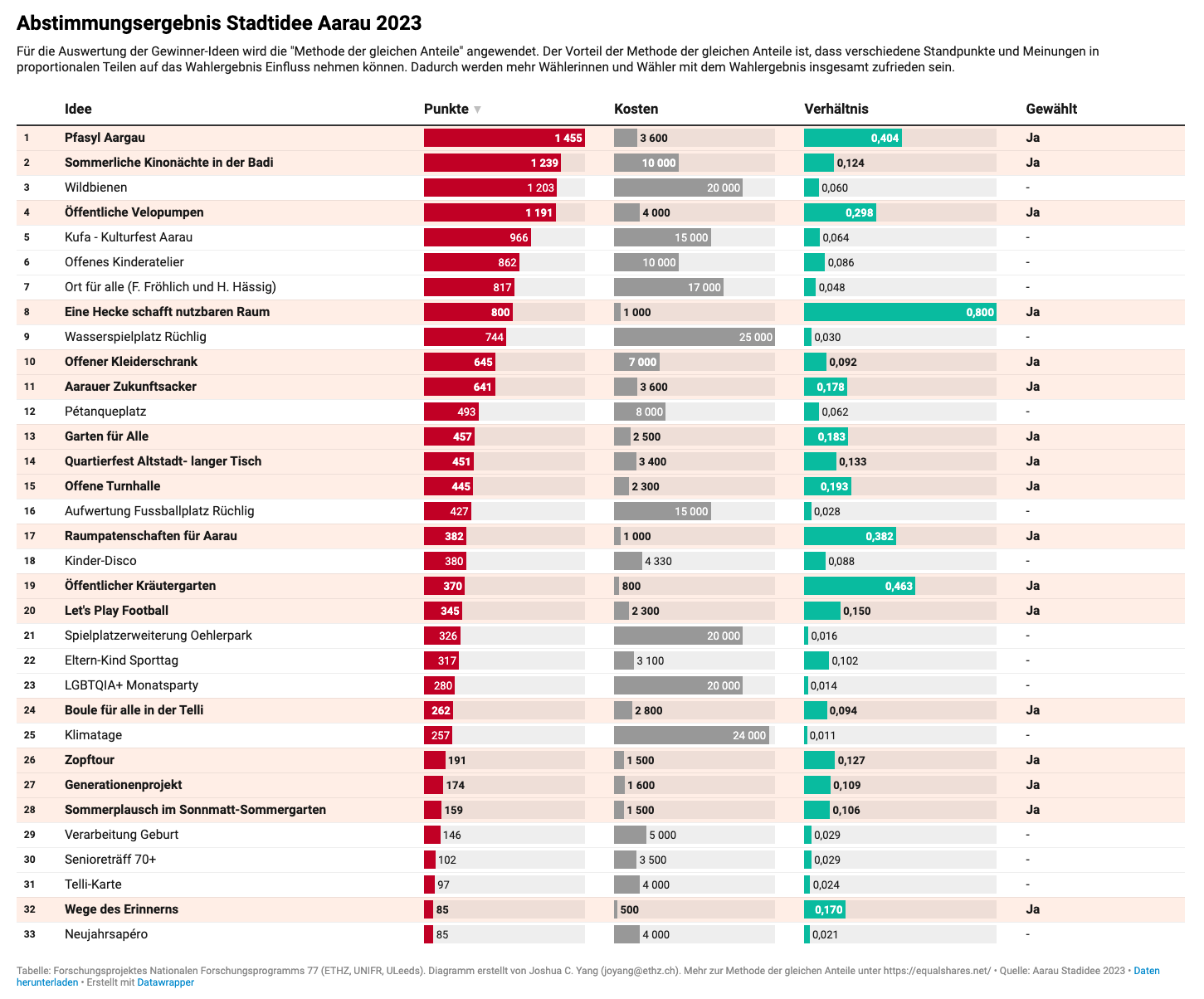}
\caption{Results of the Aarau city voting 2023, illustrating the idea proposals, their corresponding points, costs, ratios, and the decisions whether they were selected ("Ja" meaning "Yes").}
\label{fig:table}
\end{figure}

\begin{figure}[h!]
\centering
\includegraphics[width=0.9\linewidth]{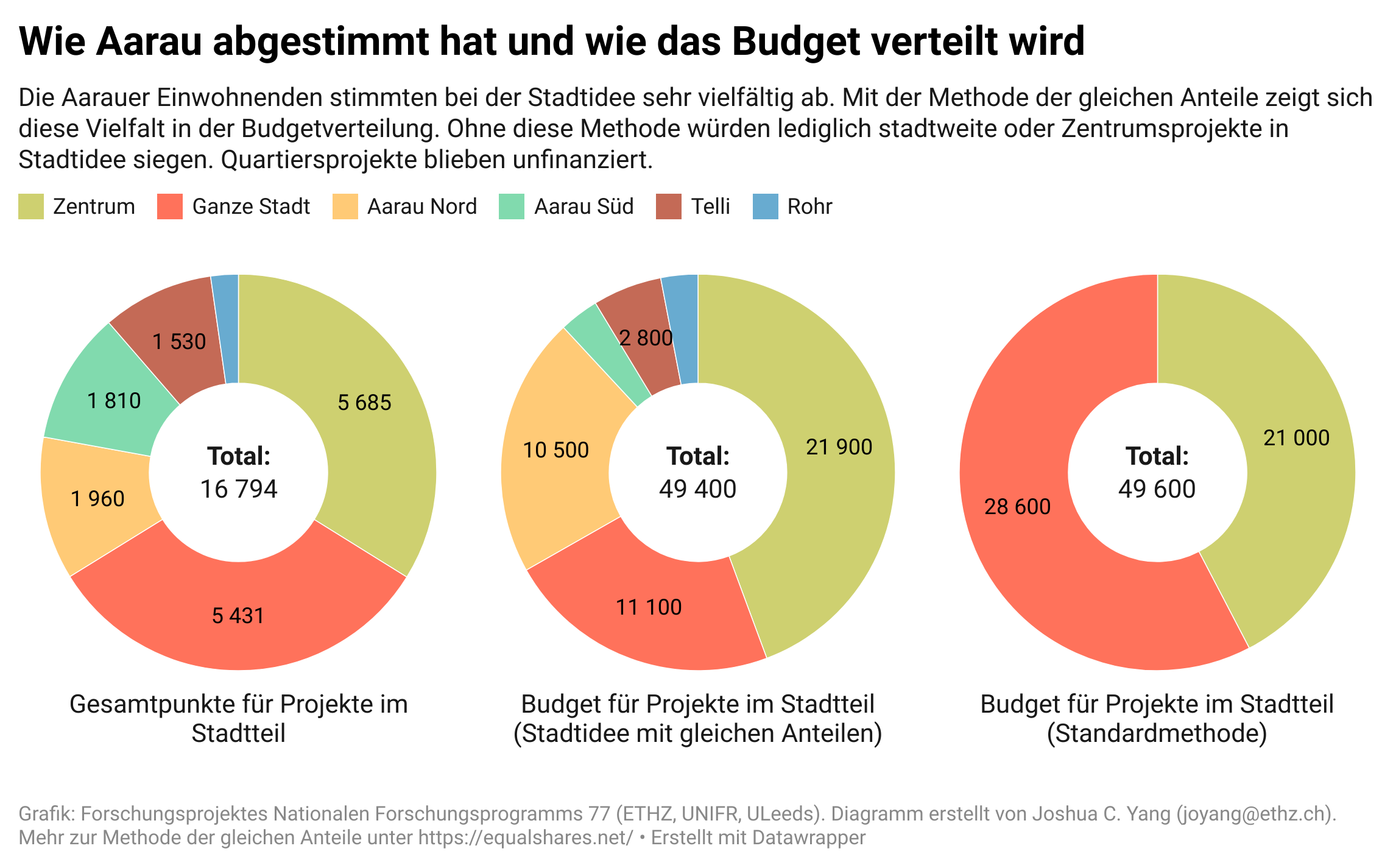}
\caption{Distribution of votes and budget allocation across various districts of Aarau, comparing the outcomes of the method of equal shares with the Standard method.}
\label{fig:district}
\end{figure}

\begin{figure}[h!]
\centering
\includegraphics[width=0.9\linewidth]{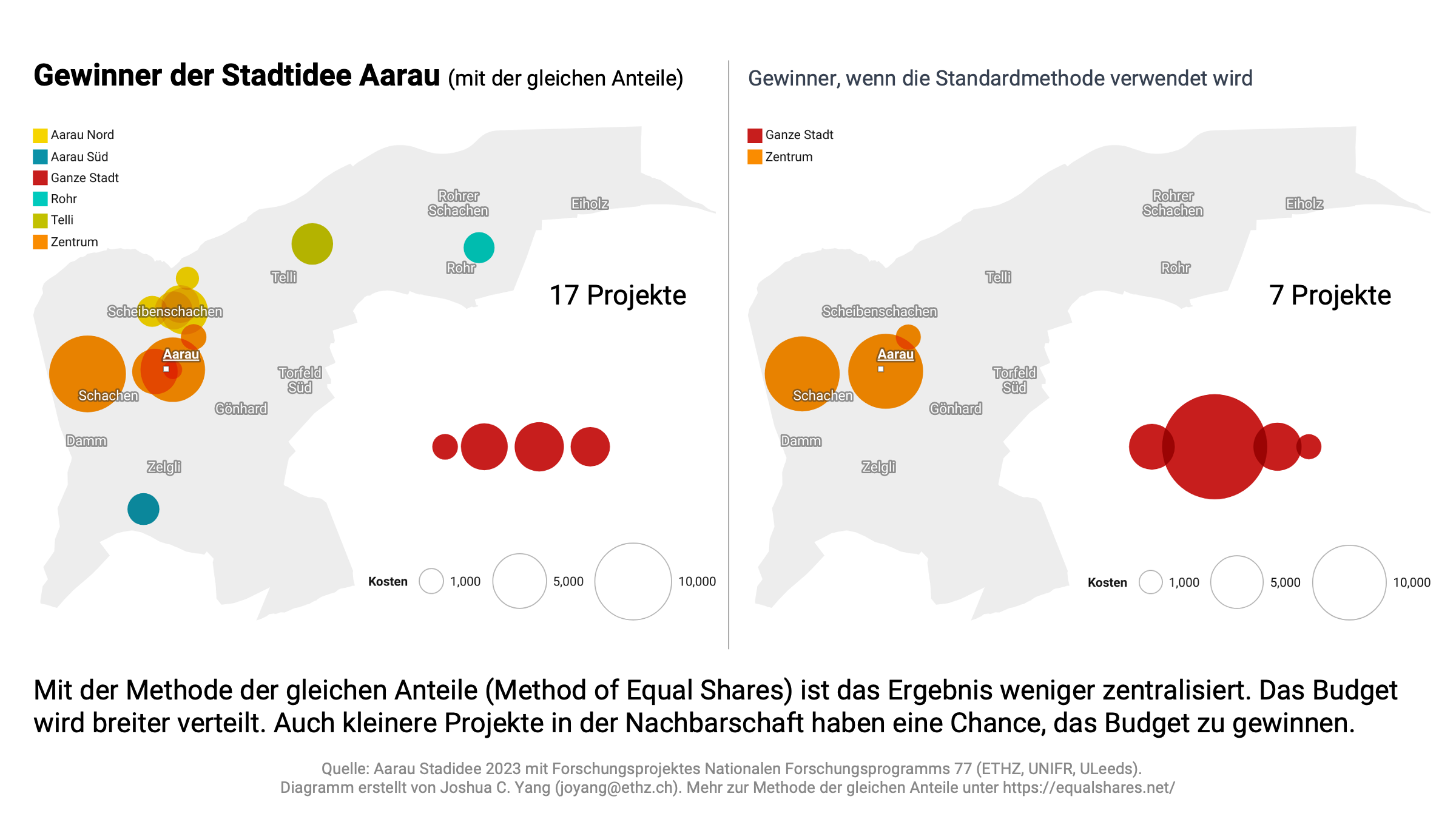}
\caption{Comparison of project winners across districts of Aarau using the method of equal shares versus the Standard method. The former results in a wider distribution, allowing smaller community projects a chance to secure funding.}
\label{fig:map}
\end{figure}

Post-vote, explanations were delivered during a media conference and made available online. Recognizing the challenges citizens might face in understanding the method of equal shares algorithm we primarily made the post-voting explanations outcome-centric. For example, Fig.~\ref{fig:table} presented an outcome table showcasing the final PB results. The table illustrated idea proposals, corresponding points, costs, ratios, and decisions regarding citizen's selections. As the method of equal shares is considered a `fairer' aggregation method, we also wanted to demonstrate the selection of winning projects as a reflection of spatial fairness. As such, Fig.~\ref{fig:district} and Fig.~\ref{fig:map} provides insights into how residents voted and how the budget was allocated across districts using the method of equal shares compared to the Standard method.\\

\clearpage	

	\bibliography{bibliography}

    \bibliographystyle{naturemag}
\makeatletter\@input{xx.tex}\makeatother